\documentclass[aps,prx,twocolumn,superscriptaddress,floatfix]{revtex4-2}
\usepackage[dvipsnames,svgnames]{xcolor}
\usepackage{adjustbox}
\usepackage{graphicx}
\usepackage{dcolumn}
\usepackage{bm}
\usepackage[inkscapearea=page]{svg}
\usepackage{soul}
\usepackage[T1]{fontenc}
\usepackage{amstext}
\usepackage{amsthm}
\usepackage{amsmath}
\usepackage{amsfonts}
\usepackage{amssymb}
\usepackage{float}
\usepackage{appendix}
\usepackage{multirow}
\usepackage{physics}
\usepackage[bookmarks,bookmarksnumbered]{hyperref}
\usepackage{bookmark}
\hypersetup{colorlinks,allcolors=black}
\usepackage{enumitem}
\usepackage{mathtools}
\usepackage[ruled,linesnumbered]{algorithm2e}
\usepackage[text]{esdiff}
\usepackage{setspace}
\usepackage[singlelinecheck=off]{subcaption}
\usepackage{siunitx}
\usepackage{xspace}
\usepackage{soul}
\setstcolor{gray}
\definecolor{bgcolor}{RGB}{241, 241, 232}
\theoremstyle{definition}
\newtheorem{definition}{Definition}
\newtheorem{notation}{Notation}

\theoremstyle{remark}
\newtheorem{example}{Example}
\newtheorem{remark}{Remark}

\newcommand{\dx}{\dd{x}}
\newcommand{\dt}{\dd{t}}
\newcommand{\dtau}{\dd{\tau}}
\newcommand{\logm}{\operatorname{logm}}
\newcommand{\spanof}{\mathrm{span}}
\newcommand{\rad}{\,\mathrm{rad}}
\newcommand{\nanosec}{\,\mathrm{ns}}
\newcommand{\iswap}{\textrm{iSWAP}\xspace}
\newcommand{\robness}{\mathcal{R}}
\newcommand{\morlet}{\operatorname{Mlt}}
\newcommand{\optalgo}{OPT\xspace}
\newcommand{\varalgo}{VAR\xspace}

\newlength{\subfiglen}
\newlength{\sboxlen}
\newlength{\mpagelen}
\newlength{\myheight}

\DeclareDocumentCommand\DD{ o g d() }{ 
	\IfNoValueTF{#2}{
		\IfNoValueTF{#3}
			{\mathrm{D}\IfNoValueTF{#1}{}{^{#1}}}
			{\mathinner{\mathrm{D}\IfNoValueTF{#1}{}{^{#1}}\argopen(#3\argclose)}}
		}
		{\mathinner{\mathrm{D}\IfNoValueTF{#1}{}{^{#1}}#2} \IfNoValueTF{#3}{}{(#3)}}
	}

\begin{document}
\preprint{APS/123-QED}

\title{Traversing Quantum Control Robustness Landscapes: A New Paradigm for Quantum Gate Engineering}

\author{Huiqi Xue}
\author{Xiu-Hao Deng}
\email{dengxh@sustech.edu.cn}
\affiliation{International Quantum Academy, Shenzhen, Guangdong 518000, China}
\affiliation{Shenzhen Institute of Quantum Science and Engineering, Southern University of Science and Technology, Shenzhen, Guangdong 518055, China}
\date{\today}

\begin{abstract}
  The optimization of robust quantum control is often tailored to specific tasks and suffers from inefficiencies due to the complexity of cost functions. Our recent findings indicate a highly effective methodology for the engineering of quantum gates by initiating the process with a robust control configuration of any arbitrary gate. We first introduce the Quantum Control Robustness Landscape (QCRL), a conceptual framework that maps control parameters to noise susceptibility. This framework facilitates a systematic investigation of equally robust controls for diverse quantum operations. By navigating through the level sets of the QCRL, our Robustness-Invariant Pulse Variation (RIPV) algorithm allows for the variation of control pulses while preserving robustness. Numerical simulations demonstrate that our single- and two-qubit gates exceed the quantum error correction threshold even with substantial noise. This methodology opens up a new paradigm for quantum gate engineering capable of effectively suppressing generic noise.
\end{abstract}
\maketitle

\section{Introduction}

Quantum control plays a pivotal role in the advancement of quantum technologies, including quantum computing, quantum communication, and quantum sensing. However, quantum systems are highly susceptible to noise and environmental disturbances, making it challenging to maintain the desired level of control~\cite{cheng2023noisy,campbellRoadsFaulttolerantUniversal2017}. This issue is particularly pronounced in the Noisy Intermediate-Scale Quantum (NISQ) era, where quantum devices operate with limited resources and are prone to various types of noise. As such, the development of robust quantum control methods is crucial to achieving fault-tolerant quantum computing~\cite{koch2022quantum}.

Traditionally, quantum control has been studied through the framework of Quantum Control Landscape (QCL)~\cite{chakrabarti2007quantum, ge2022optimization}, which maps control parameters to objective functions like fidelity. The exploration of level sets has provided a valuable framework for optimizing control fields considering multiple merits including fidelity, gate time~\cite{rothmanExploringLevelSets2006,dominyExploringFamiliesQuantum2008,chen2015near}. Applying optimal control on realistic systems is challenged by noise that includes disturbance from the environment, parameter uncertainty, crosstalk, control imperfection, and so on. Robust quantum control, which aims to minimize the impact of noise while maintaining operational accuracy, has been a key area of research to address this challenge. Various techniques have been developed to enhance control robustness, such as dynamical decoupling~\cite{souza2011robust}, composite pulse sequences~\cite{genov2014correction}, geometric gates~\cite{zhang2023geometric} and dynamically corrected gates~\cite{khodjasteh2009dynamically, zengGeneralSolutionInhomogeneous2018}. However, the focus on fidelity optimization often overlooks the importance and the independency of noise robustness. And there has been limited investigation into the landscape properties of robustness itself~\cite{dong2023learning,cao2024robust,kosut2022robust}.

In this work, we introduce the concept of Quantum Control Robustness Landscape (QCRL), a novel framework that emphasizes robustness over fidelity. Unlike traditional QCL based on the fidelity to an ideal gate, QCRL maps control parameters to the robustness of quantum operations against noise. This shift in focus is particularly relevant in the NISQ era, where noise significantly limits the performance of quantum systems. In this way, QCRL provides a new way to assess and optimize the performance of quantum gates and control pulses in the presence of noise.

\begin{figure}[bt]
  \centering
  \includegraphics[width=0.35\textwidth]{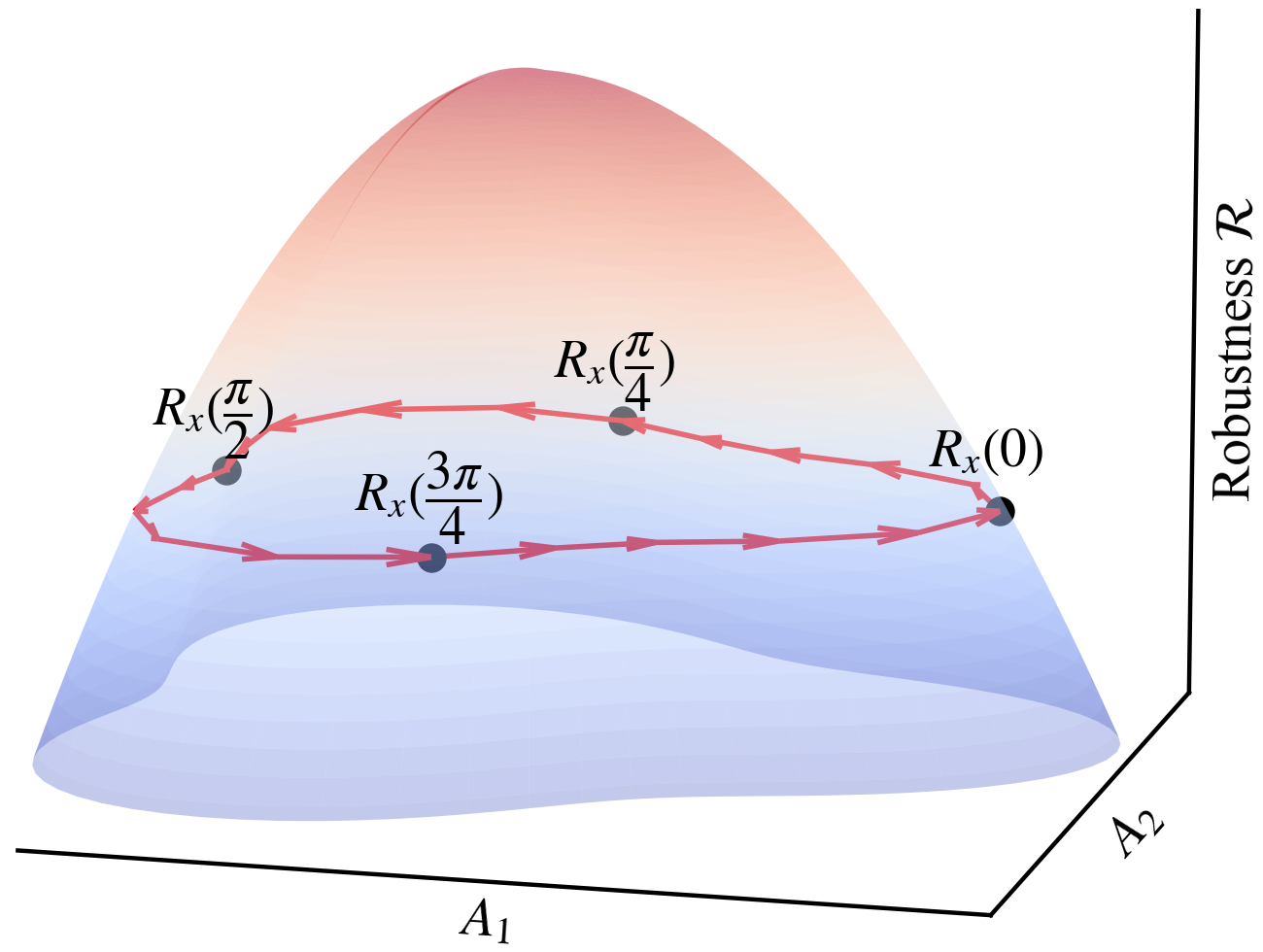}
  \caption{Schematic illustration of a level set exploration on a toy robustness landscape. \(A_1\) and \(A_2\) represent control parameters, while \(\mathcal{R}\) represents robustness to the system's noise, a function of the control.}
  \label{fig:toy-landscape}
\end{figure}

We also propose an algorithm called Robustness-Invariant Pulse Variation (RIPV), which traverses a \emph{level set} of a QCRL, a subset of control pulses that yield the same robustness.
By traversing the level sets of a QCRL, we can find robust control pulses far more efficiently than traditional methods.
An oversimplified illustration is provided in \autoref{fig:toy-landscape}, where \(A_1\) and \(A_2\) are two control parameters and the level set is represented by red arrows.
Our QCRL framework allows us to traverse the level set to find equally robust control pulses for different quantum gates. Starting from the control implementing \(R_x(0)\), we move step by step (each red arrow representing one step) to find robust controls for other \(R_x(\theta)\) gates up to \(R_x(2\pi)\). This method is more efficient than traditional QCL frameworks, as it propagates the robustness of \(R_x(0)\) to other gates. Most importantly, it enables a once impossible task -- optimizing a gate family.

\subparagraph*{Advantages of the QCRL framework:}
\begin{itemize}
  \item \textit{Noise-centered perspective.} The QCRL is defined by a noise model, rather than an ideal gate. It is more realistic as a mathematical tool for the NISQ era.
  \item \textit{Unified treatment.} In the QCRL, we can deal with all controllable gates in a unified setting without changing the objective.
  \item \textit{Suitable for the control of gate family.} Using the RIPV algorithm, we can implement robust control of gate families (such as parametric gates) with optimized control parameters.
  \item \textit{Generalizable algorithm design.} It does not assume the physical properties of the system. In principle, the RIPV algorithm can be easily generalized to multi-qubit gates and more complex noise models.
  \item \textit{Decoupling multipleobjectives.} RIPV is not limited to maintaining robustness. For any multi-objective optimization problem, it can optimize one objective without undermining the others, effectively decoupling them.
\end{itemize}

In the following sections, we present the mathematical formalism of the QCRL, introduce the RIPV algorithm, and provide numerical examples to demonstrate its effectiveness. This work aims to open new possibilities in quantum control by offering a robust and flexible approach to pulse engineering in noisy quantum systems.

\section{Preliminaries}

Our QCRL is built upon two fields: QCL and robust quantum control, which we introduce briefly in this section. Then, we discuss an important control task that is made possible for the first time by our work: the robust control of gate families.

\subsection{Quantum control landscape}

The study of QCL began in the late 1990s when researchers discovered that optimizing the objectives in quantum control was surprisingly easy as the algorithms rarely got stuck in a local extremum. Rabitz \textit{et al.}~\cite{chakrabarti2007quantum} formalized the idea of QCL as a framework that maps control parameters (e.g., pulse amplitudes, phases, or durations) to an objective function, such as gate fidelity, observable expectation, or population of state transfer. While other metrics, such as robustness, leakage, etc., are considered as constraints during the optimization on the QCL. This mapping creates a ``landscape'' in which the input space is determined by control parameters, and the output space is determined by the value of the objective function. The structure of this landscape is crucial for understanding and optimizing quantum control.

The prerequisite for studying a QCL is the \textit{controllability} of the underlying quantum system. A quantum system is \emph{fully controllable} if it can be steered to any desired state from any given initial state. As one of the most important results, Ramakrishna et al.~\cite{ramakrishna1995controllability} stated that an \(N\)-level system is fully controllable if the Lie algebra generated by the system and control Hamiltonians has dimension \(N^2\). More thorough investigation is provided by Fu et al.~\cite{fu2001complete}. There are many other discussions such as the controllability of multiple transitions~\cite{schirmer2001complete} and classification of uncontrollable systems~\cite{polack2009uncontrollable}.

A key finding is that \emph{local traps}, i.e. local extrema, are rarely encountered when optimizing quantum control objectives in practical applications. This contrasts sharply with classical optimization and has driven research into QCLs~\cite{rabitzQuantumOptimallyControlled2004}, particularly regarding trap-free conditions. It has been proven that QCLs for regular controls, characterized by local surjectivity onto \(U(N)\), are trap-free~\cite{hoLandscapeUnitaryTransformations2009}. The regularity of controls can also be described as being able to access the entire \(U(N)\) group within a finite time \(T\)~\cite{brifControlQuantumPhenomena2010}. However, extreme conditions -- such as forbidden level transitions~\cite{pechenAreThereTraps2011} or optimization with constraints~\cite{defouquieresCloserLookQuantum2013,birteaConstraintOptimizationSU2022} -- can lead to critical points that are not globally optimal. Nonetheless, these critical points do not contradict the general conclusion that regular controls are trap-free~\cite{rabitzCommentAreThere2012}, nor do they manifest as local traps in practical settings; rather, they typically appear as saddle points~\cite{wuSingularitiesQuantumControl2012,russellControlLandscapesAre2017}.

Beside those properties, the \emph{level sets} of QCLs played an important role. A level set consists of all the control fields yielding the same value of the objective function. Level sets are especially useful for refining control solutions according to additional criteria. For example, the D-MORPH algorithm~\cite{rothmanExploringLevelSets2006} and its unitary variant~\cite{dominyExploringFamiliesQuantum2008} allow for systematic exploration of level sets, enabling the optimization of secondary objectives such as gate time while maintaining transition probability or unitary gate. Pechen et al.~\cite{pechen2008control} demonstrated that the level sets in two-level quantum systems are connected, further validating the efficacy of level set exploration algorithms for quantum control tasks. This ability to navigate level sets makes QCL a versatile framework for addressing multi-objective optimization problems in quantum systems.

Real-world applications involve multiple objectives (e.g., fidelity, robustness, gate time), but multi-objective optimization doesn't have a single optimal solution. Instead, the goal is to approach the Pareto front, where improving one objective harms another. While direct calculation of the Pareto front is difficult, iterative updates can move the solution closer. Aggregating objectives, such as taking the sum of objectives, can complicate the landscape and prevent convergence. By traversing the level set of a landscape of primary interest while optimizing additional criteria simultaneously, progress towards the Pareto front can still be made.

\subsection{Robust quantum control}

Robust quantum control is an important discipline within quantum information science that focuses on maintaining the fidelity of quantum operations in the presence of various types of noise and uncertainties inherent in quantum systems~\cite{wu2019learning}. As quantum technologies advance, particularly in platforms such as superconducting qubits and solid-state spins~\cite{clarkeSuperconductingQuantumBits2008,krantz2019quantum}, the challenge of effectively managing noise - ranging from field (charge, flux, photon, etc.) fluctuations~\cite{kochChargeinsensitiveQubitDesign2007} to uncertain disturbances (crosstalk, unwanted couplings, etc.)~\cite{sarovar2020detecting,rudinger2021experimental,yi2024robust} - has become increasingly critical. Robust quantum control techniques aim to design control protocols that can withstand these disturbances, ensuring reliable performance of quantum gates and operations. Methods such as dynamical decoupling~\cite{Bylander2011}, composite pulse sequences~\cite{brownArbitrarilyAccurateComposite2004,shi2024supervised}, and advanced optimization strategies have been developed to enhance the resilience of quantum systems against noise~\cite{zhang2024smolyak,li2024experimental}. Furthermore, the introduction of a geometric framework for robust quantum control provides a powerful tool for visualizing and analyzing the robustness of quantum operations~\cite{zengGeneralSolutionInhomogeneous2018,barnesDynamicallyCorrectedGates2022,dong2021doubly,yi2024robust}. This framework not only facilitates a deeper understanding of the landscape of control parameters but also aids in the systematic design of control strategies that can effectively mitigate the impact of generic noise~\cite{haiUniversalRobustQuantum2023}.

Other than investigation on physical properties, robust quantum control is also realized by optimization algorithms~\cite{koch2022quantum}. The geometric framework can be incorporated into the gradient descent algorithm to obtain robust pulses~\cite{haiUniversalRobustQuantum2023}. Multi-stage optimization algorithms have been proposed for addressing multi-objective optimization problems, encompassing factors such as robustness~\cite{kosut2022robust}. A triobjective QCL framework is utilized to obtain the Pareto front of robustness and gate time~\cite{cao2024robust}.

\subsection{Search of robust gate families}
\label{sec:search-of-robust-gate-families}

\begin{figure}[ht]
  \centering

  \setlength{\subfiglen}{0.45\textwidth}

  \setlength{\mpagelen}{\subfiglen-13.34pt}
  \setlength{\sboxlen}{0.45\textwidth-13.34pt}
  \sbox0{\includegraphics[width=\sboxlen]
  {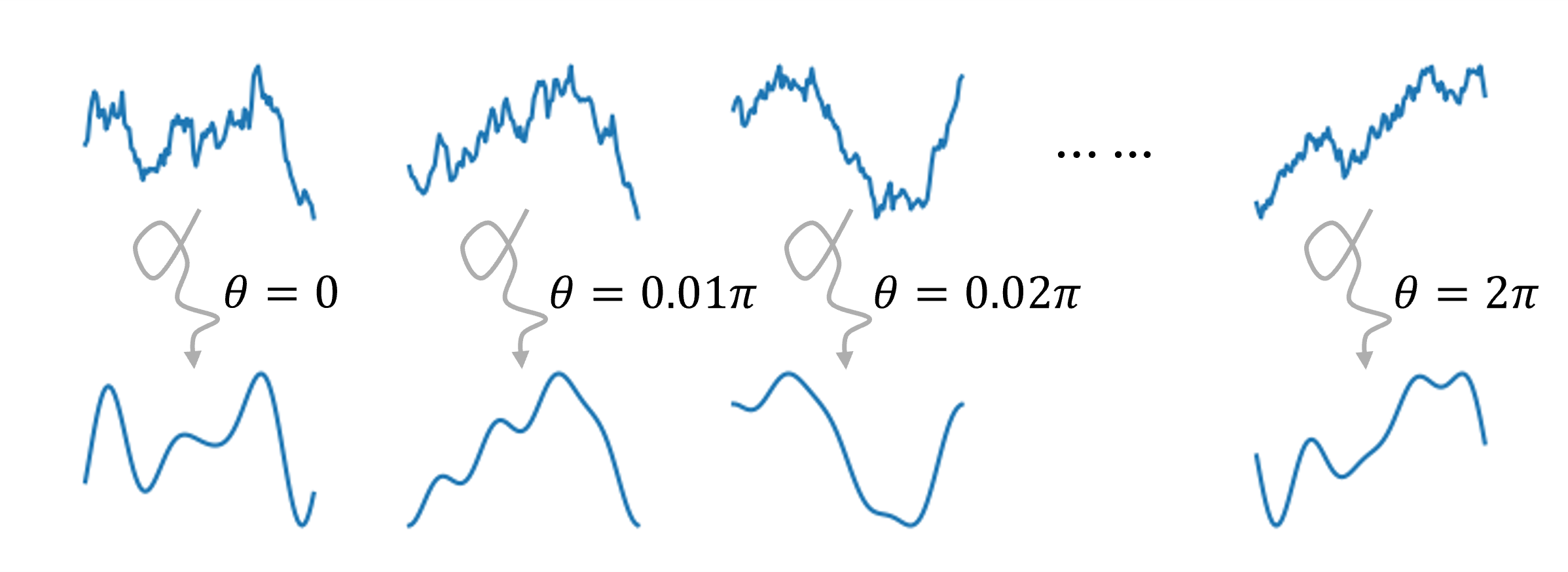}}
  \centering
  \begin{subfigure}{\subfiglen}
    \begin{minipage}[t]{13.34pt}
      \raggedleft
      \raisebox{\dimexpr \ht0 - \topskip}{(a)}
    \end{minipage}%
    \begin{minipage}[t]{\mpagelen}
      \usebox0
    \end{minipage}
  \end{subfigure}
  \setlength{\mpagelen}{\subfiglen-13.34pt}
  \setlength{\sboxlen}{0.45\textwidth-13.34pt}
  \sbox0{\includegraphics[width=\sboxlen]
  {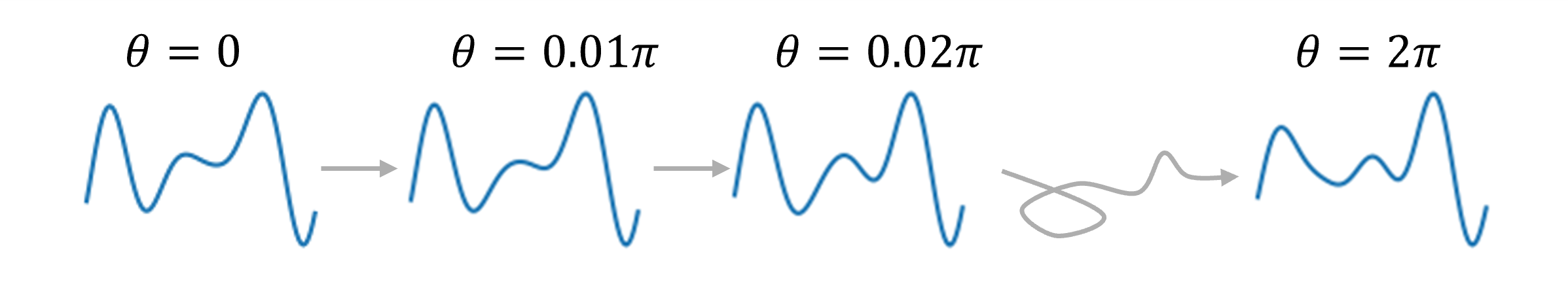}}
  \centering
  \begin{subfigure}{\subfiglen}
    \begin{minipage}[t]{13.34pt}
      \raggedleft
      \raisebox{\dimexpr \ht0 - \topskip}{(b)}
    \end{minipage}%
    \begin{minipage}[t]{\mpagelen}
      \usebox0
    \end{minipage}
  \end{subfigure}
  \caption{Comparison of different approaches to obtain robust control pulses (blue curve) for parametric gates: (a) Traditionally, optimizing the control pulse for each \(U(\theta)\) requires many runs, as shown by curved gray arrows. (b) Our RIPV algorithm varies control pulses (straight gray arrows) by traversing a level set to obtain control pulses for all \(U(\theta)\) in one run, as shown by the only curved gray arrow).}
  \label{fig:conti_param_control}
\end{figure}

In robust quantum control, existing optimization algorithms are designed to improve the robustness of \emph{one} quantum gate or state-transfer probability. However, optimizing the robustness of a gate family -- such as a series of parametric gates \(U(\theta)\) -- is impractical. This is illustrated in \autoref{fig:conti_param_control}(a), where each blue curve represents a control pulse parameterized by \(\mathbf{A}\). Since each optimization (represented by the cursive gray arrows) is applied to a single \(U(\theta)\), we have to run the optimization infinitely many times for each value of \(\theta\). One might solve this problem by saving beforehand a discrete sample of control parameters \(\{\mathbf{A}_i\}_{i=1}^{N}\) for \(\{U(\theta_i)\}_{i=1}^{N}\), and then interpolating on \(\mathbf{A}_i\) to get a continuous function \(\mathbf{A}(\theta)\). But in quantum control, there are usually multiple configurations of \(\mathbf{A}_i\) that produce the same \(U(\theta_i)\). Therefore, two consecutive parameters \(\mathbf{A}_i\) and \(\mathbf{A}_j\) are not guaranteed to stay close to each other, rendering interpolation impossible.

The key to solving this problem is ensuring the continuous dependence of \(\mathbf{A}\) on \(\theta\). For example, Sauvage et al.~\cite{sauvage2022optimal} realized this idea using neural networks (NNs), where the NN attempts to minimize the averaged cost of a continuous family of control functions. However, the computational cost of this collective optimization is unnecessarily high, and it is difficult to fine-tune the controls, since NNs are black boxes.

We ensure the continuous dependence mentioned earlier by traversing the level set of the QCRL using the RIPV algorithm, as illustrated in \autoref{fig:conti_param_control}(b). In the preparation stage, we optimize the beginning pulse to ensure it is sufficiently robust. Then we apply RIPV to vary this robust beginning pulse, generating a series of controls for \(U(\theta)\). As briefly introduced in \autoref{fig:toy-landscape}, the RIPV algorithm modifies \(\mathbf{A}\) in arbitrarily small steps (each step represented by a straight arrow), thereby ensuring the continuous dependence of \(\mathbf{A}\) on \(\theta\). Moreover, RIPV guarantees that \(\mathbf{A}\) remains within the level set of the QCRL, thereby preserving the robustness of the beginning pulse. In this manner, we obtain robust control parameters \(\{\mathbf{A}_i\}_{i=1}^{N}\) for all \(\{U(\theta_i)\}_{i=1}^{N}\) in a single run of RIPV (as shown by the cursive arrow in \autoref{fig:conti_param_control}(b)). Furthermore, the parameters \(\{\mathbf{A}_i\}_{i=1}^{N}\) can be interpolated to generate robust control for any \(U(\theta)\).

\section{Quantum Control Robustness Landscape}

In this section, we define robustness for arbitrary types of noise that can be written as a Hamiltonian with stochastic parameters, and use it to define QCRL. First, we take a look at what we are going to achieve from a toy model. Then, we define integral robustness and asymptotic robustness. Integral robustness, defined as an integral over the noise parameters, characterizes the system's resilience to noise across all noise strengths. Asymptotic robustness, defined using a local expansion, describes the system's response to noise at small noise strengths. Finally, we define QCRL as the map from pulse parameters to robustness, and we discuss some methods for pulse parametrization, which is vital to the success of algorithms about QCRL.

\subsection{A first glance into QCRL}

Enhancing the performance of quantum computers in the NISQ era involves two primary tasks: (1) increasing the fidelity of quantum operations, and (2) enhancing the robustness of these operations against noise. These two tasks are often coupled together in a way that when one is optimized, the other is compromised. The key reason to study QCRLs is that by traversing a level set (inputs that yield equal outputs) of a QCRL, one obtains a series of equally robust controls implementing different quantum gates. This effectively decouples the two objectives of quantum control.
\autoref{fig:toy-landscape} illustrates this idea on a simple toy landscape, which does not correspond to any realistic robustness model. In this figure, we plot a 2D landscape defined by the surface
\begin{align*}
  \robness &= -r^2 \\
  A_1 &= r \left( 2 + \cos^2 3\theta / 2 \right) \cos\theta \\
  A_2 &= r \left( 2 + \cos^2 3\theta / 2 \right) \sin\theta \:,
\end{align*}
where \(\robness\) represents the control's robustness, and \(A_1\) and \(A_2\) are control parameters.
The red arrows represent the small steps that we take to vary the inputs \(x\) and \(y\) while maintaining \(z=-16\) constant.

This toy landscape is unrealistically simple, since a 2D surface can only have 1D level sets which only allow movement in one direction. Real applications typically involve far more than two parameters, leading to higher-dimensional level sets and more complex traversal paths. In the toy landscape, a path in a level set appears as a closed loop in this simplified case. However, in higher-dimensional level sets, paths can be far more flexible and need not form closed loops. Therefore, usually in the parameter space, the point for \(R_x(0)\) does not implement \(R_x(2\pi)\). With extra dimensions, we have more degrees of freedom to choose a path in the level set so that the path encompasses all the gates we need.

To rigorously define QCRL, we propose a quantitative and general definition of quantum control robustness. We then present the mathematical formalism of QCRL and elucidate several key concepts.

\subsection{Integral Robustness}

\subsubsection{Notations}

First of all, we introduce a useful notation of propagator that allows us to deal with time-dependent and time-independent Hamiltonians at once. Note that throughout this paper, we set \(\hbar = 1\). 

\begin{notation}
  Let's denote by \(U^H(t)\) the propagator generated by a (whether time-dependent or time-independent) Hamiltonian \(H\) in the time period \([0,t]\).
\end{notation}

In a more common notation, for a time-independent Hamiltonian \(H\), we have \(U^H(t) = e^{-iHt}\), and for a time-dependent \(H(t)\), we have \(U^H(t) = U^{H(\tau)}(t) = \mathcal{T} e^{-i \int_0^t H(\tau) \dd{\tau}} \) with \(\mathcal{T}\) being the time-ordering operator. Pay attention to \(\tau\) in the exponent of our notation \(U^{H(\tau)}(t)\). It is only used to indicate we are dealing with a time-dependent Hamiltonian, and is essentially the variable of integration in the formal exponent \(\int_0^t H(\tau) \dd{\tau}\).

Throughout this paper, we assume a system dictated by Hamiltonian
\begin{align*}
  H = H_\text{s} + H_\text{c} + H_\text{n} \:,
\end{align*}
with \(H_\text{s}\) the system Hamiltonian, \(H_\text{c}\) the control Hamiltonian, \(H_\text{n}\) the noise Hamiltonian introducing stochastic perturbations. We omit the explicit time variable in \(H_\text{n}\) and \(H_\text{c}\) to emphasize the entire map, rather than a specific instance of \(H_\text{n}(t)\) at any given \(t\). Upon fixing a coordinate frame and time \(t\), \(H_\text{n}(t)\) is represented as a matrix. Let's denote by \(\mathcal{H}_\text{n}\) the set of matrices from which the instant noise operator \(H_\text{n}(t)\) might take value. It is a subalgebra of \(n\times n\) Hermitian matrices. Subsequently, \(H_\text{n}: t \mapsto H_\text{n}(t)\) represents a map residing within the exponential set
\begin{align}
  \label{eq:def-Hn0T}
  {\mathcal{H}_\text{n}}^{[0,T]} := \{\text{continuous } H_\text{n} : [0,T] \to \mathcal{H}_\text{n} \} \:,
\end{align}
i.e., the set of all continuous maps from \([0,T]\) to \(\mathcal{H}_\text{n}\).
If we view \(\mathcal{H}_\text{n}\) as a linear space of dimension \(n^2\), then the image of the continuous map \(H_\text{n}\) is a path (strictly speaking, a bounded 1D submanifold) when the time evolves from \(0\) to \(t\).
In this way, \({\mathcal{H}_\text{n}}^{[0,T]}\) can be viewed as a space of paths, where each \(H_\text{n}\) is a path in \(\mathcal{H}_\text{n}\). In the following discussions, we will use ``path'' as a synonym of ``time-dependent noise operator''.

\subsubsection{Path integral perspective}

In this part of the discussion, we are going to see that what we need for defining a general robustness metric is mathematically a path integral. We will start from a intuitive construction and then work our way to a rigorous definition.

To understand the system's robustness to stochastic noise, we must take into account all possible noise operators in \({\mathcal{H}_\text{n}}^{[0,T]}\) when examining the system's performance. In other words, we might need to evaluate an expression in the form of \(\int J[H_\text{n}] \, \dd{H_\text{n}}\), where \(J: {\mathcal{H}_\text{n}}^{[0,T]} \to \mathbb{R}\) is some functional of \(H_\text{n}\) that assesses the system's performance.

Since \(H_\text{n}\) is a path as we discussed in the end of the previous section, mathematically speaking, \(\int J[H_\text{n}] \, \dd{H_\text{n}}\) is a form of \emph{path integral}. It is fundamentally different from ordinary integrals over \(\mathbb{R}^n\). A common mistake is to write \(\dd{H_\text{n}} = \pdv{H_\text{n}(t)}{t} \dt\), wrongly treating it as integrating over all instant Hamiltonians \(H_\text{n}(t)\). Instead, it is important to recognize that we are integrating over all paths \(H_\text{n}\), rather than individual matrices \(H_\text{n}(t)\). To see it from another perspective, \(J[H_\text{n}]\), as a functional of \(H_\text{n}\), does not make sense when we ask the value of \(J\) at a specific time \(t\), not to mention integrating it over \(t\). To make the integral well-defined, the differential \(\dd{H_\text{n}}\) should be the measure of a neighborhood of \(H_\text{n}\). In other words, we need a \emph{measure} on the space \({\mathcal{H}_\text{n}}^{[0,T]}\). To emphasize its contrast to the ``ordinary'' measure \(\dx\), we denote this \emph{measure of paths} with \(\DD{H_\text{n}}\). To summarize, we want to define a path integral of the form
\begin{align*}
  \int_{H_\text{n} \in {\mathcal{H}_\text{n}}^{[0,T]}} J[H_\text{n}] \DD{H_\text{n}} \:.
\end{align*}

To compute the path integral, it is imperative to first define a measure \(\mu\) on the space of paths \({\mathcal{H}_\text{n}}^{[0,T]}\), or equivalently the space of maps. Defining a measure in such a space poses significant challenges owing to its large \emph{cardinality} (size of an infinite set). Nevertheless, for practical scenarios, a parametrization of \(H_\text{n}\) induces a measure on \({\mathcal{H}_\text{n}}^{[0,T]}\). 

Let us see how the measure on \({\mathcal{H}_\text{n}}^{[0,T]}\) can be induced by the measure of \(\mathbb{R}^n\).
Once we parameterize the noise Hamiltonian \(H_\text{n}\) by \(\mathbf{B} \in \mathbb{R}^n\), formally
\begin{align*}
  H_\text{n}(t) = H_\text{n}(t; \mathbf{B}) \:,
\end{align*}
we obtain a map from parameters to paths, \(H_\text{n}: \mathbb{R}^n \to {\mathcal{H}_\text{n}}^{[0,T]}\).
Intuitively, \(H_\text{n}\) is indeed such a map because a parameter vector \(\mathbf{B}\) is mapped to a time-dependent noise operator in \({\mathcal{H}_\text{n}}^{[0,T]}\).
To be more rigorous, contemplate the notation \(H_\text{n}(\_; \mathbf{B})\) where \(\mathbf{B}\) is fixed but \(t\) is to be determined. It defines a path in \({\mathcal{H}_\text{n}}\), because whenever we insert a time instant \(t\), we obtain a matrix in \(\mathcal{H}_\text{n}\) (see definition in \autoref{eq:def-Hn0T}). Therefore, \(H_\text{n}\) is such a map that when we feed it a parameter vector \(\mathbf{B} \in \mathbb{R}^n\), we get a path \(H_\text{n}(\_; \mathbf{B}) \in {\mathcal{H}_\text{n}}^{[0,T]}\).
Moving on to subsets, if \(\mathcal{B} \subset \mathbb{R}^n\) is a measurable neighborhood of \(\mathbf{B}\), we can denote by \(H_\text{n}(\_; \mathcal{B}) \subset {\mathcal{H}_\text{n}}^{[0,T]}\) the subset of operators parameterized by all \(\mathbf{B} \in \mathcal{B}\), and the measure of \(H_\text{n}(\_; \mathcal{B})\) is thus defined as
\begin{align}
  \label{eq:def-measureH-measureB}
  \mu\big(H_\text{n}(\_; \mathcal{B})\big) := \mu_L(\mathcal{B}) \:,
\end{align}
where \(\mu_L(\mathcal{B})\) denotes the Lebesgue measure of \(\mathbb{R}^n\). In other words, when we calculate the path integral, we convert back to \(\mathbb{R}^n\) as
\begin{align}
  \label{eq:def-DHn-dB}
  \int J[H_\text{n}] \DD{H_\text{n}}
  := \int J[H_\text{n}(\_; \mathbf{B})] \dd{\mathbf{B}} \:.
\end{align}

In this manner, we have defined a measure and therefore the integral on the parametrizable subset of \(\mathcal{H}_\text{n}^{[0,T]}\), though lacking some mathematical rigor. We discuss these issues as follows.
\begin{enumerate}
  \item \textit{Symmetric treatment on parameters.} By using a rotational symmetric measure in \(\mathbb{R}^n\), we presume subjectively, that each component \(B_i\) in vector \(\mathbf{B}\) has equal influence on \(H_\text{n}\). To allow different effects \(B_i\) has on \(H_\text{n}\), we need a Jacobian-like coefficient. This is addressed by inserting a probability function in the formal definition, writing down something like \(\int J[H_\text{n}] p(H_\text{n}) \DD{H_\text{n}}\).
  \item \textit{Additivity.} To ensure the induced measure to be additive, two sufficient conditions should be posited: (1) the map \(H_\text{n}\) is continuous over both \(t\) and \(\mathbf{B}\); (2) the map \(H_\text{n}: \mathbb{R}^n \to {\mathcal{H}_\text{n}}^{[0,T]}\) acts as an injection nearly everywhere, with the exception of a null set (a subset of zero measure). Given these sufficient conditions, we can ascertain that
  \begin{align*}
    & \mu \big(H_\text{n}(\_; \mathcal{B}_1) \big) + 
    \mu \big(H_\text{n}(\_; \mathcal{B}_2) \big) \\
    & = \mu \big(H_\text{n}(\_; \mathcal{B}_1 \cup \mathcal{B}_2) \big) \:,
  \end{align*}   
  and the induced measure is thus well defined.
  \item \textit{Invariance under frame transformation.} It is noteworthy that the measure \(\mu(H_\text{n})\) should remain \textbf{invariant} under the transformation  of any frame \(U(t)\). However, addressing the property of invariance proves to be challenging, and consequently, computations are typically executed numerically within a fixed coordinate frame.
\end{enumerate}

To summarize, by defining the measure of \(H_\text{n}(\_; \mathcal{B})\) to be the measure in \(\mathcal{B}\) as in \autoref{eq:def-measureH-measureB}, we can safely replace \(\DD{H_\text{n}}\) by \(\dd{\mathbf{B}}\) as in \autoref{eq:def-DHn-dB}. Keep in mind that this replacement does \textbf{not} introduce the Jacobian in the way \(\DD{H_\text{n}} = \abs{\pdv{H_\text{n}}{\mathbf{B}}} \dd{\mathbf{B}}\), because, as we discussed earlier, \(\DD\) is the measure of the paths rather than the differential of the matrix \(H_\text{n}(t; \mathbf{B})\).

Given a well-defined differential \(\DD{H_\text{n}}\), it is reasonable to further postulate that the noise adheres to a probability distribution characterized by the density function \(p(H_\text{n})\). Consequently, the probability that the noise resides within a small neighborhood around \(H_\text{n}\) is denoted by \(p(H_\text{n}) \DD{H_\text{n}}\). In numerical computations, the specific probability of a time-dependent Hamiltonian is frequently unknown; however, we can hypothesize the probability distribution of the noise parameters. Therefore, it becomes feasible to substitute \(p(H_\text{n})\) with \(p(\mathbf{B})\) and perform an integration over \(\mathbf{B}\). Equipped with these methodologies, we can formally characterize the robustness of a control with respect to the system's noise. In the subsequent definition, temporal dependence is presumed throughout, yet omitted for clarity unless explicitly necessary.

\subsubsection{Formal definition}

\begin{definition}[Robustness]
  \label{def:full-robustness}
  Given a control \(H_\text{c}(t)\) applied to a system, its \emph{robustness} \(\boxed{ \robness[H_\text{c}] }\) against a stochastic noise Hamiltonian \(H_\text{n}\) is defined as the mathematical expectation of the gate fidelity between the noiseless propagator
  \(U_\text{sc} = U^{H_\text{s} + H_\text{c}}(T)\)
  and the actual noisy propagator
  \(U_\text{scn} = U^{H_\text{s} + H_\text{c} + H_\text{n}}(T)\), averaged over \(H_\text{n}\). Mathematically,
  \begin{align}
    \label{eq:integral-robustness-dHn}
    & \robness[H_\text{c}]
    := \int_{ \mathrlap{ \raisebox{-0ex}{\(_{H_\text{n} \in {\mathcal{H}_\text{n}}^{[0,T]}}\)} } } \qquad
    F\big(U_\text{sc}(T), U_\text{scn}(T)\big) \, p(H_\text{n}) \DD{H_\text{n}} \:,
  \end{align}
  where \(p(H_\text{n})\) is non-zero only at the noise operators that might affect the system.
  If we only consider noise operators parametrizable by \(\mathbf{B}\), and denote \(U_\text{scn}(T; \mathbf{B}) = U^{H_\text{s} + H_\text{c} + H_\text{n}(\tau; \mathbf{B})}(T)\)
  \begin{align}
    \label{eq:integral-robustness-dB}
    \robness[H_\text{c}]:= & \int_{\mathbf{B} \in \mathbb{R}^n} F\big(U_\text{sc}(T), U_\text{scn}(T; \mathbf{B})\big) \, p(\mathbf{B}) \dd{\mathbf{B}} \:.
  \end{align}
\end{definition}

Based on the definition of fidelity, what we defined here is a number \(0 \le \robness \le 1\).
In the following text, we occasionally refer to such \(\robness[H_\text{c}]\) as the \emph{integral robustness} to distinguish it from the more practical \(n\)-th order asymptotic robustness that is to be introduced.

\begin{remark}
  It is important to highlight that this definition is NOT an ``average of fidelity'' in the conventional sense, as ``fidelity'' typically involves a comparison between the propagator \(U_\text{sc}\) and the ideal gate \(U_\text{ideal}\). In contrast, robustness is determined by comparing the noiseless propagator \(U_\text{sc}\) with the noisy propagator \(U_\text{scn}\). Robustness \(\robness\) is influenced exclusively by two competing factors: the noise \(H_\text{n}\) that generates errors and the control \(H_\text{c}\) employed to suppress these errors.
\end{remark}

\begin{remark}
It should be noted that the integral robustness can be directly defined as the integral over the noise parameters \(\mathbf{B}\) without invoking the concept of path integral. However, the measure of paths need not be derived from these parameters; hence, it has been conceptualized through the measure on the space of paths \({\mathcal{H}_\text{n}}^{[0,T]}\). Should a more generalized measure be introduced, this definition of robustness could be further refined and rendered more rigorous.
\end{remark}

\subsubsection{Error evolution}


Transitioning to the interaction picture with \(U_\text{sc}(t) = U^{H_\text{s} + H_\text{c}(\tau)}(t)\), we will see that the accumulation of errors induced by the noise can be effectively suppressed through the application of control fields. This accumulated error is characterized by an evolution operator.
By analyzing the error evolution in the interaction picture, we can simplify and formalize the definition of robustness. In this picture, the noise Hamiltonian \(H_\text{n}\) transforms as:
\begin{align}
  H_\text{n}^\text{sc}(t) & = 
  U^{- H_\text{s} - H_\text{c}(\tau)}(t)
  \cdot H_\text{n}(t) \cdot
  U^{H_\text{s} + H_\text{c}(\tau)}(t) \\
  & = U_\text{sc}^\dagger H_\text{n} U_\text{sc}^{\vphantom{\dagger}} \:.
\end{align}
Notably, the noise operator \(H_\text{n}\) may exhibit either time-dependent or time-independent characteristics, a distinction that is irrelevant in our discussion.

\begin{definition}[Error evolution]
  The \emph{error evolution} or \emph{error propagator} is defined as the unitary evolution
  \begin{align*}
    U_\text{n}^\text{sc}(t)
    = U^{H_\text{n}^\text{sc}(\tau)}(t)
    = U^{U_\text{sc}^\dagger H_\text{n} U_\text{sc}}(t) \:,
  \end{align*}
  where \(H_\text{n}^\text{sc}\) is the noise Hamiltonian under the interaction picture.
\end{definition}

From this definition, we see that \(U_\text{n}^\text{sc}(T) = I\) means the error is canceled exactly at time \(T\). The propagators in two pictures are related by \(U_\text{scn} = U_\text{sc} U_\text{n}^\text{sc}\).

\begin{remark}[Integral robustness by error evolution]
  \label{remark:error-evolution}
  With the definition of error evolution and the interaction picture, the fidelity between \(U_\text{sc}\) and \(U_\text{scn}\) becomes,
  \begin{align*}
    & \hphantom{{}={}} F\big(U_\text{sc}(T), U_\text{scn}(T)\big) \\
    & = \Tr\big(U_\text{sc}^\dagger(T) \cdot U_\text{scn}(T)\big) / d \\
    & = \Tr\big(U_\text{sc}^\dagger(T) \cdot U_\text{sc}(T) \cdot U_\text{n}^\text{sc}(T)\big) / d \\
    & = \Tr\big(U_\text{n}^\text{sc}(T)\big) / d \\
    & = F\big(U_\text{n}^\text{sc}(T), \mathrm{I}\big) \:,
  \end{align*}
  where \(d\) is the dimension of the Hilbert space.
  Hence, the integral robustness can also be written as
  \begin{align}
    \label{eq:integralrob-trace}
    \robness[H_\text{c}]
    & = \int_{H_\text{n} \in {\mathcal{H}_\text{n}}^{[0,T]}} \frac{1}{d} \Tr \big( U_\text{n}^\text{sc}(T) \big) p(H_\text{n}) \DD{H_\text{n}} \\
    \label{eq:integralrob-FofUI}
    & = \int_{ H_\text{n} \in {\mathcal{H}_\text{n}}^{[0,T]} } F(U_\text{n}^\text{sc}(T), I) p(H_\text{n}) \DD{H_\text{n}} \:.
  \end{align}
  This expression aligns with our intuition that, within the interaction picture, a robust system characterized by a larger \(\mathcal{R}\) should exhibit an error evolution \(U_\text{n}^\text{sc}\) that closely resembles the identity evolution \(I\). In this way, the information of the robustness of a control is encapsulated in the error evolution \(U_\text{n}^\text{sc}\).
\end{remark}

\subsection{Asymptotic robustness}
\label{sec:asymptotic-robustness}

While the integral robustness is elegant in both construction and interpretation, it may be impractical for many applications, as it requires simulating for fidelity under all possible noise operators. Instead, we can take a more practical approach by focusing on robustness to small noise. In this section, we define asymptotic robustness metrics that capture the system's resilience as the noise strength approaches zero.

Let us first look at a simpler case, where \(H_\text{n} = \delta H_{\text{n}, 0}\) is quasi-static noise, with \(\delta\) an unknown constant.
\begin{align*}
  H_\text{n}^\text{sc}(t) & = 
    U^{- H_\text{s} - H_\text{c}(\tau)}(t)
    \cdot \delta H_{\text{n}, 0} \cdot
    U^{H_\text{s} + H_\text{c}(\tau)}(t) \\
  & = \delta U_\text{sc}^\dagger(t) H_{\text{n}, 0} U_\text{sc}^{\vphantom{\dagger}}(t) \:.
\end{align*}
The error evolution is \(U_\text{n}^\text{sc} = U^{H_\text{n}^\text{sc}}(t) = U^{\delta U_\text{sc}^\dagger H_{\text{n}, 0} U_\text{sc}^{\vphantom{\dagger}}}(t)\).
Since noise should be relatively small compared to control field strength, we can assume \(\delta\) is very small. Then the error evolution can be approximated, to the first order, as
\begin{align*}
  U_\text{n}^\text{sc}(T)
  & = \mathcal{T} e^{ -i \int_{0}^{T} \delta U_\text{sc}^\dagger(\tau) H_{\text{n}, 0} U_\text{sc}^{\vphantom{\dagger}}(\tau) \dd{\tau} } \\
  & \approx I - i \delta \int_{0}^{T} U_\text{sc}^\dagger(\tau) H_{\text{n}, 0} U_\text{sc}^{\vphantom{\dagger}}(\tau) \dd{\tau} \:.
\end{align*}
Then the overlap between \(U_\text{n}^\text{sc}\) and \(I\) all boils down to the norm of the matrix integral \(\int_{0}^{T} U_\text{sc}^\dagger(\tau) H_{\text{n}, 0} U_\text{sc}^{\vphantom{\dagger}}(\tau) \dd{\tau}\). The value of \(\mathcal{R}\) is mainly determined by the matrix value when \(\delta\) is small enough. We can generalize this quantity to higher orders.

In general, by using tools like the Magnus expansion or high-order derivatives, we can define another type of robustness that holds in the limit of \(\delta \to 0\), where \(\delta\) is the strength of some noise (not necessarily the strength of quasi-static noise).

\begin{example}[\(n\)-th order robustness by Magnus expansion]
  \label{example:magnus-robustness}
  In previous example, we have shown that the robustness can be calculated by \(\robness[H_\text{c}] = \int_{H_\text{n}} \Tr(U_\text{n}^\text{sc}) / d \DD{H_\text{n}}\). Furthermore, we can use Magnus expansion to obtain a more useful quantity:
  \begin{align}
    & U^{H_\text{n}^\text{sc}}(T)\\
    = & \exp \Bigg(
      - i \delta \int_0^T \frac{H_\text{n}^\text{sc}(t)}{\delta} \dt \\
      & - \frac{1}{2} \delta^2 \int_{0}^{T} \left[
      \frac{H_\text{n}^\text{sc}(t)}{\delta},
      \int_{0}^{t} \frac{H_\text{n}^\text{sc}(\tau)}{\delta} \dtau \right] \dt
      + O(\delta^3)
    \Bigg) \\
    \label{eq:magnus-terms}
    =: & \exp \left(
      - i \delta M_1(T) - \frac{1}{2} \delta^2 M_2(T) + O(\delta^3)
    \right) \:,
  \end{align}
  where \(M_k\) denotes the \(k\)-th order term in the Magnus expansion.
  The 1st-order term  \(M_1\) tells us, to the 1st order, the amount of error induced by a unit of noise (i.e., \(\delta=1\)) under the amplification or suppression of the control.
  We refer to its norm as the \emph{1st-order noise susceptibility} \(\mathcal{S}^1_\text{(M)}\):
  \begin{align}
    \label{eq:def-s1}
    \mathcal{S}^1_\text{(M)} =
    \norm{M_1(T)} =
    \norm{ \int_0^T \frac{H_\text{n}^\text{sc}(t)}{\delta} \dt } \:.
  \end{align}
  The subscript \(\square_\text{(M)}\) indicates it is defined by the Magnus expansion, and will be omitted when the context is clear. We always assume this Magnus-expansion-based definition throughout this paper.
  Similarly, we can define the \emph{2nd-order noise susceptibility}:
  \begin{align}
    \mathcal{S}^2_\text{(M)} & =
    \norm{M_2(T)} \\
    \label{eq:def-s2}
    & = \norm{
      \int_{0}^{T} \left[
      \frac{H_\text{n}^\text{sc}(t)}{\delta},
      \int_{0}^{t} \frac{H_\text{n}^\text{sc}(\tau)}{\delta} \dtau \right] \dt
    } \:.
  \end{align}
  Naturally, we define \(n\)-th order noise susceptibility as
  \begin{align*}
    \mathcal{S}^n_\text{(M)} = \norm{ M_n(T) } \:.
  \end{align*}
  Because we mainly consider its relative magnitude, the norms used here can be any norm, as long as it bounds every component of the matrix. For example, any element-wise \(p\)-norm works fine, but trace norm is unfavorable since it does not bound off-diagonal components.

  Nevertheless, the magnitudes of \(\mathcal{S}^n\) are dependent on gate time \(T\) and \(n\) (the order of \(\delta\)). Another drawback is that \(\mathcal{S}^n\) is negatively correlated with the control's robustness. Hence, from the definitions of \(n\)-th order noise susceptibilities \(\mathcal{S}^n\), we can define the \(n\)-order robustness \(\robness^n\).
  To get rid of the relation to \(T\) and \(n\), we take the \(n\)-th root and then divide it by \(T\),  which gives \(\frac{\sqrt[n]{\mathcal{S}^n}}{T}\). To make this quantity positively correlated to robustness, we take its reciprocal.
  We finally take logarithm of base 10 to make the numbers compact and to manifest the order of error suppression. In summary, we define \(n\)-th order robustness (by Magnus expansion) as
  \begin{align*}
    \robness^n_\text{(M)} & = \log_{10} \left( {\frac{T}{\sqrt[n]{\mathcal{S}^n_\text{(M)}}}} \right) \\
    & = \log_{10} T - \frac{1}{n} \log_{10} \mathcal{S}^n_\text{(M)} \:.
  \end{align*}
  This robustness is more intuitive. Theoretically, it could exceed the computer precision when \(\mathcal{S}^n \to 0\), but this hardly happens in realistic applications.
  Since the robustness and susceptibility are monotonically related, we mainly use susceptibility in this paper for its computational simplicity.
\end{example}

\begin{example}[\(n\)-th order robustness by derivatives]
  We can also define asymptotic noise susceptibility by higher derivatives,
  \begin{align*}
    \mathcal{S}^n_\text{(D)} & = \norm{\pdv[n]{U_\text{n}^\text{sc}}{\delta}} \:.
  \end{align*}
  And we define asymptotic robustness by derivatives similarly,
  \begin{align*}
    \robness^n_{\text{(D)}} & = \log_{10} T - \frac{1}{n} \log_{10} \mathcal{S}^n_\text{(D)} \:.
  \end{align*}
  Note that the 1st-order noise susceptibility defined by either derivatives or the Magnus expansion is the same, but they disagree on higher-order susceptibilities (hence also on robustness):
  \begin{align*}
    \pdv{U_\text{n}^\text{sc}}{\delta} = - M_1 
    & \Rightarrow \mathcal{S}^1_\text{(D)} = \mathcal{S}^1_\text{(M)} \\
    \pdv[2]{U_\text{n}^\text{sc}}{\delta} = - \frac{1}{2} M_1^2 + iM _2
    & \Rightarrow \mathcal{S}^2_\text{(D)} \le \frac{1}{2} \left(\mathcal{S}^1_\text{(D)}\right)^2 \! + \mathcal{S}^2_\text{(M)} \:,
  \end{align*}
  where \(M_k\) is a shorthand for \(M_k(T)\) in the Magnus expansion at time \(T\).
  
\end{example}

\medskip

By induction, having vanishing asymptotic robustness under the two definitions is equivalent. If the lower-than-\(n\)-th order susceptibilities \(\mathcal{S}^k_\text{(M)}\) all vanish, then that of susceptibilities \(\mathcal{S}^k_\text{(D)}\) would also vanish. Precisely,
\begin{gather*}
  \mathcal{S}^k_\text{(M)} = 0 \quad \forall k = 1, \dots, n,
  \\
  \Updownarrow
  \\
  \mathcal{S}^k_\text{(D)} = 0 \quad \forall k = 1, \dots, n,
\end{gather*}
Particularly, the two definitions of noise susceptibility agree on the 1st order, which we simply denote by \(\mathcal{S}^1\) (and the robustness by \(\mathcal{R}^1\)):
\begin{align*}
  \mathcal{S}^1 := \mathcal{S}^1_\text{(D)} = \mathcal{S}^1_\text{(M)} \:.
\end{align*}

\begin{example}[Multiple noise sources]
  If there are more than one noise source, i.e., \(H_\text{n} = \sum_k H_{\text{n}, k}\) where each \(H_{\text{n}, k}\) is an independent quasi-static noise, we can easily find out that the 1st-order term in the Magnus expansion for noise \(H_\text{n}\) is simply the sum as following
  \begin{align*}
    M_1(T;H_\text{n})
    = \int_0^T \frac{\sum_k H_{\text{n}, k}^\text{sc}(t)}{\delta} \dt
    = \sum_k M_1(T;H_{\text{n}, k}) \:.
  \end{align*}
  Then, by the triangle inequality of matrix norms, the 1st-order noise susceptibility is bounded by the sum
  \begin{gather*}
    \norm{M_1(T;H_\text{n})}
    = \mathcal{S}^1[H_\text{n}] \\
    \le \sum_k \mathcal{S}^1[H_{\text{n}, k}]
    = \sum_k \norm{M_1(T;H_{\text{n}, k})} \:.
  \end{gather*}
Thus, the first-order robustness concerning a single noise source can be readily extended to encompass multiple noise sources. Higher-order robustness could similarly be generalized to the scenario involving multiple noise sources; however, this would entail a remarkably complex array of mixed products of different \(H_{\text{n},k}\)'s. Consequently, such formulations are not presented here.
\end{example}

\begin{remark}[Comparison to optimization on QCL]
  In QCL, our objective is either the state fidelity, the gate fidelity, or the expectation of an observable \(\mathcal{O}\). The objective is always a function of the final time propagator \(U_\text{sc}(T)\). With \(\eta\) being the ideal final state and \(G\) being the ideal gate, the objectives are,
  \begin{align*}
    F_\text{state} & = \Big( \Tr \sqrt{ \sqrt{\eta} U_\text{sc}(T) \rho_0 U_\text{sc}^\dagger(T) \sqrt{\eta} } \Big)^2 \:, \\
    F_\text{gate} & = \Tr\left( U_\text{sc}^\dagger(T) G \right) / d \:, \\
    \expval{\mathcal{O}} & = \Tr \left( U_\text{sc}^\dagger(T) \mathcal{O} U_\text{sc}^{\vphantom{\dagger}}(T) \right) \:.
  \end{align*}
  On the other hand, in QCRL, if we optimize the 1st-order susceptibility \(\mathcal{S}^1\), our objective is an integral of all propagators \(U_\text{sc}(t)\) for \(t \in [0,T]\), which is
  \begin{align*}
    \mathcal{S}^1 & = \norm{ \int_0^T \frac{H_\text{n}^\text{sc}(t)}{\delta} \dt } \\
    & = \norm{ \int_0^T U_\text{sc}^\dagger(t) H_{\text{n}, 0} U_\text{sc}^{\vphantom{\dagger}}(t) \dt } \:.
  \end{align*}
  Notice the difference in \(U_\text{sc}(t)\) and \(U_\text{sc}(T)\).These definitions indicate that the landscape of noise susceptibility (hence also robustness) is fundamentally different from that of the fidelity-based QCLs.
\end{remark}

\subsection{Definition of QCRL}

As elucidated previously, QCL defines the map from control parameters to the fidelity of the ideal quantum gate. Since now a metric of robustness is defined as \autoref{eq:integral-robustness-dHn}, there emerges a novel landscape distinct from QCL, which facilitates the characterization of quantum control performance, particularly with respect to robustness.

Hereinafter, we suppose the control Hamiltonian is parameterized by a vector \(\mathbf{A}\in \mathbb{R}^n\), formally
\begin{align*}
  H_\text{c}(t) = H_\text{c}(t; \mathbf{A}) \:.
\end{align*}

\begin{definition}[QCRL]
  The QCRL is defined as the map \(\robness: \mathbf{A} \mapsto \robness(\mathbf{A})\) from control parameters \(\mathbf{A}\) to the control's robustness \(\robness\) against certain types of noise.
\end{definition}

In practical quantum systems, reference \(\mathbf{A}\) typically pertains to the parameters of control fields, which include waveforms of electronic or magnetic fields, as well as the envelope and frequency of microwave fields, among others. In this work, we assume the general form of control
\begin{align*}
  H_\text{c}(t; \mathbf{A}) & := \sum_k \Omega_k(t; \mathbf{A}) H_{\text{c}, k},
\end{align*}
where time-dependence is reflected in \(\Omega_k(t)\) and \(H_{\text{c}, k}\) are time-independent operators. We refer to \(\Omega_k(t)\) as \emph{pulses}, which are not necessarily the physically applied pulse signals.

\begin{remark}[Decomposition of QCRL]
  If we look into this \(\robness(\mathbf{A})\), it is actually composed of the following maps. In this paper, we use \([\dots]\) to denote functional dependence, and \((\dots)\) to denote time dependence.
  \begin{enumerate}
    \item A vector of maps \(\vec{\Omega}=\{ \Omega_k \}_k\) from control parameters \(\mathbf{A}\) to control pulses \(\Omega_k(t; \mathbf{A})\), each corresponding to one of the control terms \(H_{\text{c}, k}\). Each \(\Omega_k(\mathbf{A})\) is a function of \(t\), denoted by \(\Omega_k(t; \mathbf{A})\):
      \begin{align*}
        \Omega_k: \mathbf{A} & \mapsto \Omega_k(\mathbf{A}), \\
        \text{s.t. } \Omega_k(\mathbf{A})(t) & = \Omega_k(t; \mathbf{A}) \:.
      \end{align*}
    \item A map \(H_\text{c}\) from pulses \(\vec{\Omega}\) to a control Hamiltonian \(H_\text{c}[\vec{\Omega}]\), where \(H_\text{c}[\vec{\Omega}]\) is a time-dependent operator:
      \begin{align*}
        H_\text{c}: \vec{\Omega} & \mapsto H_\text{c}[\vec{\Omega}], \\
        \text{s.t. } H_\text{c}[\vec{\Omega}](t) & = \sum_k \Omega_k(t) H_{\text{c}, k} \:.
      \end{align*}
    \item A map \(U_\text{sc}\) from control Hamiltonian \(H_\text{c}\) to the noiseless propagator \(U_\text{sc}[H_\text{c}]\), where \(U_\text{sc}[H_\text{c}]\) is a time-independent operator:
      \begin{align*}
        U_\text{sc}: H_\text{c} & \mapsto U_\text{sc}[H_\text{c}], \\ 
        \text{s.t. } U_\text{sc}[H_\text{c}](t) & = U^{H_\text{s} + H_\text{c}(\tau)}(t) \:. 
      \end{align*}
    \item A map from the propagator \(U_\text{sc}(t)\) to any metric of robustness \(\robness\) (integral robustness, asymptotic robustness, etc.):    
      \begin{align*}
        \robness: U_\text{sc} \mapsto \robness[U_\text{sc}] \:.
      \end{align*}
  \end{enumerate}

  Finally, we obtain the robustness map from control parameters \(\mathbf{A}\) to some metric of robustness \(\robness\) by composing all the maps,
  \begin{align}\label{eq:decomposeRA-elewise}
    \robness:
    \mathbf{A}
    \mapsto \vec{\Omega}(\mathbf{A})
    \mapsto H_\text{c}[\vec{\Omega}]
    \mapsto U_\text{sc}[H_\text{c}]
    \mapsto \robness[U_\text{sc}(\mathbf{A})] \:.
  \end{align}
  If we write out the time dependence explicitly, the map then looks like
  \begin{align*}
    \robness:
    & \mathbf{A}
    \xmapsto{\vec{\Omega}} \{\Omega_k(t; \mathbf{A})\}_k
    \xmapsto{H_\text{c}} \sum_{k} \Omega_k(t; \mathbf{A}) H_{\text{c}, k} \\
    & \xmapsto{U_\text{sc}} U^{H_\text{s} + \sum_{k} \Omega_k(\tau; \mathbf{A}) H_{\text{c}, k}} (t)
    \xmapsto{\robness} \robness[U_\text{sc}] \:.
  \end{align*}

  Among the four maps above, only the first one \(\vec{\Omega}\) is a function of numbers, i.e., the control parameters \(\mathbf{A}\). The second map \(H_\text{c}\) maps the functions \(\vec{\Omega}(t)\) to a time-dependent Hamiltonian \(H_\text{c}(t)\). The rest two maps, \(U_\text{sc}\) and \(\robness\), have either time-ordered exponential or integrals. So they are all functionals that depend on the whole input functions defined on the time interval \([0, T]\). This decomposition into four maps will aid us in further discussions on topics such as critical points.
\end{remark}

\medskip

\begin{remark}[Functional dependence on \(U_\text{n}^\text{sc}\)]
  We have assumed in previous discussions that the robustness \(\robness\) is always a functional of \(U_\text{sc}(t)\), i.e., they depend on all the values of \(U_\text{sc}(t)\) on \(t\in [0,T]\).
  Let's check that the definitions of robustness we have defined before always fit this assumption. The integral robustness is defined as
  \begin{align*}
    \robness[U_\text{sc}] & = \int_{H_\text{n} \in \mathcal{H}_\text{n}} F(U_\text{n}^\text{sc}, I) p(H_\text{n}) \dd{H_\text{n}} \\
    & = \int_{H_\text{n} \in \mathcal{H}_\text{n}} F\left( U^{ U_\text{sc}^\dagger(\tau) H_\text{n} U_\text{sc}^{\vphantom{\dagger}}(\tau) } (T), I \right) \; p(H_\text{n}) \dd{H_\text{n}} \:.
  \end{align*}
  Hence, the integral robustness is indeed a functional of \(U_\text{sc}(t)\).
  By definition, the first order robustness is also a functional of \(U_\text{sc}(t)\), since
  \begin{align*}
    \mathcal{S}^1 = \mathcal{S}^1[U_\text{sc}]
    & = \norm{ \int_{0}^{T} \frac{H_\text{n}^\text{sc}(t)}{\delta} \dt } \\
    & = \norm{ 
        \int_0^T U_\text{sc}^\dagger(t)
        \, H_{\text{n}, 0} \,
        U_\text{sc}(t) \dt
    } \:.
  \end{align*}
  As for higher order robustness defined by Magnus expansion, notice that they are all commutators and integrals of \(\frac{H_\text{n}^\text{sc}(t)}{\delta}\), which is \(U_\text{sc}^\dagger(t) H_{\text{n}, 0} U_\text{sc}^{\vphantom{\dagger}}(t)\).
\end{remark}

\medskip

While this marks the first dedicated discussion of QCRL, the topics of interest are similar to those of QCL, namely controllability, local traps, level sets, etc.
In this work, we emphasize the level sets, based on which we will later introduce an algorithm.

\begin{definition}
  In a landscape, a \emph{level set} is all inputs yielding the same output. A \emph{critical point} has a zero derivative. A \emph{local trap} is a local extremum that is not the global extremum.
\end{definition}


Although we do not provide proof regarding the existence of local traps, they were not encountered in our numerical experiments. This may indicate a trap-free property of the QCRL, but it could also be due to the simple structure of \(SU(2)\) used in our numerical experiments. Further investigation is needed.

Among the four maps in the decomposition of \(\robness(\mathbf{A})\) in \autoref{eq:decomposeRA-elewise}, the two in the middle are defined by physical implementations: \(\vec{\Omega}
\mapsto H_\text{c}[\vec{\Omega}]\) and \(H_\text{c}
\mapsto U_\text{sc}[H_\text{c}]\). The final map \(U_\text{sc}
\mapsto \robness[U_\text{sc}]\) is determined by the mathematical definition of robustness. We have limited freedom of choice over the physical implementations and mathematical definition in optimization. Therefore, the most important factor in shaping the landscape is the parametrization of the pulses, i.e., the map \(\mathbf{A} \mapsto \vec{\Omega}(\mathbf{A})\).

\subsection{Pulse parametrization}

Quantum systems are often controlled through modulated microwave or optical pulses. The parametrization of control pulses plays an important role in shaping the structure of the QCL and QCRL, as discussed in the last section. We showcase two common parametrization methods: \emph{piecewise constant} and \emph{functional basis} parametrization.

Let \(\Omega(t;\mathbf{A})\) denote the pulse parametrized by \(\mathbf{A}\). Piecewise constant parametrization divides the pulse into piecewise constant segments, using amplitudes at each segment as parameters. Functional basis parametrization uses a finite number of parametrized basis functions to compose the pulse.

\subsubsection{Piecewise constant parametrization.}
We first discretize the time interval \([0,T]\) into \(N\) pieces, namely
\begin{align*}
  t_0=0, t_1=\frac{T}{N}, \dots, t_k=\frac{kT}{N}, \dots, t_N=T \:.
\end{align*}
The pulse is then defined by the parameter vector \(\mathbf{A} = (A_0, \dots, A_N)\) as
\begin{align*}
  \Omega(t) = \begin{cases}
    0 & t\notin [0,T] \\
    A_k & t \in [t_{k}, t_{k+1}].
  \end{cases}
\end{align*}
The parametrization is straightforward to implement, but applying constraints could be challenging. In practice, we often require the pulse to be continuous, smooth (with a continuous derivative), and to smoothly vanish at the beginning and end times. While it is possible to impose these constraints on time-sliced pulses, doing so requires significant effort.

\subsubsection{Functional basis parametrization.}

There are many functional bases to choose. Two common methods are the Taylor expansion and Fourier expansion, where the pulse parameters are the coefficients of the expansion terms.
Besides, wavelets are among the most useful functional bases for constructing pulses, for they are finitely supported and have various merits by construction.

Since the pulses are composed of smooth functions, they naturally satisfy the smoothness requirements.
To satisfy the vanishing boundary conditions,  the pulse is constructed to inherently satisfy these conditions, regardless of the parameter values.

\paragraph{Taylor expansion.}
The simplest parametrization is to multiply the Taylor expansion at \(t=0\) with that at \(t=T\).
The pulse parametrized by \(\mathbf{A} = (a_1, \dots,\allowbreak a_N,\allowbreak b_1, \dots,\allowbreak b_N)\) is given by
\begin{align*}
  \Omega(t; \mathbf{A}) = \left(
    \sum_{k=1}^{N} a_k t^k
  \right) \cdot \left(
    \sum_{k=1}^{N} b_k (T - t)^k
  \right) \:.
\end{align*}
If we expect the beginning and ending points of the pulse to vanish up to \(K\)-th order, we only need to set \(a_k = b_k = 0\) for all \(k=0, 1, \dots, K\).

\paragraph{Fourier expansion.}
Using Fourier expansion, the pulse parametrized by \(\mathbf{A} = (a_0, a_1, \dots, a_N, b_1, \dots, b_N)\) is given by
\begin{align*}
  \Omega(t; \mathbf{A}) = W(t) \Bigg( a_0
    & + \sum_{k=1}^{N} a_k \cos(2\pi k \cdot \frac{t}{T}) \\
    & + \sum_{k=1}^{N} b_k \sin(2\pi k \cdot \frac{t}{T})
  \Bigg) \:,
\end{align*}
where \(W(t)\) is a window (envelope) function that, along with its derivative, vanishes outside \([0,T]\).

Useful window functions include
\begin{gather*}
  \sin \left( \pi \frac{t}{T} \right), \quad
  \sin^2 \left( \pi \frac{t}{T} \right), \quad
  \frac{1}{\sigma\sqrt{2\pi}} e^{ - \frac{\left( t - \frac{T}{2} \right)^2}{2\sigma^2} } \:.
\end{gather*}
The \(\sin\) envelope ensures zero amplitude on both sides, while the \(\sin^2\) envelope further ensures continuous transitions on both sides. The Gaussian envelope actually gives rise to Morlet wavelets, which will be introduced in the next part.

\paragraph{Wavelets.}
Pulses can also be synthesized by employing wavelets. In simple terms, a wavelet is a function characterized by finite support, typically modulated by time scaling and frequency modulations. Functions that approximate zero beyond a finite interval are likewise regarded as wavelets. The continuous wavelet transform (CWT), unlike the Fourier transform, offers simultaneous insights into both temporal and frequency information. Consequently, the composition of pulses using wavelets intrinsically facilitates precise modulation across both the time and frequency domains.

There are many wavelets to choose from, one should make a choice based on specific desired properties of the composed pulse. Of particular interest is the Morlet wavelet, which offers an optimal balance between temporal and frequency resolutions. The spectrum of control signals is important if we want to avoid crosstalk and coupling to noise. Therefore, using Morlet-like wavelets naturally generates a clean spectrum compared to other wavelets.

The standard definition of Morlet wavelets is essentially a complex exponential multiplied by a Gaussian envelope, not particularly tailored to our needs. As quantum control pulses, we hope the pulse to be as fast as possible, necessitating truncation within a short time interval. To ensure that the truncated pulse closely approximates zero smoothly at both ends, we adopt an alternative definition for the Morlet wavelet basis,
\begin{align*}
  \morlet_k(t) = c e^{-2r^2 \cdot \tau^2} \cos \big( (2k+1) \pi \tau \big) \:,
\end{align*}
where \(t\in [0,T]\) is gate operation time, \(\tau = \left( \frac{t - T/2}{T} \right) \in [-\frac{1}{2},\frac{1}{2}]\) is normalized time, \(c\) is the normalization constant such that \(\int_{0}^{T} \morlet_k(t) \dt = 1\), and \(r\) is the ratio \(\frac{T/2}{\sigma}\) with \(\sigma\) being the standard deviation of the Gaussian envelope. The \(k\)-th order Morlet wavelet \(\morlet_k(t)\) has a central frequency approximately equal to (in fact, a little higher than) \(\frac{2k+1}{T}\).
The pulse parametrized by \(\mathbf{A} = (A_1, \dots, A_N)\) is defined as
\begin{align*}
  \Omega(t; \mathbf{A}) := \sum_{k=1}^{N} A_k M_k(t).
\end{align*}

\section{Level set exploration}
\label{sec:level-set-exploration}

In this section, we introduce the Robustness-Invariant Pulse Variance (RIPV) algorithm, which is designed to modify control pulses while maintaining robustness.
We assume the Hamiltonian governing the system is
\begin{align*}
  H(t) = H_\text{s} + H_\text{c}(t) + H_\text{n}(t) \:,
\end{align*}
the control Hamiltonian is
\begin{align*}
  H_\text{c} = \sum_k \Omega_k(t; \mathbf{A}) H_{\text{c}, k} \:,
\end{align*}
and the noise Hamiltonian is
\begin{align*}
  H_\text{n} = \sum_j \delta_j H_{\text{n}, j} \:.
\end{align*}

The algorithm achieves this robustness by ensuring that the control remains within the same level set of a robustness function. Starting with robust pulses \(\vec{\Omega}(t; \mathbf{A}_{\theta_0})\) for a parametric gate \(U(\theta_0)\), which is robust against \(K\) noise sources \(H_{\text{n}, k}\) (\(k=1, \dots, K\)), the algorithm systematically varies \(\mathbf{A}\) to identify a series of robust pulses \(\vec{\Omega}(t; \mathbf{A}_{\theta_i})\) for each parametric gate \(U(\theta_i)\). Throughout this process, it ensures that the robustness functions \(\robness_k\) for all noise sources \(H_{\text{n}, k}\) remain unchanged, thus preserving robustness. The RIPV algorithm is inspired by the unitary D-MORPH algorithm~\cite{dominyExploringFamiliesQuantum2008}, which explores a level set in a quantum control landscape, but they differ in both approach and application.

Recall that we want to interpolate between pulses, as introduced in \autoref{sec:search-of-robust-gate-families}. Since there are different continuous paths of \(\mathbf{A}\) that implement all \(U(\theta)\), we must ensure that the control parameters \(\mathbf{A}\) stay in the same continuous path. So, the sequence \(\{\mathbf{A}_{\theta_i}\}_{i=1}^N\) generated by RIPV must be continuous in the following sense.

\begin{definition}[Continuous sequence]
\label{def:continuous-sequence}
  Denote by \(U[\mathbf{A}]\) the gate implemented by \(\mathbf{A}\).
  Assume the sequence \(\{{\theta_i}\}_{i=1}^N\) is spaced with a constant interval \(\Delta \theta = \theta_{i+1} - \theta_{i}\).
  A sequence of parameters \(\{\mathbf{A}_{\theta_i}\}_{i=1}^N\) is \emph{continuous} if, for any \(\mathbf{A}_{\theta_i}\) and \(\mathbf{A}_{\theta_{i+1}}\), there is a continuous function \(\mathbf{A}(\kappa)\) for \(\kappa\in [0,1]\) such that
  \begin{align*}
    U[\mathbf{A}(\kappa)] & = U\big(\kappa \theta_{i} - (1-\kappa) \theta_{i+1}\big) \:.
  \end{align*}
  In particular,
  \begin{align*}
    U[\mathbf{A}(0)] & = U(\theta_i) \:, \\
    U[\mathbf{A}(1)] & = U(\theta_{i+1}) \:.
  \end{align*}
\end{definition}

\subsection{Simplest RIPV preview}
\label{sec:ripv-single-control-preview}

Suppose we expect to implement \(U(\theta)\) robustly, where \(\theta\) represents a gate parameter, typically the Bloch sphere's rotation angle around some axis.
We temporarily assume single control
\begin{align*}
  H_\text{c} = \Omega(t; \mathbf{A}) H_{\text{c}, 0} \:.
\end{align*}
By assuming that \([H_\text{s}, H_{\text{c}, 0}] = 0\), we focus on the simplest case so that we do not generate rotation on other axes without the influence of noise. We aim to find an \(\mathbf{A}_\theta\) that implements the quantum gate \(U(\theta)\) for each value of \(\theta\) in \([\theta_L, \theta_R]\) at intervals of \(\Delta \theta\). Occasionally, we will refer to \(\mathbf{A}\) as the ``pulse'' in this context when the meaning is clear. 
The procedure to implement \(U(\theta)\) with equal robustness \(\robness\) to one noise source is as follows.

\begin{enumerate}
  \item (Initialization.)
    Starting from an arbitrary pulse \(\mathbf{A}_\text{init}\) (called the \emph{initial pulse}), use any optimization algorithm to optimize the robustness function \(\robness(\mathbf{A})\). After optimization, we obtain a pulse \(\mathbf{A}_0\) and its corresponding gate parameter \(\theta_0\). Note that we only record but do not designate the value \(\theta_0\).
  \item (Variation step.) Starting from \(\mathbf{A}_0\) (called the \emph{beginning pulse}), we add a small vector \(\Delta \mathbf{A}\) each time. But we choose \(\Delta \mathbf{A}\) smartly so that \(\robness\) stays the same but \(\theta\) changes by \({\Delta \theta}\). That is, we require, in each step,
    \begin{align}
      \label{eq:rob-not-change}
      \robness(\mathbf{A} + \Delta \mathbf{A}) & = \robness(\mathbf{A}) \\
      \theta(\mathbf{A} + \Delta \mathbf{A}) & = \theta(\mathbf{A}) \pm {\Delta \theta} \:.
    \end{align}
    The sign before \(\Delta \theta\) is determined by whether one expects to increase \(\theta\) (when \(\theta_0 \le \theta_L\)) or decrease it (when \(\theta_0 \ge \theta_R\)).
    After variation, we record the value of \(\mathbf{A}_1 = \mathbf{A}_0 + \Delta \mathbf{A}\) and \(\theta_1 = \theta(\mathbf{A}_1)\).
  \item (Termination condition.)
    Repeat the variation step for \(M\) iterations. We get controls \(\mathbf{A}_0,\allowbreak \mathbf{A}_1, \dots\allowbreak,\allowbreak \mathbf{A}_M\) with \(\theta_0,\allowbreak \theta_1, \dots, \theta_M\).
    Terminate when \([\theta_0, \theta_M]\) covers the desired range \([\theta_L,\theta_R]\).
    In the desired range, we obtain \(N+1\) pulses where \(N=\left\lceil \frac{\theta_R - \theta_L}{\Delta \theta} \right\rceil\).
    To aid discussions, it is sufficient to focus on the case where the variation of \(\theta\) starts exactly at \(\theta_L\) and ends at \(\theta_R\), which is the assumption from now on.
  \item (Relabel and output.)
    We relabel the \(N\) pulses associated with \(\theta \in [\theta_L, \theta_R]\)by \(0,\allowbreak 1, \dots,\allowbreak N\). Finally, we obtain \(N+1\) pulses \(\mathbf{A}_0,\allowbreak \mathbf{A}_1, \dots,\allowbreak \mathbf{A}_N\), along with gate parameters \(\theta_L=\theta_0, \theta_1, \dots, \theta_N=\theta_R\).
\end{enumerate}
Here we implicitly assumed only one noise source, hence only one robustness function \(\robness\) to optimize and maintain. If there are multiple noise sources, we simply require \autoref{eq:rob-not-change} for every robustness function \(\robness_k\) against \(H_{\text{n}, k}\) .

\subsection{Gradient Orthogonal Variation}

How can we ``smartly'' choose \(\Delta \mathbf{A}\) so that \(\robness\) does not change but \(\theta\) increases or decreases by a proper amount that fits our needs? We employ the \emph{Gradient Orthogonal Variation} (GOV) algorithm for this purpose. The goal of GOV is to vary the input vector \(\mathbf{A}\) while keeping certain functions of \(\mathbf{A}\) unchanged.
It is basically the orthogonal version of the Gradient Descent (GD) algorithm.
These unchanged functions, referred to as ``constraints,'' are not limited to robustness and are denoted by a different font, \(R_1(\mathbf{A}), \dots, R_n(\mathbf{A})\). It is important to note that these ``constraints'' differ slightly from the traditional meaning, as they are not required to satisfy specific bounds or inequalities but instead must remain invariant.

\subsubsection{Variation vs. optimization}

Before diving into GOV, we emphasize that GOV is \textbf{not} an optimization algorithm.

Traversing a level set, such as when implementing a robust parametric gate, is fundamentally different from an optimization algorithm (referred to as ``\optalgo'' in this section) and cannot naturally be formulated as one. The goal is not to optimize anything, but rather to vary the solution. Therefore, we refer to this type of algorithm as a \emph{variation algorithm} (referred to as ``\varalgo'' in this section).

There are at least three key differences between variation and optimization algorithms.

\paragraph*{\bf Goal.} \optalgo aims to find the optimal solution, whereas \varalgo seeks a set of solutions that share the same constraint values (e.g., robustness) as the beginning solution, regardless of optimality.

\paragraph*{\bf Role of constraints.} The term ``constraint'' has a slightly different meaning. In \varalgo, constraints must remain unchanged, serving as the goal of the algorithm. In \optalgo, constraints must be equal to or less than some predefined values, serving as obstructions to our goal -- optimization.

\paragraph*{\bf Role of intermediate solutions.} In \optalgo, we solve equations or inequalities only once to find the optimal solution, such as with Lagrange multipliers. Even in gradient descent, we only need the final result. In contrast, \varalgo collects all the intermediate solutions, requiring the variation condition to be solved multiple times. It is more comparable to solving differential equations, where each intermediate step provides one slice of the final solution function.

\subsubsection{Variation condition}

Let's reformulate \autoref{eq:rob-not-change} into a more useful condition. Recall that in calculus, we have this simple equation
\(\dd{y} = \dv{y}{x} \dd{x}\)
that states the infinitesimal change of \(y\) as a function of \(x\) is equal to the infinitesimal change of \(x\) multiplied by the derivative \(\dv{y}{x}\). This holds true approximately when we numerically change \(x\) by a small enough \(\Delta x\).

Now in RIPV, we need to ``smartly'' choose \(\Delta \mathbf{A}\) so that \(\Delta R = R(\mathbf{A} + \Delta \mathbf{A}) - R(\mathbf{A}) = 0\). Suppose we choose an infinitesimal \(\dd{\mathbf{A}}\). In the tangent space, this amounts to choosing \(\dd{\mathbf{A}} \neq 0\) so that \(\dd{R} = 0\). Hereinafter, we denote the change of \(\mathbf{A}\) by \(\dd{\mathbf{A}}\) to indicate the change is sufficiently small that it approximates the tangent space. For the robustness function \(R(\mathbf{A})\), we have similarly
\begin{align*}
  \dd{R} = \pdv{R}{\mathbf{A}} \cdot \dd{\mathbf{A}} \:.
\end{align*}
In this equation, the derivative \(\pdv{R}{\mathbf{A}}\) is taken with regard to a vector \(\mathbf{A}\), which actually means taking the gradient \(\nabla R\). We stick to partial derivative notation to clearly indicate the variable with respect to which the gradient is taken.
Since we want \(\dd{\mathbf{A}} \neq 0\) and, in general, \(\pdv{R}{\mathbf{A}} \neq 0\) holds, the only way to get \(\dd{R} = 0\) is to enforce the condition
\begin{align}
  \label{eq:variation-condition}
  \text{Variation condition:} \quad \dd{\mathbf{A}} \perp \pdv{R}{\mathbf{A}} \:.
\end{align}

This condition shows that GOV is the orthogonal counterpart of GD. Their objectives are orthogonal: one maintains a quantity constant, while the other seeks to maximize a quantity. Their methods are orthogonal: one moves in a direction perpendicular to the gradient, while the other moves parallel to it.

The variation condition can be satisfied by simply selecting an arbitrary vector \(\dd{\mathbf{A}}^\text{pre}\) as what we call ``pre-variation'', and then performing the Gram-Schmidt process to get the component \(\dd{\mathbf{A}}^\text{pre}_\perp\) that is perpendicular to \(\pdv{R}{\mathbf{A}}\). This gives us the direction of \(\dd{\mathbf{A}}\), after which we still have to adjust its length. In order to understand the significance of the pre-variation during GOV, we first discuss orthogonalization.

\subsubsection{Orthogonalization}

In short, the Gram-Schmidt process orthogonalizes a vector \(\mathbf{v}\) to a group of vectors \(\mathbf{u}_1, \mathbf{u_2, \dots, \mathbf{u}}_N\). The process is mathematically equivalent to decomposing \(\mathbf{v}\) as
\begin{align*}
  \mathbf{v} = \mathbf{v}_\perp + \mathbf{v}_\parallel \:,
\end{align*}
where \(\mathbf{v}_\perp\) (or \(\mathbf{v}_\parallel\)) is orthogonal to (or inside)
\(\spanof\{\mathbf{u}_i\}_i = \spanof\{\mathbf{u}_1, \dots, \mathbf{u}_N\}\).

To maintain all of \(R_1(\mathbf{A}), \dots, R_n(\mathbf{A})\) constant, we apply the Gram-Schmidt process to get \(\dd{\mathbf{A}}^\text{pre}_\perp\) (or \(\dd{\mathbf{A}}^\text{pre}_\parallel\)) that is orthogonal to (or inside) \(\spanof\{\pdv{R_i}{\mathbf{A}}\}_i\). We then vary \(\mathbf{A}\) along the direction \(\dd{\mathbf{A}}^\text{pre}_\perp\). The tangent subspace \(\left( \spanof\{\pdv{R_i}{\mathbf{A}}\}_i \right)^\perp\), which is orthogonal to \(\spanof\{\pdv{R_i}{\mathbf{A}}\}_i\), is the subspace in which our variation happens. We refer to it as the \emph{variation subspace} and denote it hereinafter by \(V\).

\subsubsection{Choice of pre-variation}

The ``official variation'' \(\dd{\mathbf{A}}\) is an orthogonal projection of \(\dd{\mathbf{A}}^\text{pre}\) into the variation subspace \(V\). Therefore, the selection of this pre-variation \(\dd{\mathbf{A}}^\text{pre}\) is very important to roughly determine the direction along which the pulse \(\mathbf{A}\) should be adjusted, since.
This choice can be tailored according to specific requirements.

The simplest choice would be a random vector. This would allow one to trace a stochastic trajectory in the space of control parameters, which lies in the level set of the QCRL. It is useful when exploring the level sets.

Another application is to maximize some objective function \(F(\mathbf{A})\), for example, the fidelity of control. The fastest direction that maximizes \(F(\mathbf{A})\) is \(\pdv{F}{\mathbf{A}}\), because
\begin{align*}
  {\dd{F}}
  = {\pdv{F}{\mathbf{A}} \cdot \dd{\mathbf{A}}}
  \le \norm{\pdv{F}{\mathbf{A}}} \cdot \norm{\dd{\mathbf{A}}} \:.
\end{align*}
The equality holds if and only if \(\pdv{F}{\mathbf{A}} \parallel \dd{\mathbf{A}}\), or equivalently \(\dd{\mathbf{A}} = \alpha \pdv{F}{\mathbf{A}}\). Setting \(\dd{\mathbf{A}} = \alpha\pdv{F}{\mathbf{A}}\) gives the classic GD algorithm.
To maximize \(F\), we can consider setting \(\dd{\mathbf{A}}^\text{pre} = \pdv{F}{\mathbf{A}}\). When we orthogonalize \(\dd{\mathbf{A}}^\text{pre}\) to \(\{\pdv{R_i}{\mathbf{A}}\}_{i=1}^{n}\) and set \( \dd{\mathbf{A}} = \alpha \dd{\mathbf{A}}^\text{pre}_\perp \), this \(\dd{\mathbf{A}}\) would generally not be the fastest direction to maximize \(F(\mathbf{A})\). However, it is still the fastest direction to maximize \(F(\mathbf{A})\) while keeping \(\{R_i\}_{i=1}^{n}\) unchanged, for the following reasons.
It maximizes \(F(\mathbf{A})\), because the Gram-Schmidt process ensures the angle \( \angle \left( \dd{\mathbf{A}}^\text{pre}_\perp, \dd{\mathbf{A}}^\text{pre} \right) \) is an acute angle, i.e., it ensures \( \dd{\mathbf{A}}^\text{pre}_\perp \cdot \pdv{F}{\mathbf{A}} \ge 0 \). Therefore,
\begin{align*}
  \dd{F}
  = \pdv{F}{\mathbf{A}} \cdot \dd{\mathbf{A}}
  = \pdv{F}{\mathbf{A}} \cdot \alpha \dd{\mathbf{A}}^\text{pre}_\perp
  > 0 \:.
\end{align*}
It is also the fastest,
because in the tangent subspace \(\left( \spanof\{\pdv{R_i}{\mathbf{A}}\}_i \right)^\perp\) that keeps \(\{R_i\}_i\) unchanged, the orthogonal projection \(\dd{\mathbf{A}}^\text{pre}_\perp\) is the vector closest to \(\dd{\mathbf{A}}^\text{pre} \! = \pdv{F}{\mathbf{A}}\), i.e., closest to the direction that \(\theta\) increases.
In summary, setting
\begin{align*}
  \dd{\mathbf{A}}^\text{pre} = \pdv{F}{\mathbf{A}}, \quad
  \dd{\mathbf{A}} = \alpha \dd{\mathbf{A}}^\text{pre}_\perp
\end{align*}
where \(\dd{\mathbf{A}}^\text{pre}_\perp \in \left( \spanof\{\pdv{R_i}{\mathbf{A}}\}_i \right)^\perp \) and with constant \(\alpha > 0\), one could maximize a function \(F(\mathbf{A})\).

Lastly, we consider our RIPV algorithm, where the goal is to implement \(U(\theta)\) for the interval \([\theta_L, \theta_R]\). To ensure \(\theta\) is increasing, we define the pre-variation as \(\dd{\mathbf{A}}^\text{pre} = \pdv{\theta}{\mathbf{A}}\). The orthogonalized component \(\dd{\mathbf{A}}^\text{pre}_\perp\) gives the direction of \(\dd{\mathbf{A}}\). However, the magnitude of \(\dd{\mathbf{A}}\) must be more carefully chosen to ensure that \(\theta\) evolves smoothly, with evenly spaced increments throughout the process. 

\subsubsection{Normalization}

Now that the \textbf{}{direction} of variation is determined as \(\dd{\mathbf{A}}^\text{pre}_\perp\), the step size of variation is to be derived. Given our objective to implement \(U(\theta)\) for every value of \(\theta\), it is imperative to ensure that \(\theta\) is uniformly distributed along the interval \([\theta_L, \theta_R]\), with the optimal step size being \({\Delta \theta}_\text{ideal}\). The linear part of variation \(\dd{\theta}\), caused by \(\dd{\mathbf{A}}\), is determined by the differential relation \({\dd{\theta}} = {{\pdv{\theta}{\mathbf{A}}} \cdot \dd{\mathbf{A}}}\). Consequently, it becomes necessary to dictate \({\dd{\theta}} = {\Delta \theta}_\text{ideal}\), i.e.
\begin{align*}
  \norm{\dd{\mathbf{A}}} & = \frac{ {\Delta \theta}_\text{ideal} } { \norm{\pdv{\theta}{\mathbf{A}}} \cdot \cos \angle \left( \dd{\mathbf{A}}^\text{pre}_\perp, \pdv{\theta}{\mathbf{A}} \right) } \\
  & = \frac{ {\Delta \theta}_\text{ideal} }{ \norm{\pdv{\theta}{\mathbf{A}}} } \cdot \frac{ \norm{\dd{\mathbf{A}}^\text{pre}_\perp} \cdot \norm{\pdv{\theta}{\mathbf{A}}} }{ \langle \dd{\mathbf{A}}^\text{pre}_\perp, \pdv{\theta}{\mathbf{A}} \rangle } \\
  & = \frac{ {\Delta \theta}_\text{ideal} \cdot \norm{\dd{\mathbf{A}}^\text{pre}_\perp} }{ \langle \dd{\mathbf{A}}^\text{pre}_\perp, \pdv{\theta}{\mathbf{A}} \rangle } \:,
\end{align*}
where \(\angle (\mathbf{u}, \mathbf{v})\) denotes the angle between two vectors \(\mathbf{u}\) and \(\mathbf{v}\).
Then the ``official variation'' should be the unit vector \(\dd{\mathbf{A}}^\text{pre}_\perp / \norm{\dd{\mathbf{A}}^\text{pre}_\perp}\) multiplied by this magnitude, which is
\begin{align*}
  \dd{\mathbf{A}} = \frac{ {\Delta \theta}_\text{ideal} }{ \langle \dd{\mathbf{A}}^\text{pre}_\perp, \pdv{\theta}{\mathbf{A}} \rangle } \dd{\mathbf{A}}^\text{pre}_\perp \:.
\end{align*}

\subsubsection{Summary of GOV}

To summarize, the goal of GOV is to maintain constraints \(\{R_i\}_i\) unchanged, for which we need to achieve the variation condition \(\dd{\mathbf{A}} \perp \pdv{R_i}{\mathbf{A}}\). The general procedure is as follows.
\begin{enumerate}[itemsep=0mm]
  \item Choose an arbitrary pre-variation vector \(\dd{\mathbf{A}}^\text{pre}\), either random or task-specific.
  \item Apply the Gram-Schmidt process to get \(\dd{\mathbf{A}}^\text{pre}_\perp \in V = \left( \spanof\{\pdv{R_i}{\mathbf{A}}\}_i \right)^\perp\).
  \item Choose an appropriate \(\alpha\) according to the task, and vary \(\mathbf{A}\) by \(\dd{\mathbf{A}} = \alpha \dd{\mathbf{A}}^\text{pre}_\perp\).
\end{enumerate}

For our RIPV algorithm, where we traverse the rotation angle \(\theta\) while maintaining several fixed criteria \(\{R_i\}_i\), we need to further specify the pre-variation \(\dd{\mathbf{A}}^\text{pre}\) and normalization factor \(\alpha\).
\begin{enumerate}[itemsep=0mm]
  \item Calculate the pre-variation \(\dd{\mathbf{A}}^\text{pre} = \pdv{\theta}{\mathbf{A}}\).
  \item  Apply the Gram-Schmidt process to achieve \(\dd{\mathbf{A}}^\text{pre}_\perp \in V = \left( \spanof\{\pdv{R_i}{\mathbf{A}}\}_i \right)^\perp\).
  \item Normalize the vector and adjust the step size to obtain \( \dd{\mathbf{A}} = \frac{ {\Delta \theta}_\text{ideal} }{ \langle \dd{\mathbf{A}}^\text{pre}_\perp, \pdv{\theta}{\mathbf{A}} \rangle } \dd{\mathbf{A}}^\text{pre}_\perp \).
\end{enumerate}

The GOV algorithm extends beyond robust quantum control aimed at countering quasi-static noise. It allows for adjusting the rotation angle while preserving robustness against any form of noise or leakage, or conversely, enhancing robustness without compromising the fidelity to a specific gate. Moreover, its applicability is not limited to quantum control tasks. Whenever parameter variation is needed without altering certain criteria, the GOV algorithm proves to be a valuable tool.

\subsection{RIPV algorithm implementation}

Knowing how to choose the variation \(\dd{\mathbf{A}}\) so that \(\robness\) (or multiple \(R_i\)'s) will not change,  we are able to finalize the RIPV algorithm. We first address certain pertinent details, and then present the algorithm's pseudocode.

\subsubsection{Extraction of rotation angle}
\label{sec:extraction-rotation-angle}

For the simplest control scheme \(H_\text{c}(t) = \Omega(t) \sigma\) where \(\sigma\) is a constant operator, the rotation angle is simply the integral \(\theta(\mathbf{A}) = \int_{0}^{T} \Omega(t; \mathbf{A}) \dt\). For example, if we expect to implement \(R_x(\theta)\) with \(\sigma_x\) control, then this integration gives the exact rotation angle \(\theta\) without the risk of extra rotation along undesired axes, because the control scheme allows only \(\sigma_x\) rotation.

However, if the control involves more than one term, the rotation angle can no longer be calculated by simple integration. An even worse problem is that two non-commuting time-dependent operators can generate rotations along a third axis. For example, the control scheme \(H_\text{c}(t) = \Omega_x(t) \sigma_x + \Omega_y(t) \sigma_y\) can even implement an \(R_z(\theta)\) gate.

To extract the rotation angle from the noiseless propagator \(U_\text{sc}(T) = U^{H_\text{s} + H_\text{c}(\tau)}(T)\), we can compute the matrix logarithm to obtain its exponent \(\eta\) and then project \(\eta\) onto the desired rotation axis \(\sigma\). The \(\sigma\) here is a Pauli matrix in single-qubit control. For a \(d\) dimensional Hilbert space, the matrix inner product \(\tr(\eta^\dagger \sigma)\), gives the (generalized) rotation angle by the axis \(\sigma\).
Additionally, we must ensure that the rotation angles \(\vartheta_j\) along all undesired axes \(\varsigma_j\) remain zero (\(\vartheta, \varsigma\) are variant forms of \(\theta, \sigma\), here denoting undesired angles/axes).

In summary, the rotation angle along an axis \(\sigma\) can be determined as
\begin{align*}
  \theta & = \Tr \left( \eta^\dagger \sigma \right),
  & \text{where} \quad \eta & = i \logm U_\text{sc}(T) \:,
\end{align*}
where \(\logm\) stands for matrix logarithm.
Meanwhile, in addition to preserving robustness throughout the variation process, we must also ensure the rotation angles \(\vartheta_j = \tr(\eta^\dagger \varsigma_j)\) along undesired axes \(\varsigma_j\), remain at zero.

\subsubsection{Calculation of derivatives}

We need to calculate the value of the derivatives \(\pdv{\theta}{\mathbf{A}}\) and \(\pdv{R_i}{\mathbf{A}}\) at data points. These could be carried out by hand and hard-coded into the program, but we adopt a more general solution, Automatic Differentiation, hereinafter referred to as autodiff.

Autodiff is fundamentally different from numerical differentiation and offers significant advantages in computational complexity. It is a technique implemented in libraries such as TensorFlow, PyTorch, and Jax, which automatically compute derivatives alongside function evaluations. Unlike numerical differentiation, which approximates derivatives by introducing small perturbations, autodiff constructs a computation graph that tracks the data flow during function evaluation. This graph enables the differentiator to retrieve exact derivative formulas and efficiently apply the chain rule to compute derivatives directly with respect to every free variable. While numerical methods require \(n\) passes to calculate derivatives for \(n\) variables, autodiff achieves this in a single pass with a bit of overhead.

To calculate the derivatives that we need, one only has to change the ordinary numbers and vectors into the traceable (or differentiable) variable types provided by the autodiff library so that it can automatically construct the computation graph and output derivatives.

\subsubsection{Single noise source}
\label{sec:single-noise-ripv}

\begin{figure*}[htbp]
  \centering
  \begin{algorithm}[H]
    \caption{RIPV for single control single noise}
    \label{algo:ripv-pseudocode}
    \SetKwData{records}{records}
    \SetKwFunction{append}{append}
    \KwData{Initial pulse with \(\mathbf{A}_0\) that implements \(\theta_0=\theta_L\), rotation angle range \([\theta_L, \theta_R]\), variance step size \({\Delta \theta}_\text{ideal}\)}
    \KwResult{An array of pairs of rotation angle and pulse parameters \records\(=\{ (\theta_k, \mathbf{A}_k) \}_k\)}
    Initialize the pulse parameter iterator \(\mathbf{A}^\text{now} \leftarrow \mathbf{A}_0\) \;
    Initialize the rotation angle \(\theta \leftarrow \theta_0\) \;
    Record the first pair \records.\append{\(\theta_0, \mathbf{A}_0\)} \;
    \While{ \( \theta^\textup{now} < \theta_R \) }{
      \setstretch{1.2}
      Calculate \(\theta^\text{now}\) and \(\robness^\text{now}\) from \(\mathbf{A}^\text{now}\) with autodiff on \;
      Calculate \(\displaystyle \dd{\mathbf{A}}^\text{pre} \leftarrow \eval{ \pdv{\theta}{\mathbf{A}} }_{\mathbf{A}^\text{now}}\) using autodiff \;
      Calculate \(\displaystyle \eval{ \pdv{\robness}{\mathbf{A}} }_{\mathbf{A}^\text{now}}\) using autodiff \;
      Perform Gram-Schmidt process
      \(\displaystyle \dd{\mathbf{A}}^\text{pre} = \dd{\mathbf{A}}^\text{pre}_\perp + \dd{\mathbf{A}}^\text{pre}_\parallel\) so that \(\displaystyle \dd{\mathbf{A}}^\text{pre}_\perp \perp \pdv{\robness}{\mathbf{A}}\) \;
      Normalize \(\dd{\mathbf{A}}^\text{pre}_\perp\) according to \({\Delta \theta}_\text{ideal}\),
      \(\displaystyle \dd{\mathbf{A}} \leftarrow \frac{{\Delta \theta}_\text{ideal}}{ \left\langle \pdv{\theta}{\mathbf{A}}, \dd{\mathbf{A}}^\text{pre}_\perp \right\rangle } \cdot \dd{\mathbf{A}}^\text{pre}_\perp\) \;
      Vary the current value of \(\mathbf{A}\),
      \(\mathbf{A}^\text{now} \leftarrow \mathbf{A}^\text{now} + \dd{\mathbf{A}}\) \;
      Record a new pair
      \records.\append{\(\theta^\textup{now}, \mathbf{A}^\textup{now}\)} \;
      \setstretch{1.0}
    }
  \end{algorithm}
\end{figure*}

The simplest RIPV algorithm can deal with one control term \(H_{\text{c}, 0}\) and one quasi-static noise source \(H_{\text{n}, 0}\), assuming the noise is correctable, i.e., \([H_{\text{c}, 0}, H_{\text{n}, 0}] \neq 0\). We only have to vary one rotation angle \(\theta\) and maintain one robustness \(\robness\). The pseudocode of this simplest scenario is shown in \autoref{algo:ripv-pseudocode}. If we use the \(n\)-th order asymptotic robustness as a constraint in RIPV, we refer to this algorithm as \(n\)-th order RIPV.

\subsubsection{Multiple noise sources}
\label{sec:multiple-noise-ripv}

To deal with multiple noise sources, we might need to modify our algorithm. We need to discuss two cases, where one of them is trivial and another one requires a modified RIPV algorithm.

\textbf{Case 1.} There is only one control, \(H_\text{c}(t) = \Omega(t) H_{\text{c}, 0}\) satisfying \([H_\text{s}, H_\text{c}] = 0\) (same assumptions as in \autoref{sec:ripv-single-control-preview}). This simplest control model will never generate any extra rotation on undesired axes in a noiseless setting. Hence, without the effects of noise, it always perfectly implements one of the parametric gates with fidelity \(1\). In this case, if we try to mitigate the effects of multiple noise sources that are all orthogonal to the control term \(H_{\text{c}, 0}\), we simply need to add the robustness function for every noise source into our constraints to keep them constant during variation. In fact, we can even use only one constraint as they are algebraically equivalent: correcting one orthogonal noise means correcting all. This simplest scenario is fundamentally the same as the single noise source case.

\textbf{Case 2.} Either \([H_\text{s}, H_\text{c}] \neq 0\) or there are multiple control terms. In this case, applying arbitrary pulses \(\{\Omega_k(t)\}_k\) to \(\{H_{\text{c}, k}\}_k\) does not guarantee to only generate the desired rotation \(\theta\) along the specified axis. Consequently, we must modify two steps in the RIPV algorithm. Assume we want to implement the rotation angle \(\theta\) on \(\sigma\), meanwhile ensuring that the rotation angle \(\vartheta_j\) on \(\varsigma_j\) remains \(0\). The two modifications are:
\begin{enumerate}
    \item Since the rotation angle is no longer a simple integration of the pulse, we have to use the projection method mentioned in \autoref{sec:extraction-rotation-angle} instead of simple integral to calculate rotation angles, including \(\theta\) and \(\vartheta_j\).
    \item \(\vartheta_j\)'s should also remain unchanged. In the orthogonalization step, we should orthogonalize \(\dd{\mathbf{A}}^\text{pre}\) to both the gradient of robustness against each noise source \(\pdv*{\robness_i}{\mathbf{A}}\) and the gradient of undesired rotations \(\pdv*{\vartheta_i}{\mathbf{A}}\).
\end{enumerate}

Consider, for instance, the implementation of \(R_x(\theta)\) which exhibits robustness against noise affecting \(\sigma_x\), \(\sigma_y\), and \(\sigma_z\). Then we have five constraints, namely \(\robness_x, \robness_y, \robness_z, \vartheta_y\) and \(\vartheta_z\). We need to maintain \(\robness_x\), \(\robness_y\) and \(\robness_z\) so that the robustness of the three operators is unchanged throughout the variation, and we also need to maintain \(\vartheta_y\) and \(\vartheta_z\) so that they remain zero throughout the variation.

The modifications in case 2 are very important because we often apply multiple controls when fighting off multiple noise sources.
The prerequisite for dynamical noise-cancellation is having an ``orthogonal control''.
Concretely, from the previous work~\cite{haiUniversalRobustQuantum2023}, we have known that, in order to correct the error in direction \(\sigma\), we need to have a control in an orthogonal direction of \(\sigma\). In other words, for each noise source \(H_{\text{n}, k}\), there should be at least one control term \(H_{\text{c}, j}\) satisfying \([H_{\text{n}, k}, H_{\text{c}, j}] \neq 0\). A notable instance arises when the aim is to mitigate quasi-static noise affecting the control term. In this scenario, the control mechanism is incapable of self-correcting its own noise perturbations, thus necessitating the deployment of at least two control terms to counterbalance the noise impacting each other.

\subsection{Numerical considerations}

In this section, we discuss some technical problems when implementing the RIPV algorithm numerically.

\subsubsection{Number of parameters}

The number of independent control parameters, equivalently the dimension of the landscape, matters. It must be sufficiently large to accommodate a viable solution and ensure that the desired solutions are in the same connected component as the beginning pulse. This is intuitively easy to understand: after all, more parameters mean more choices. We delineate the specifics below.

First, it is imperative to ascertain at precisely which stage a sufficiently large dimensionality becomes requisite and to determine the magnitude of this dimensionality. Remember that after we chose a pre-variation \(\dd{\mathbf{A}}^\text{pre}\), we then have to orthogonalize it to the set of constraints \(\{R_i\}_i\). Generally, if we have \(m\) constraints \(\{R_1, \dots, R_m\}\), their gradients would span an \(m\)-dimensional space \(W=\spanof\{\pdv{R_i}{\mathbf{A}}\}_{i=1}^{m}\). Now we want the ``official'' \(\dd{\mathbf{A}}\) to be in \(V={W}^\perp\), then the vector \(\mathbf{A}\) should be at least \((m+1)\)-dimensional. However, only one extra dimension would render us having no freedom over choosing the direction of variation. In conclusion, we must have at least \(m+2\) independent parameters in order to uphold \(m\) constraints.

\begin{figure}[htbp]
  \centering
  \includegraphics[width=0.3\textwidth]{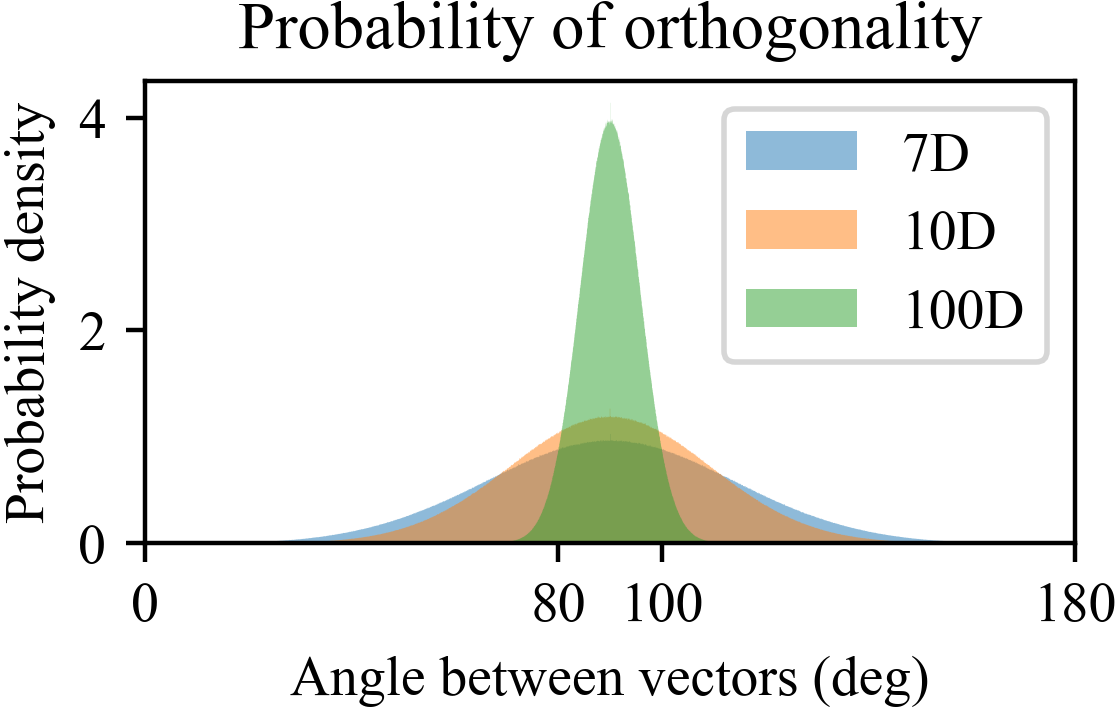}
  \caption{Probability of two vectors being orthogonal in an \(n\)-dimensional parameter space (\(n=7, 10, 100\)). This is a histogram of \(10^5\) randomly generated vectors.}
  \label{fig:probability-orthogonality}
\end{figure}

Second, having more than \(m+1\) parameters is still beneficial. In essence, we are looking for a direction that is orthogonal to \(W\), so it is best if the ``probability of being orthogonal'' is higher. If we consider strict orthogonality, then, in a \(n\)-dimensional parameter space, the probability of two vectors being orthogonal is simply \(\frac{n-1}{n}\). However, since we are running a numerical algorithm, where everything has a numerical precision limit, we actually only need approximate orthogonality where the angle between two vectors is \(\frac{\pi}{2} \pm \epsilon\) for very small \(\epsilon\). A corollary from the Johnson-Lindenstrauss lemma states that for a fixed \(\epsilon\), the probability of ``\(\frac{\pi}{2} \pm \epsilon\) orthogonal'' grows exponentially with the dimension \(n\). In \autoref{fig:probability-orthogonality}, we show a simple simulation where we take 5000 random vectors and ask for the distribution of the angle between each pair of vectors. Even raising the dimension from 7 to 10 results in a visible rise in probability. If we have 100 parameters (though unlikely), we can see that ``almost all'' pairs of vectors have an angle between them in \([80,100]\) degrees. In summary, we would have a better chance of locating a direction orthogonal to \(W\) if our parameter space has a higher dimension.

\subsubsection{Error from linear approximation}
\label{sec:extra-correcting-step}

The RIPV algorithm inevitably introduces error when it takes finite-length steps in the tangent space of the level set instead of the actual level set. No matter how small a step we choose for variation, we can only slow down the accumulation of error but not eliminate it. This problem can be dealt with by an extra correcting step between two consecutive variation steps where we bring our solution back to the level set. The unitary D-MORPH~\cite{dominyExploringFamiliesQuantum2008} and the work by Chen et al.~\cite{chen2015near} proved this correcting step to be useful.

In our algorithm, since we use autodiff, we can insert a gradient-based optimization procedure after taking the variation step in order to keep the parameters on the level set. Suppose we are at a point on the level set of \(\robness = \robness_0\) with \(\theta = \theta_k\), and a variation in the parameters moves \(\theta\) to \(\theta_{k+1}\), resulting in a deviation to a neighboring level set \(\robness = \robness_0 + \epsilon\). The tricky point here is that we cannot change \(U(\theta)\) in the optimization procedure. One way is to apply the simplest gradient descent to optimize the sum of the fidelity to \(U(\theta_{k+1})\) and the robustness deviation \(\robness - \robness_0\). A more complex way is to ensure we only deviate to a more robust level set. I.e., if \(\epsilon < 0\), we can apply a mini GOV procedure, where we maintain the fidelity to \(U(\theta_{k+1})\) while enhancing robustness.

In our numerical examples, these steps are not implemented because linear approximation is good enough. Also, this extra correcting step takes a lot of both coding effort and execution time.

\subsubsection{Viability of variation direction}

Even if we used a lot of free parameters, there is still a chance that we cannot find a viable direction for \(\dd{\mathbf{A}}\). This problem might appear when we orthogonalize \(\dd{\mathbf{A}}^\text{pre}\) to \(W=\spanof\{\pdv{R_i}{\mathbf{A}}\}_{i=1}^{m}\).
If we chose the pre-variation \(\dd{\mathbf{A}}^\text{pre} = \pdv{\theta}{\mathbf{A}}\) and found that \(\pdv{\theta}{\mathbf{A}} \in W\), then orthogonalization is impossible, which means we would not be able to increase or decrease \(\theta\) without changing at least one \(R_i\).

Let us examine the implications of this phenomenon and explore potential strategies to avoid it.
Suppose we are currently at a point \(\mathbf{A}_0\) in \(\mathbb{R}^n\), the space of \(n\) real parameters \(\{A_i\}_{i=1}^{n}\). Normally we can expect
\begin{align*}
  \dim \left(
    \spanof \left\{
      \pdv{\theta}{\mathbf{A}},
      \pdv{R_1}{\mathbf{A}},
      \dots,
      \pdv{R_m}{\mathbf{A}},
    \right\}
  \right) = m+1 \:,
\end{align*}
which means \(\pdv{\theta}{\mathbf{A}}\) cannot be spanned by \(\pdv{R_i}{\mathbf{A}}\)'s.
We call \(\mathbf{A}_0\) an \emph{irregular point} if we find out \(\pdv{\theta}{\mathbf{A}} \in W\), i.e., if \(\pdv{\theta}{\mathbf{A}}\) has no component orthogonal to \(\{\pdv{R_i}{\mathbf{A}}\}_{i=1}^{m}\).

An irregular point appears in one of the following cases.
\begin{enumerate}[label=(\arabic*)]
  \item The objective \(\theta(\mathbf{A})\) is irregular  at every point. In this case, no matter which value of \(\mathbf{A}\) we try, \(\pdv{\theta}{\mathbf{A}}\) is always a linear combination of \(\pdv{R_i}{\mathbf{A}}\)'s, as
    \begin{align*}
      \pdv{\theta}{\mathbf{A}} & = f_1(\mathbf{A}) \pdv{R_1}{\mathbf{A}} + \dots + f_m(\mathbf{A}) \pdv{R_m}{\mathbf{A}} \:.
    \end{align*}
    It means they are fundamentally dependent to each other and cannot be decoupled as objective and constraints. 
    For instance, one possible cause is that \(\theta\) being a function of \(R_i\)'s, with \(\pdv{\theta}{R_i}\) serving as \(f_i\):
    \begin{align*}
      \pdv{\theta}{\mathbf{A}} & 
      = \pdv{\theta}{R_1} \pdv{R_1}{\mathbf{A}} + \dots + \pdv{\theta}{R_m} \pdv{R_m}{\mathbf{A}} \:.
    \end{align*}
    In this case, we have to change the objective function or constraints and check our physical model.
  \item The objective \(\theta(\mathbf{A})\) is irregular at a subset containing \(\mathbf{A}_0\). In this case, we are just at an unlucky point \(\mathbf{A}_0\). One solution is that, we can move \(\mathbf{A}\) into a different, maybe more robust level set, using a correcting step. Another solution is that, we can start off from a different beginning pulse in the hope of getting rid of this irregular point.
\end{enumerate}

In our numerical experiments, none of the above happens. We presume the reason is the large number of parameters (18 when using both X and Y control for one qubit) makes the variation subspace large enough to accommodate a viable variation direction \(\dd{\mathbf{A}}\).

\subsubsection{Interpolation of gate parameters}

As mentioned in \autoref{sec:search-of-robust-gate-families}, we need to interpolate between the pulses after RIPV.
After all, we can only generate a discrete sequence of pulses, instead of a genuinely continuous function. Since our algorithm is a local gradient-based algorithm, which means it traces a continuous path in the parameter space, the parameter sequence \(\{\mathbf{A}_i\}_{i=0}^{N}\) is a continuous sequence defined in \autoref{def:continuous-sequence}. This means we can interpolate between two pulses \(\vec{\Omega}_{\theta_k}(t)\) and \(\vec{\Omega}_{\theta_{k+1}}(t)\) to get a pulse \(\vec{\Omega}_{\theta}(t)\) with any intermediate \(\theta \in [\theta_k, \theta_{k+1}]\).

Different to the interpolation on \(\theta(\mathbf{A})\) introduced in \autoref{sec:search-of-robust-gate-families}, it is more convenient to interpolate a function \(\mathbf{A}(\theta)\).
More precisely, if we denote by \(\mathbf{A}_i\) that implements \(U(\theta_i)\), we can interpolate on the \(N\) pairs \(\{ (\theta_i, \mathbf{A}_i) \}_{i=1,\dots, N}\) obtained by RIPV to get a continuous function \(\mathbf{A}(\theta)\). In this way, we obtain a map from rotation angles to control pulses \(\theta \mapsto \vec{\Omega}_\theta = \vec{\Omega}(t; \mathbf{A}(\theta))\). To implement \(U(\theta)\) in experiments, we simply extract \(\mathbf{A}(\theta)\) from the interpolated function.

Having a continuous function \(\mathbf{A}(\theta)\) is also meaningful for calibration process, because when numerical simulations and physical experiments disagree, we can calibrate \(\theta\) by adjusting \(\mathbf{A}\) values while keeping its robustness.

\subsection{Comparison to unitary D-MORPH}

We mentioned earlier in \autoref{sec:level-set-exploration} that our RIPV algorithm is inspired by the unitary D-MORPH algorithm by J. Dominy and H. Rabitz~\cite{dominyExploringFamiliesQuantum2008} with some important modifications.

\paragraph*{\bf Elementary description.}
RIPV specifically avoids the descriptive language of differential manifolds. Because, despite being extremely accurate and enlightening, it requires a lot of prior knowledge on the subject of manifolds, which is not necessarily familiar to quantum computing scientists and engineers. Instead, RIPV adopts the language used in optimization algorithms such as gradient descent.


\paragraph*{\bf Versatility.}
The RIPV, or more generally the GOV algorithm, is not limited to maintaining robustness. It makes no assumption on the \emph{structure} or the \emph{number} of the constraint functions. Calculations of the unitary D-MORPH were based on the goal to keep either the gate or the gate time fixed while changing the pulse. It cannot keep both fixed. But GOV can keep any number of constraints unchanged, regardless of the type of constraints, while improving one merit of the control or just randomly exploring the parameter space.

\section{Numerical examples}

In order to substantiate the capability to generate a continuous sequence of control pulses for parametric gates, several numerical experiments are conducted. Prior to examining these experiments, we introduce a tool for better visualization.

In this section, the notation \(\mathcal{S}^k\) means susceptibility defined by the Magnus expansion, i.e., the \(\mathcal{S}^k_\text{(M)}\) defined in \autoref{sec:asymptotic-robustness}.

\subsection{Quantum Error Evolution Diagram}

When displaying our numerical results, we use a diagram called Quantum Error Evolution Diagram (QEED)~\cite{barnes2022generating,haiUniversalRobustQuantum2023}, to manifest how error is accumulated dynamically during the quantum operation. To put it simply, for a two-level quantum system subject to noise on operator \(\sigma\), the \(M_1(t)\) in the Magnus expansion of the error evolution (see \autoref{eq:magnus-terms}) can be mapped to a 3D curve \(\mathbf{r}_\sigma(t)\) parametrized by \(t\), referred to as the \emph{error curve}. The distance from \(\mathbf{r}_\sigma(t)\) to the origin represents the first-order susceptibility \(\mathcal{S}_\text{(M)}^1\). In the presence of three noise sources within this two-level system, namely \(\sigma_x\), \(\sigma_y\), and \(\sigma_z\), there are three distinct error curves, \(\mathbf{r}_{\sigma_x}\), \(\mathbf{r}_{\sigma_y}\), and \(\mathbf{r}_{\sigma_z}\), each being a 3D curve.

The curves \(\mathbf{r}_j(t)\) (where \(j=\sigma_x, \sigma_y, \sigma_z\)) are defined by the 1st-order term of the Magnus expansion of the error evolution \(U_\text{n}^\text{sc}(t)\) associated with noise on \(H_\text{n} = \delta_x \sigma_x, \delta_y \sigma_y, \delta \sigma_z\). Since \(U_\text{n}^\text{sc}(t)\) encodes information about the applied control, the QEED and the control can be mutually reconstructed from one another. Additionally, we can directly read the asymptotic robustness, as defined by the Magnus expansion, from the QEED. Concretely, if the control corrects quasi-static noise \(\delta \sigma\) to the first order, i.e., \(\mathcal{S}^1 = 0\), then the curve \(\mathbf{r}_\sigma\) should form a closed loop. Furthermore, correcting \(\delta \sigma\) to the second order, \(\mathcal{S}^2 = 0\), implies that the projections of \(\mathbf{r}_\sigma\) onto the three coordinate planes would have a net-zero area.

When considering a single noise source and a single control, such as \(H_\text{n} = \delta \sigma_z\) and \(H_\text{c} = \frac{\Omega(t)}{2} \sigma_x\), the resulting error curve \(\mathbf{r}_{\sigma_z}\) is confined to a 2D subspace. This subspace, spanned by \(\sigma_z\) and \(\sigma_y\), lies perpendicular to the control direction. In this simpler case, we can directly read one more piece of information from the QEED, that the rotation angle \(\theta = \int_{0}^{T} \Omega(t) \, dt\) is the angle between the curve's tangent vectors at the starting and end points. Furthermore, if the curve \(\mathbf{r}_{\sigma_z}\) forms a closed loop, this indicates that the control is first-order robust to \(\sigma_z\) noise. A zero net area enclosed by the curve -- such as in an ``8-shaped'' trajectory -- implies second-order robustness. Hence, in this scenario, we will leverage the QEEDs, in combination with the control pulses, to directly demonstrate the robustness of the control to the readers.

\subsection{Single qubit gates}

In the rotation picture \(e^{-i t \omega_d \sigma_z}\) at the drive frequency,
the most commonly used single qubit dynamic model is \(H(t) = \frac{\Delta}{2} \sigma_z + \frac{\Omega_x(t)}{2} \sigma_x + \frac{\Omega_y(t)}{2} \sigma_y\), where \(\Delta = \omega_d - \omega_q\) is the detuning between drive frequency \(\omega_d\) and qubit frequency \(\omega_q\).
We assume on-resonance control, where \(\Delta=0\) and the dynamics of the qubit is governed by \(H(t) = \frac{\Omega_x(t)}{2} \sigma_x + \frac{\Omega_y(t)}{2} \sigma_y\).

\subsubsection{Single noise source}

In this part, we assume the noise is on \(\sigma_z\). According to Hai et al.~\cite{haiUniversalRobustQuantum2023}, we need only one control on either one of the perpendicular axes, which means control on either \(\sigma_x\) or \(\sigma_y\). We choose \(\sigma_x\) as our control term to implement robust \(R_x(\theta)\). 
To summarize, we assume
\begin{align*}
  H_\text{c}(t) & = \frac{\Omega(t)}{2} \sigma_x \\
  H_\text{n} & = \delta \sigma_z \:,
\end{align*}
where \(\delta\) is constant during the gate operation.
The pulse \(\Omega(t)\) is parameterized by an \((2n+1)\)-dimensional vector \(\mathbf{A} = (a_0, \allowbreak a_1, \allowbreak \dots, \allowbreak a_n, \allowbreak \phi_1, \allowbreak \dots, \allowbreak \phi_n)\) as
\begin{align}
  \label{eq:pulse-parametrization}
  \Omega(t; \mathbf{A}) = \sin(\pi\tilde{t}) \left(
    a_0 + \sum_{k=1}^{n} a_k \cos (2\pi k \tilde{t} + \phi_k)
  \right) \:,
\end{align}
where \(\tilde{t} = t/T\) is time normalized by gate time \(T = 50\nanosec\).
The beginning pulse parameters used in these simulations are all from ref.~\cite{haiUniversalRobustQuantum2023}.
We will denote the pulse that implements \(R_x(\theta)\) as \(\Omega_\theta(t)\), its parametrization as \(\Omega_\theta(t; \mathbf{A}_\theta)\).

\paragraph{1st order robust \(R_x(\theta)\).}
In the first demonstration, we use 1st-order RIPV, where \(\mathcal{S}^1\) defined in \autoref{eq:def-s1} is the only constraint. We start from a 2nd order robust pulse \(\Omega_{2\pi}(t)\) implementing \(R_x(2\pi)\), parameterized by \autoref{eq:pulse-parametrization} with the values of \(\mathbf{A}_{2\pi}\) given by \(R_{\text{ex;}\perp}^{2\pi}\) in the supplementary materials of Hai's paper~\cite{haiUniversalRobustQuantum2023}.
We then vary the pulse \(\Omega_{2\pi}(t)\) with pre-variation \[\dd{\mathbf{A}^\text{pre}} = \pdv{\theta}{\mathbf{A}} \:,\]
to increase the rotation angle \(\theta\) until it reaches \(4\pi\). The step size is set to be
\[{\Delta \theta}_\text{ideal} = 0.001 \rad \:,\]
which means we need approximately \(2\pi / 0.001 \approx 6283\) iterations. With these settings, we obtain a series of pulses \(\Omega_\theta(t) = \Omega(t; \mathbf{A}_\theta)\) with \(\theta\) almost evenly distributed in \([0, 2\pi]\) with an interval \(0.001 \rad\). Dropping the global phase \(e^{-i\pi}\) (let \(\theta\) modulo \(2\pi\)), we effectively acquire control pulses capable of implementing \(R_x(\theta)\) for \(\theta\in [0, 2\pi]\).In our following discussions, this global phase will always be neglected and the beginning pulse is denoted as \(\Omega_0(t)\).

\begin{figure*}[thbp]
  \centering
  \begin{subfigure}{\textwidth}
    \setlength{\subfiglen}{0.65\textwidth}
    \sbox0{\includegraphics[width=\subfiglen-13.34pt]{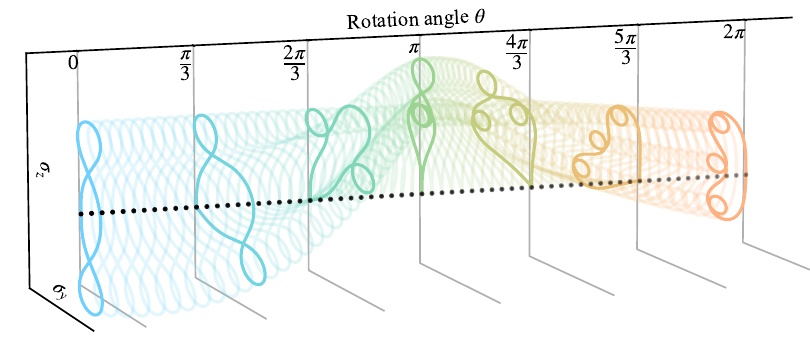}}
    \setlength{\myheight}{\ht0}
    \centering
    \begin{subfigure}[h]{\subfiglen}
      \begin{minipage}[t]{13.34pt}
        \raggedleft
        \raisebox{\dimexpr \ht0 - \topskip}{(a)}
      \end{minipage}%
      \begin{minipage}[t]{\subfiglen-13.34pt}
        \usebox0
      \end{minipage}
    \end{subfigure}
    \hspace{1pt}
    \begin{subfigure}[h]{0.3\textwidth}
      \vspace{-1pt}
      \setlength{\subfiglen}{\textwidth}
      \sbox0{\includegraphics[width=\subfiglen-13.34pt,height=0.45\myheight,keepaspectratio]{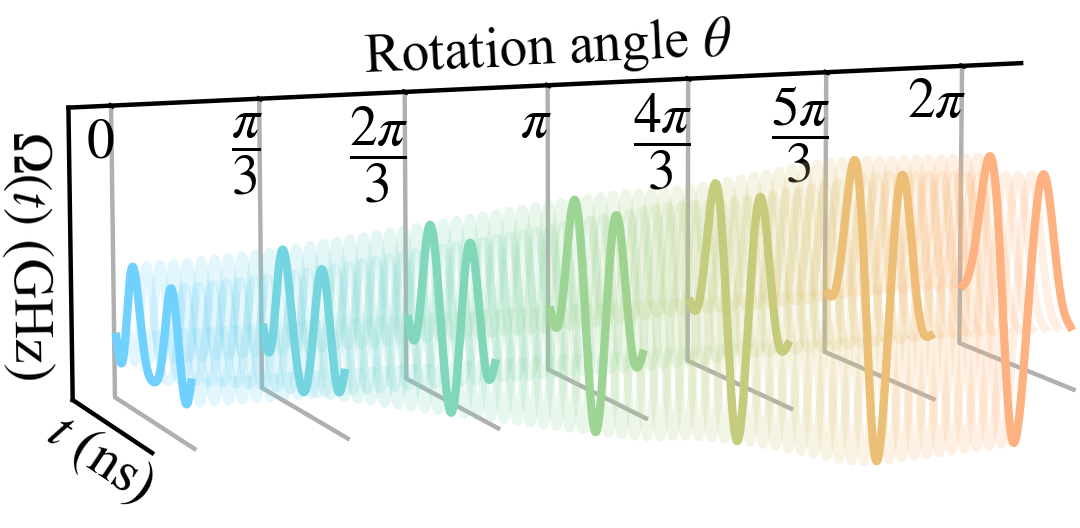}}
      \centering
      \begin{subfigure}{\subfiglen}
        \begin{minipage}[t]{13.34pt}
          \raggedleft
          \raisebox{\dimexpr \ht0 - \topskip}{(b)}
        \end{minipage}%
        \begin{minipage}[t]{\subfiglen-13.34pt}
          \usebox0
        \end{minipage}
      \end{subfigure}
      \setlength{\subfiglen}{\textwidth}
      \sbox0{\includegraphics[width=\subfiglen-13.34pt,height=0.55\myheight,keepaspectratio]{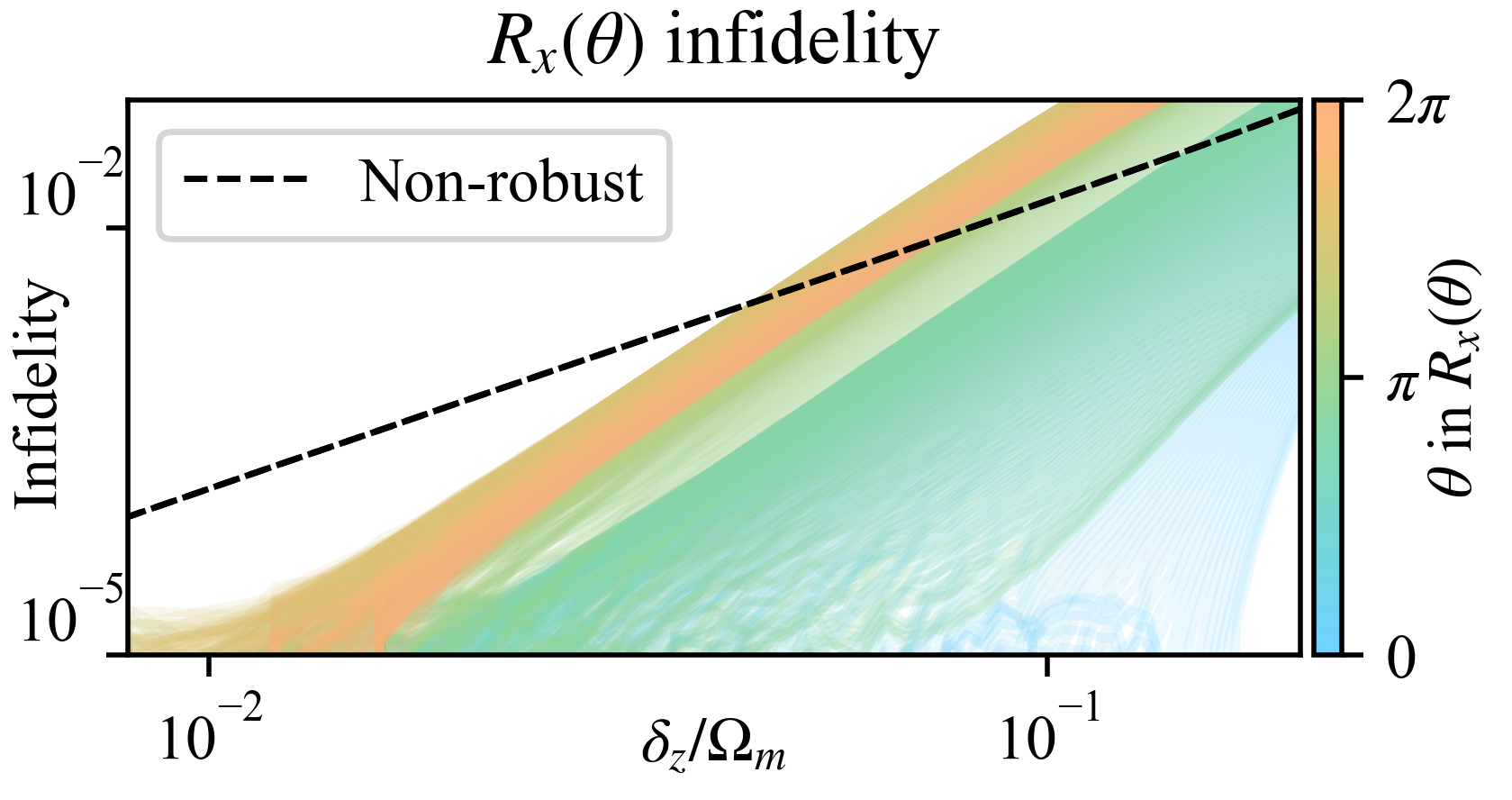}}
      \centering
      \begin{subfigure}{\subfiglen}
        \begin{minipage}[t]{13.34pt}
          \raggedleft
          \raisebox{\dimexpr \ht0 - \topskip}{(c)}
        \end{minipage}%
        \begin{minipage}[t]{\subfiglen-13.34pt}
          \usebox0
        \end{minipage}
      \end{subfigure}
    \end{subfigure}
  \end{subfigure}
  
  \begin{subfigure}{\textwidth}
    \setlength{\subfiglen}{0.65\textwidth}
    \sbox0{\includegraphics[width=\subfiglen-13.34pt]{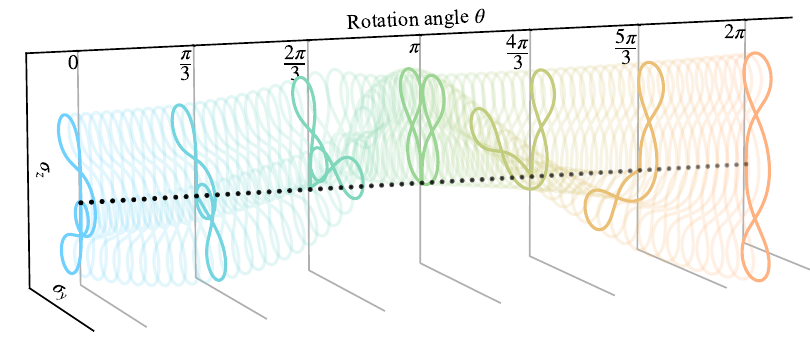}}
    \setlength{\myheight}{\ht0}
    \centering
    \begin{subfigure}[h]{\subfiglen}
      \begin{minipage}[t]{13.34pt}
        \raggedleft
        \raisebox{\dimexpr \ht0 - \topskip}{(d)}
      \end{minipage}%
      \begin{minipage}[t]{\subfiglen-13.34pt}
        \usebox0
      \end{minipage}
    \end{subfigure}
    \hspace{1pt}
    \begin{subfigure}[h]{0.3\textwidth}
      \vspace{-1pt}
      \setlength{\subfiglen}{\textwidth}
      \sbox0{\includegraphics[width=\subfiglen-13.34pt,height=0.45\myheight,keepaspectratio]{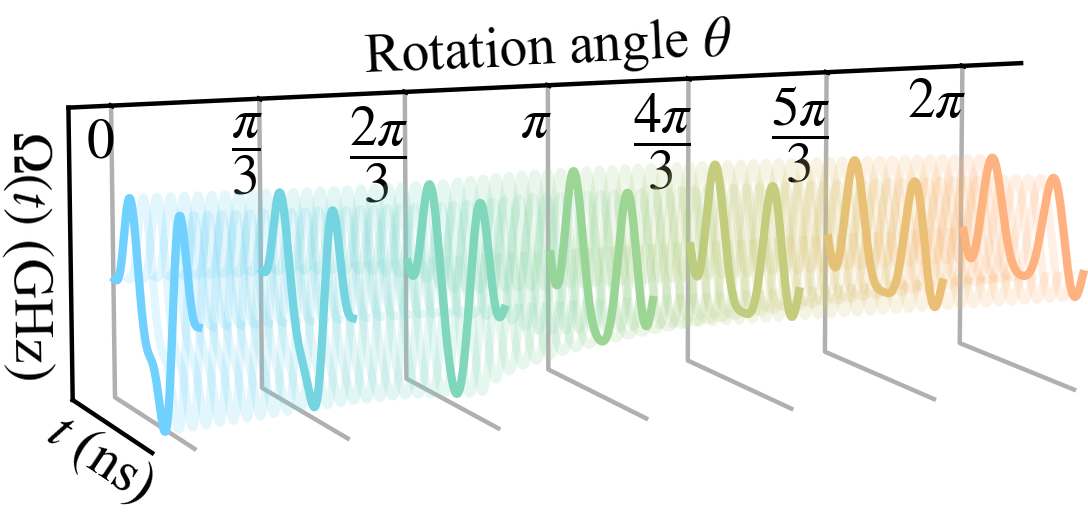}}
      \centering
      \begin{subfigure}{\subfiglen}
        \begin{minipage}[t]{13.34pt}
          \raggedleft
          \raisebox{\dimexpr \ht0 - \topskip}{(e)}
        \end{minipage}%
        \begin{minipage}[t]{\subfiglen-13.34pt}
          \usebox0
        \end{minipage}
      \end{subfigure}
      \setlength{\subfiglen}{\textwidth}
      \sbox0{\includegraphics[width=\subfiglen-13.34pt,height=0.55\myheight,keepaspectratio]{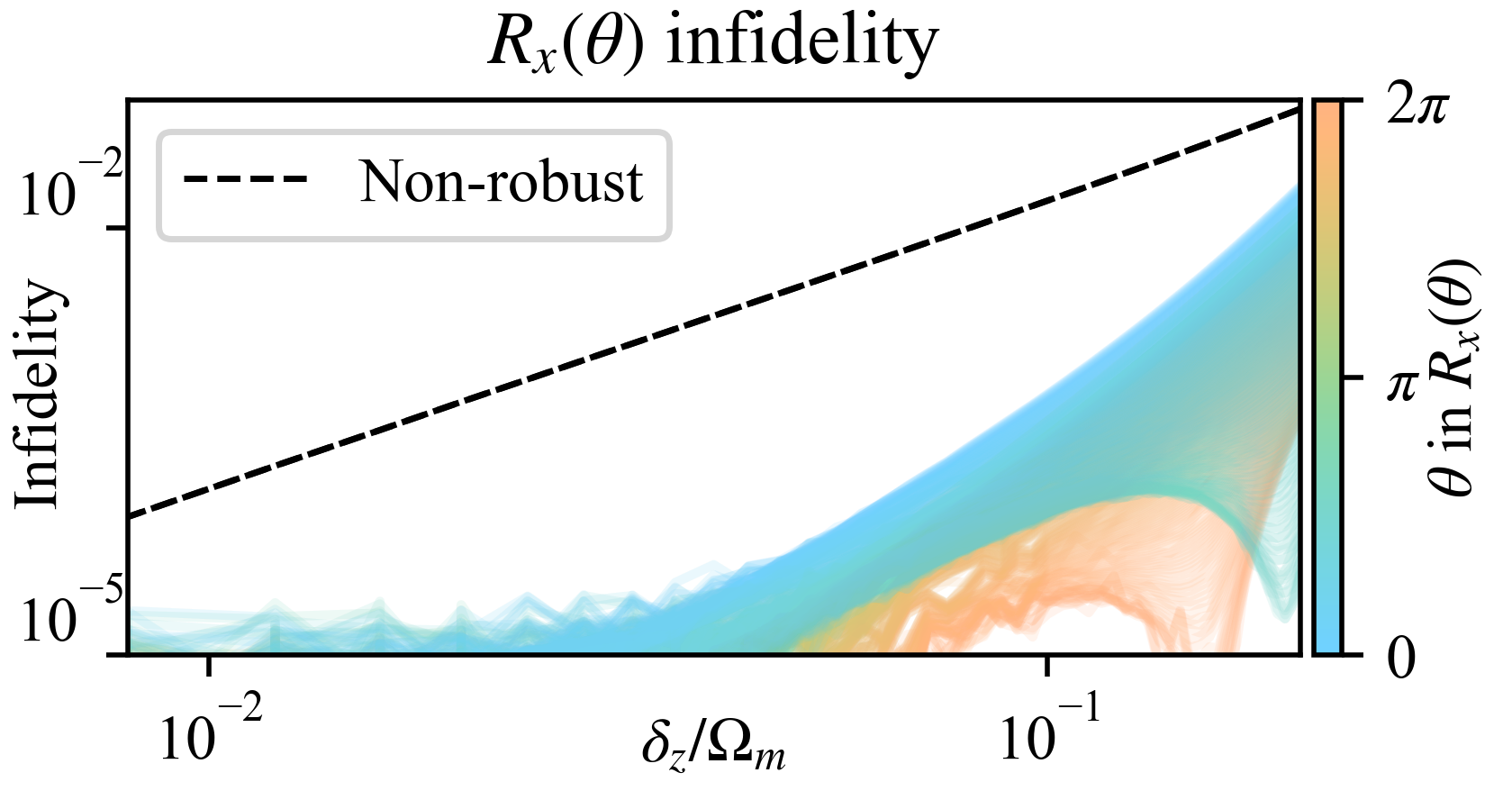}}
      \centering
      \begin{subfigure}{\subfiglen}
        \begin{minipage}[t]{13.34pt}
          \raggedleft
          \raisebox{\dimexpr \ht0 - \topskip}{(f)}
        \end{minipage}%
        \begin{minipage}[t]{\subfiglen-13.34pt}
          \usebox0
        \end{minipage}
      \end{subfigure}
    \end{subfigure}
  \end{subfigure}
  \caption{
    Robust \(R_x(\theta)\) pulses for \(\theta \in [0, 2\pi]\), each having nine parameters, using (a-c) 1st order RIPV with positive \(\Delta \theta\) in and (d-f) 2nd order with negative \(\Delta \theta\), all variations started from \(R_x(2\pi)\). In (a-c), global phase \(2\pi\) is neglected so \(R_x(\theta)\) means precisely \(R_x(\theta+2\pi)\). In each group, on the left we show the QEEDs generated by the pulses. On the right, we show the pulses on the top and fidelity vs noise on the bottom, where \(\Omega_m\) means the maximum amplitude of the pulse. The fidelity and noise strength are both in logarithm scale with negative \(\delta\) mirrored to positive half. Fidelity for curves of different \(\theta\) are stacked with transparency. The dashed lines show that for the non-robust \(\cos\) pulse implementing \(R_x(2\pi)\). In all graphs, for visualizing purposes, only about \(1/100\) of all pulses are displayed and colors are used uniformly to represent \(\theta\).
  }
  \label{fig:robustRx}
\end{figure*}

In \autoref{fig:robustRx}(a-c), we showed respectively the pulses obtained by the RIPV algorithm, the corresponding QEEDs, and the infidelity-noise graph, all colored according to the value of \(\theta\) from \(0\) to \(2\pi\).
In the QEED (a) and pulse (b) figures, since it is not easy to print an animation, we plotted the pulse series in a 3D graph for about 1 out of every 100 pulses, among which we highlight those with \(\theta = \frac{n\pi}{3}\) for \(n=0,1,2,3,4,5,6\). 
The infidelity-noise graph (c), on the other hand, is made half transparent and stacked in a 2D graph for better comparison. The noise strength is shown relative to the maximum amplitude of the pulse \(\Omega_m = \max_{t\in [0,50]} \Omega(t)\).
We also plot the infidelity for a non-robust pulse \(\Omega(t) = \sin \pi t/T\) with a dashed black line.

We can draw several conclusions by observing these graphs.
From the QEED graph \autoref{fig:robustRx}(a), we can see clearly that they all closed at the origin (black dot), indicating that \(\mathcal{S}^1\) was kept near zero for all \(\Omega_\theta(t)\) pulses.
From \autoref{fig:robustRx}(b), we see that this series of \(R_x(\theta)\) pulses is continuously changing with regard to \(\theta\). This means we can interpolate between any two pulses to get an intermediate \(\theta\) value for \(R_x(\theta)\).
In \autoref{fig:robustRx}(c), as the variation progresses (with the color changing from blue to orange-red), the fidelity plateau, although narrowing, remains above \(0.999\) despite a \(5\%\) relative noise. This plateau is ensured because, throughout the variation, \(1.414 < \mathcal{S}^1 < 1.425\). In contrast, the baseline non-robust pulse has a much higher value of \(\mathcal{S}^1 = 21.433\). Furthermore, while the robust pulses maintain a flat response for small \(\delta_z\) in the log-log infidelity plot, the non-robust pulse shows almost a linear growth.

An examination of the areas enclosed by the curves reveals a decline in 2nd-order robustness since we used 1st-order RIPV.
The beginning pulse \(\Omega_{0}(t)\) (the left-most blue one in \autoref{fig:robustRx}(a)) is 2nd-order robust to \(\sigma_z\) noise. It has \(\mathcal{S}^2 = 10.69\), which is small compared to \(T^2 = 2500\), resulting in visually zero net area. More directly, if we calculate the 2nd order robustness \(\robness^2 = 1.18\), we know this pulse cancels approximately \(1.18\) digits of \(\delta\) to the second order.
In \autoref{fig:robustRx}(c), the first few blue curves near \(\theta=0\) maintain robustness close to \(\Omega_0(t)\), and possess a similar plateau.
As the variation progressed, the 2nd order susceptibility \(\mathcal{S}^2\) becomes worse. As a result, the error curves \autoref{fig:robustRx}(a) gained larger positive net area and their infidelity rose fast as \(\theta\) goes to \(2\pi\) (color changing from blue to orange-red).

\paragraph{2nd order robust \(R_x(\theta)\).}
In the second demonstration, we aim to maintain both 1st and 2nd order robustness. We start from \(\Omega_{2\pi}(t)\) and vary backwards to get \(\Omega_\theta(t)\) for \(\theta \in [0, 2\pi]\).
The parameters \(\mathbf{A}_{2\pi}\) for the beginning pulse are still given by \(R_{\text{ex;}\perp}^{2\pi}\) in the original paper~\cite{haiUniversalRobustQuantum2023}.
With the same \(H_\text{c}\) and \(H_\text{n}\), we only change \(\dd{\mathbf{A}}^\text{pre}\) and \(\Delta\theta_\text{ideal}\) to negative values of previous settings.
Notice that since we are varying backward this time, we are not neglecting any global phase in the following discussions.

We can then see from \autoref{fig:robustRx}(d) that the error curves not only form closed loops but also maintain visually zero net area. Quantitatively, all of the pulses maintained \(1.415 < \mathcal{S}^1 < 1.440\) and \(10.45 < \mathcal{S}^2 < 10.7\), both very small compared to \(T=50\) and \(T^2 = 2500\). In \autoref{fig:robustRx}(f), compared to the results of 1st order RIPV, we are able to keep a much wider fidelity plateau as the variation progressed (with color changing from orange-red to blue, since we are decreasing \(\theta\) here). The fidelity remains above \(0.999\) under \(10\%\) noise relative to \(\Omega_m\). In comparison, the baseline non-robust pulse had \(\mathcal{S}^1 = 21.433\) and \(\mathcal{S}^2 = 127.7775\) which led to bad fidelity and almost linear growth in log-log scale.

A little compromise of keeping 2nd order robustness, though harmless, is that the dynamics become very complicated, as we can see from the error curves when the pulses are varied towards \(\theta=0\). This complicated error curve implies that the trajectories traced by qubits on the Bloch sphere would be convoluted. This complication cannot be overcome because the continuous variation of curves (or homotopy, in mathematical terms) does not change the topology of the error curve. Provided the objective is to maintain the net area close to zero, these trajectories are inherently constrained to evolve into three circular configurations as they approach \(\theta=0\).

\subsubsection{Multiple noise sources}
\label{sec:results-multiple-noise-sources}

If there are multiple noise sources, as we discussed in \autoref{sec:multiple-noise-ripv}, the algorithm is slightly different.
To suppress quasi-static noise from all three directions, we need at least 2 orthogonal controls.
The rotation angle \(\theta_j\) on direction \(\sigma_j\) where \(j=x,y,z\) is no longer a simple integral of the pulse; instead, it must be calculated by
\begin{align*}
  \theta_j=\tr \Big(\sigma_j^\dagger \big(-i \logm U_\text{sc}(T)\big) \Big) \:,
\end{align*}
where \(\logm\) stands for matrix logarithm.
Assuming the goal is to implement \(R_x(\theta)\) for practical purposes, it becomes necessary to keep five constraints constant, specifically \(\robness_x, \robness_y, \robness_z, \vartheta_y\) and \(\vartheta_z\). Thus, we maintain the integrity of robustness across three axes while keeping unwanted rotations constant (utilizing variant font \(\vartheta\) for these undesired rotations).

In this demonstration, we implement \(R_x(\theta)\) with noises on all three axes and control on two axes, namely
\begin{align*}
  H_\text{c}(t) & = \frac{\Omega_x(t)}{2} \sigma_x + \frac{\Omega_y(t)}{2} \sigma_y \\
  H_\text{n} & = \delta_x \sigma_x + \delta_y \sigma_y + \delta_z \sigma_z \:.
\end{align*}
Each pulse is parametrized as \autoref{eq:pulse-parametrization}.
We denote \(\vec{\Omega}(t) = (\Omega_x(t), \Omega_y(t))\).
The beginning pulse \(\vec{\Omega}_\pi\), from which we start the variation, implements \(R_x(\pi)\). The pulse data were given in ref.~\cite{haiUniversalRobustQuantum2023}, denoted by \(R_{1;\text{all}}^{\pi}\) in the original paper.
We used two runs to generate \(\vec{\Omega}_\theta\) for \(\theta\in [0,2\pi]\): one for \(\theta\in [\pi,2\pi]\) with \(\Delta \theta_\text{ideal} > 0\) and another for \(\theta\in [0,\pi]\) with \(\Delta \theta_\text{ideal} < 0\).
The step size was set to
\[ \abs{\Delta \theta_\text{ideal}} = 5\times 10^{-4} \rad \:. \]

\begin{figure}[thbp]
  \centering
  
  \setlength{\subfiglen}{0.45\textwidth}

  \setlength{\mpagelen}{\subfiglen-13.34pt}
  \setlength{\sboxlen}{0.37\textwidth-13.34pt}
  \sbox0{\includegraphics[width=\sboxlen]
  {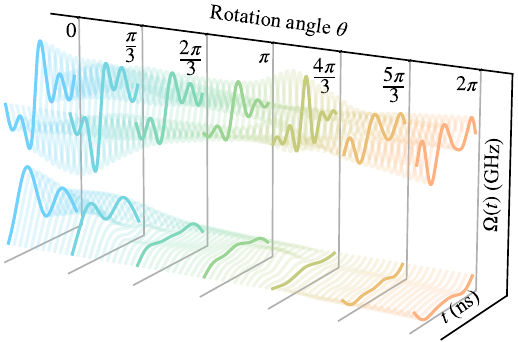}}
  \centering
  \begin{subfigure}{\subfiglen}
    \begin{minipage}[t]{13.34pt}
      \raggedleft
      \raisebox{\dimexpr \ht0 - \topskip}{(a)}
    \end{minipage}%
    \begin{minipage}[t]{\mpagelen}
      \usebox0
    \end{minipage}
  \end{subfigure}
  
  \setlength{\mpagelen}{\subfiglen-13.34pt}
  \setlength{\sboxlen}{0.43\textwidth-13.34pt}
  \sbox0{\includegraphics[width=\sboxlen]
  {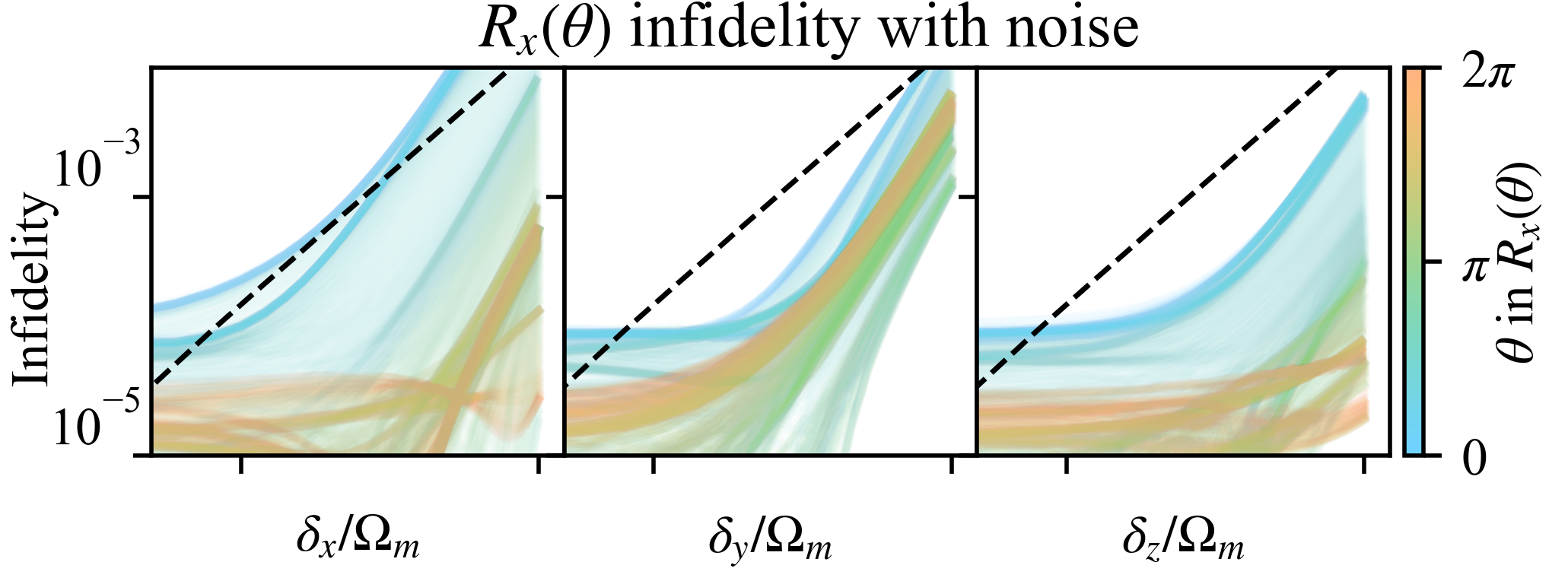}}
  \centering
  \begin{subfigure}{\subfiglen}
    \begin{minipage}[t]{13.34pt}
      \raggedleft
      \raisebox{\dimexpr 4pt+\ht0 - \topskip}{(b)}
    \end{minipage}%
    \begin{minipage}[t]{\mpagelen}
      \usebox0
    \end{minipage}
  \end{subfigure}

  \vspace{4pt}

  \begin{subfigure}[h]{\subfiglen}
    \setlength{\subfiglen}{0.49\textwidth}
    \setlength{\mpagelen}{\subfiglen-13.34pt}
    \setlength{\sboxlen}{\subfiglen}%
    \sbox0{\includegraphics[width=\sboxlen]%
    {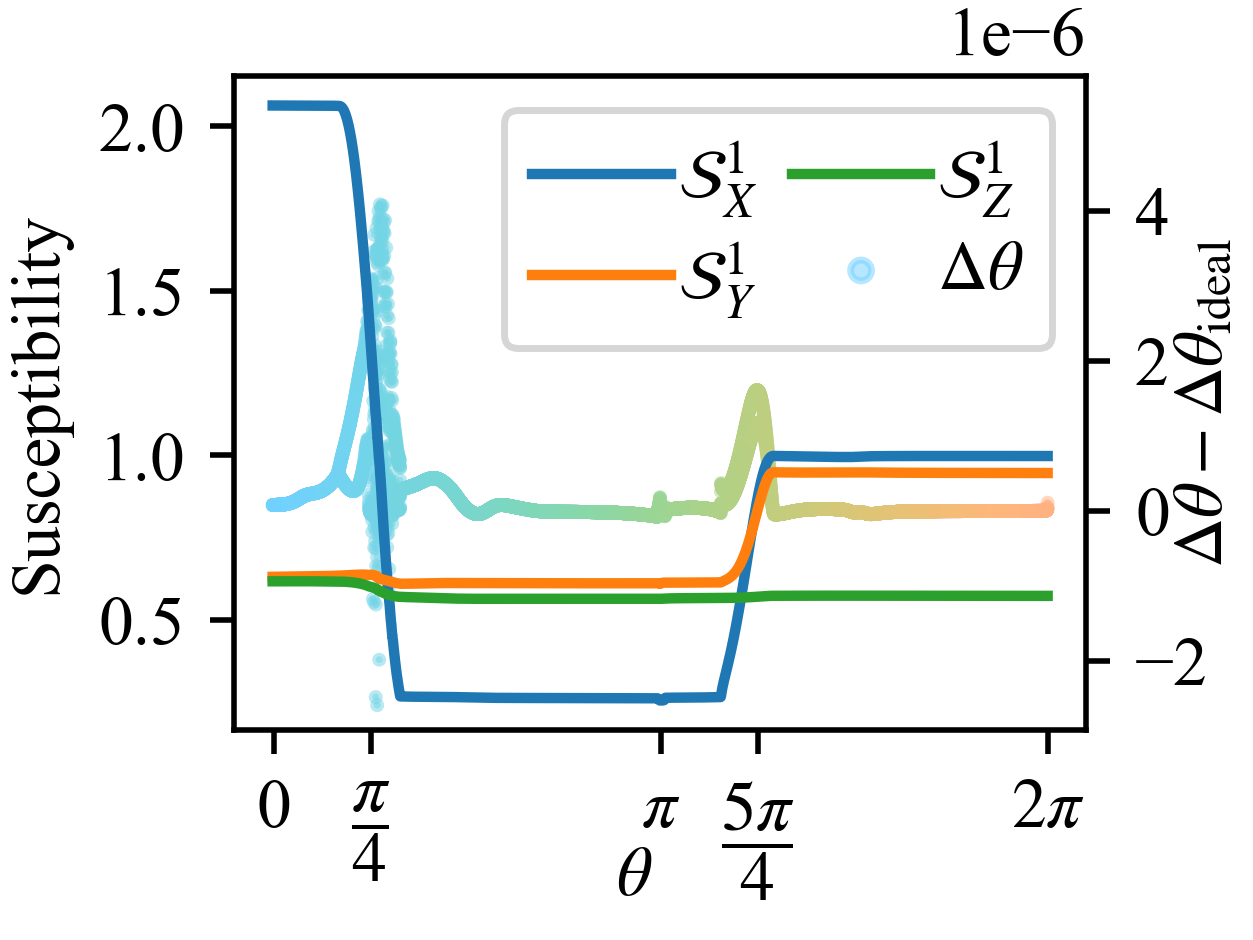}}%
    \begin{subfigure}{\subfiglen}
      \usebox0
      \put(-\wd0,\dimexpr 4pt+\ht0 - \topskip){\makebox(0,0)[lb]{(c)}}
    \end{subfigure}%
    \hfill%
    \setlength{\subfiglen}{0.49\textwidth}%
    \setlength{\mpagelen}{\subfiglen-13.34pt}%
    \setlength{\sboxlen}{\subfiglen}%
    \sbox0{\includegraphics[width=\sboxlen]%
    {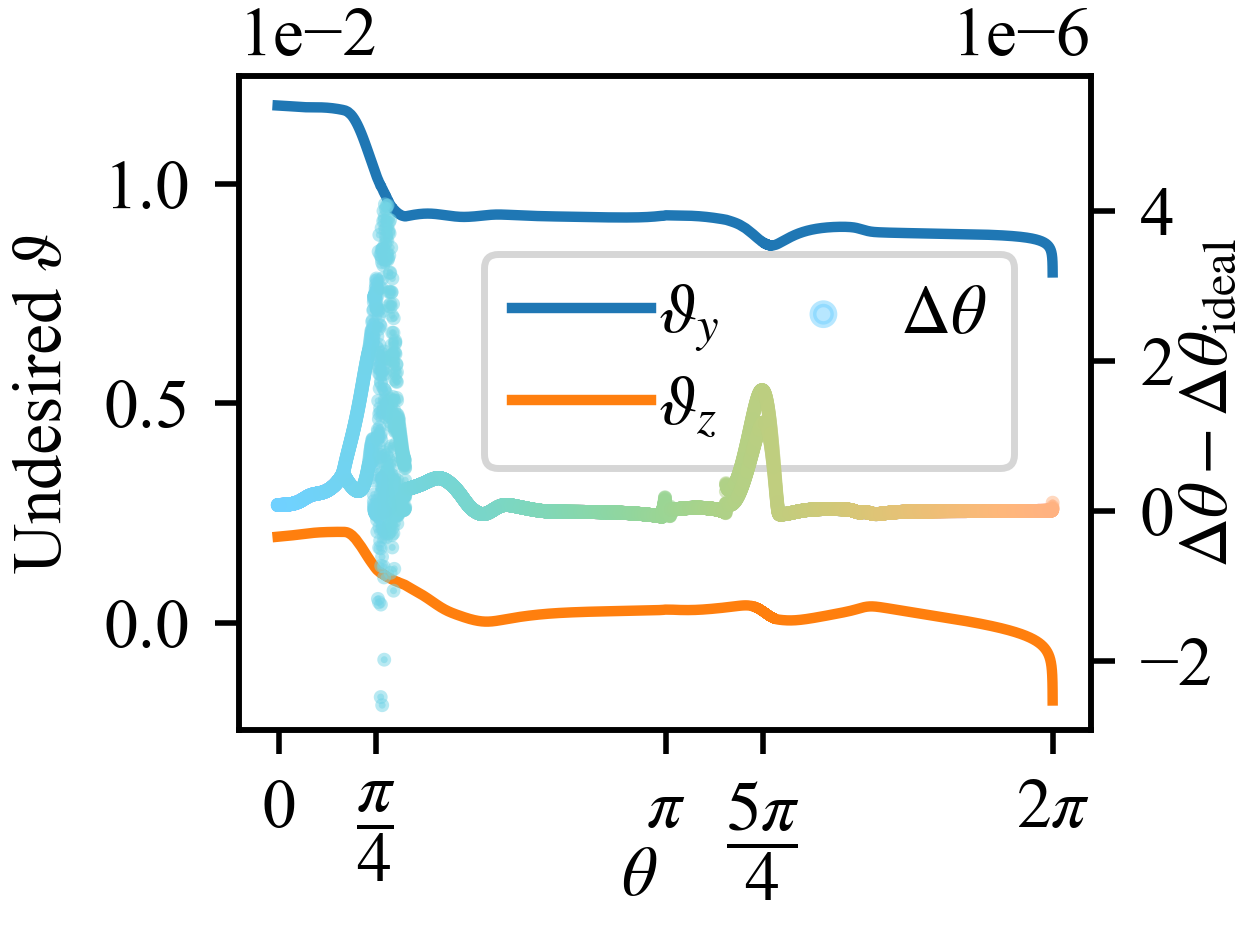}}%
    \begin{subfigure}{\subfiglen}
      \usebox0
      \put(-\wd0,\dimexpr 4pt+\ht0 - \topskip){\makebox(0,0)[lb]{(d)}}
    \end{subfigure}
  \end{subfigure}

  \caption{
    Robust \(R_x(\theta)\) pulses against noise on \(\sigma_x, \sigma_y\) and \(\sigma_z\), with control \(H_\text{c} = \frac{\Omega_x(t)}{2} \sigma_x +  \frac{\Omega_y(t)}{2} \sigma_y\). (a) Pulses as \(\theta\) goes from \(0\) to \(2\pi\). (b) Fidelity vs. noise on three directions, in logarithm scale on both axes and all stacked with transparency. The positive and negative \(\delta\) are stacked with the same color, with two ticks at \(10^{-2}\) and \(10^{-1}\). The black dashed line represents that of the baseline \(\sin\) pulse. \(\Omega_m\) is the maximum pulse amplitude. (c, d) The change of noise susceptibility on three directions and two undesired rotations. The colorful scattered plot (using the y-axis on the right) shows the change of finite difference \(\Delta \theta\) on \(\sigma_x\), relative to the ideal value \(\Delta \theta_\text{ideal} = 5\times 10^{-4}\). In all graphs, for visualizing purposes, only about \(1/100\) of all pulses are displayed and colors are used uniformly to represent \(\theta\) of \(R_x(\theta)\).
  }
  \label{fig:robust-xyz-noise}
\end{figure}

For visualization, we only show the pulses and the infidelity-noise graphs in \autoref{fig:robust-xyz-noise}(a, b), because the error curves are too many to visualize. There are now 3 curves for each pulse, each curve being 3D, which means if we project every 3D curve to three 2D axial planes, we would obtain \(3\times3=9\) curves for each pulse.

In \autoref{fig:robust-xyz-noise}(a), we see again that the pulses \(\Omega_x(t)\) (higher) and \(\Omega_y(t)\) (lower) changed continuously.
In \autoref{fig:robust-xyz-noise}(b), the infidelity-noise graphs on the three directions, we see that these pulses indeed preserved robustness very well to \(\sigma_x\), \(\sigma_y\) and \(\sigma_z\). As for \(\delta_x \sigma_x\) noise source, they are robust when \(\theta\) is larger than \(\pi/4\) (corresponding to colors other than blue), but the fidelity plateau shrank fast as the variation processes towards \(\theta=0\) (colored blue). The worst part is that when there is no noise, the ``robust pulses'' behaved worse than the non-robust pulse.
However, these issues are not a flaw of the algorithm but rather a precision problem, which can be mitigated. We discuss these phenomena below.

\textit{Why the fidelity plateau on \(\sigma_x\) noise shrank very fast?}
If we plot the 1st order noise susceptibility \(\mathcal{S}^1_j\) on three directions \(j=x,y,z\) as a function of \(\theta\) in \autoref{fig:robust-xyz-noise}(c), we can see that at around \(\theta = \pi/4\), the error susceptibility \(\mathcal{S}^1_x\) rose very fast when \(\theta\) decreases (recall that for \(\theta<\pi\), we \emph{decrease} \(\theta\) during variation). The rise in \(\mathcal{S}^1_x\) theoretically suggests that the fidelity plateau is \emph{likely} to shrink. The emphasis on \emph{likely} reflects the uncertainty, since \(\mathcal{S}^1_x\) is only an asymptotic local property defined in the limit \(\delta_x \to 0\). Nonetheless, we can say that the shrinking of the fidelity plateau is caused by the increase of \(\mathcal{S}^1_x\).

\textit{Why \(\mathcal{S}^1_x\) rose so fast?}
To understand what happened on the landscape, we plot the values of \(\Delta \theta - \Delta \theta_\text{ideal}\), the deviation of the finite difference \(\Delta \theta = \theta(\mathbf{A} + \dd{\mathbf{A}}) - \theta(\mathbf{A})\) from its ideal value \(\Delta \theta_\text{ideal}\), with scatter points and stack them in \autoref{fig:robust-xyz-noise}(c). Observe closely the range of \(\theta\) where at least one of the 1st order robustness changes, i.e., around \(\theta=\pi/4\) and \(\theta=5\pi/4\). It is no coincidence that whenever the 1st order robustness changes abruptly, the variation \(\Delta \theta\) also deviates sharply (albeit small, at the order of \(10^{-6}\)) around its ideal value \({\Delta \theta}_\text{ideal}\). Recall the definition of difference
\begin{align*}
    \Delta \theta & = \theta(\mathbf{A} + \dd{\mathbf{A}}) - \theta(\mathbf{A}) \\
    & = \dd{\mathbf{A}} \pdv{\theta}{\mathbf{A}} + O(\dd{\mathbf{A}}^2) \\
    & = \dd{\theta} + O(\dd{\mathbf{A}}^2) \:.
\end{align*}
Since in RIPV we require \(\dd{\theta} = \Delta \theta_\text{ideal}\), the large deviation of \(\Delta \theta\) signifies a fact that the landscape \(\theta(\mathbf{A})\) is highly nonlinear around those points, i.e., \(\Delta \theta - \dd{\theta} \gg 0\). Geometrically, the fast oscillation implies that the level set of the landscape \(\theta(\mathbf{A})\) is oscillating about its tangent space.

\textit{Why does the infidelity at \(\delta \approx 0\) increase as \(\theta\) decreases?}
We calculate the undesired rotations \(\vartheta_y\) and \(\vartheta_z\) to see why the infidelity rises. We see again in \autoref{fig:robust-xyz-noise}(d) that the regions where undesired rotations increase abruptly coincide with those where \(\Delta \theta\) significantly deviates from \(\Delta \theta_\text{ideal}\), both being around \(\theta = \frac{\pi}{4}\). This means, in the small noise region, the drop of fidelity as \(\theta\) approaches \(0\) is due to the highly nonlinear nature of the landscapes \(\vartheta_y(\mathbf{A})\) and \(\vartheta_z(\mathbf{A})\).

In summary, the drop in robustness and the increase of infidelity are not flaws of algorithm design, but are rooted in linear approximation. It can be solved by using smaller \(\Delta \theta_\text{ideal}\), or by adding an extra optimizing step described in \autoref{sec:extra-correcting-step} to move the parameter point back on the level set.
  
\subsection{Two qubit gates}

Implementing two-qubit parametric gates is vital to near-term quantum devices, as it provides a threefold reduction in circuit depth as compared to a standard decomposition~\cite{demoContinuousTwoQubitGates}. We can also use RIPV to get a continuous series of robust two-qubit gates, with only a change of the calculation of rotation angle and adding more undesired rotations to constraints. Yet, we show a simpler solution here.

If we apply the pulse sequence used for \(R_x(\theta)\) with single \(\sigma_x\) control to the two-qubit operator \(\sigma_x\sigma_x + \sigma_y\sigma_y\), we can achieve a parametric gate generalized from the iSWAP gate, namely the \(R_{XY}(\theta)\) gate defined as
\begin{align*}
    R_{XY} (\theta) = \exp( -i \frac{\theta}{2} \cdot \frac{\sigma_x \sigma_x + \sigma_y \sigma_y}{2} ) \:.
\end{align*}
In particular, we have \(R_{XY}(\frac{\pi}{2}) = \iswap\).

The Hamiltonian for our two-qubit system is
\begin{align*}
  H_0 = \omega_1 \sigma_z I + \omega_2 I \sigma_z + g(\sigma_x \sigma_x + \sigma_y \sigma_y) \:,
\end{align*}
where \(\omega_j\) is the frequency of qubit \(j\).
Here we only deal with the computational subspace because we would like to separate the effects of leakage from that of coherent noise.
To implement an \iswap gate on superconducting qubits, one could use tunable couplers to make the coupling strength \(g\) a driving term \(g(t)(\sigma_x \sigma_x + \sigma_y \sigma_y)\).
Hence, we set the control and the noise Hamiltonian as
\begin{align}
  \label{eq:Hc-for-RXY}
  H_\text{c}(t) & = \frac{\Omega(t)}{2} \cdot \frac{\sigma_x \sigma_x + \sigma_y \sigma_y}{2} \\
  H_\text{n} & = \frac{\delta}{2} (\sigma_z I - I \sigma_z) \:.
\end{align}
This \(H_\text{n}\) signifies the noise on the detuning of two qubits due to unknown frequency shift.

\begin{figure}[htbp]
  \centering
  \includegraphics[width=0.3\textwidth]{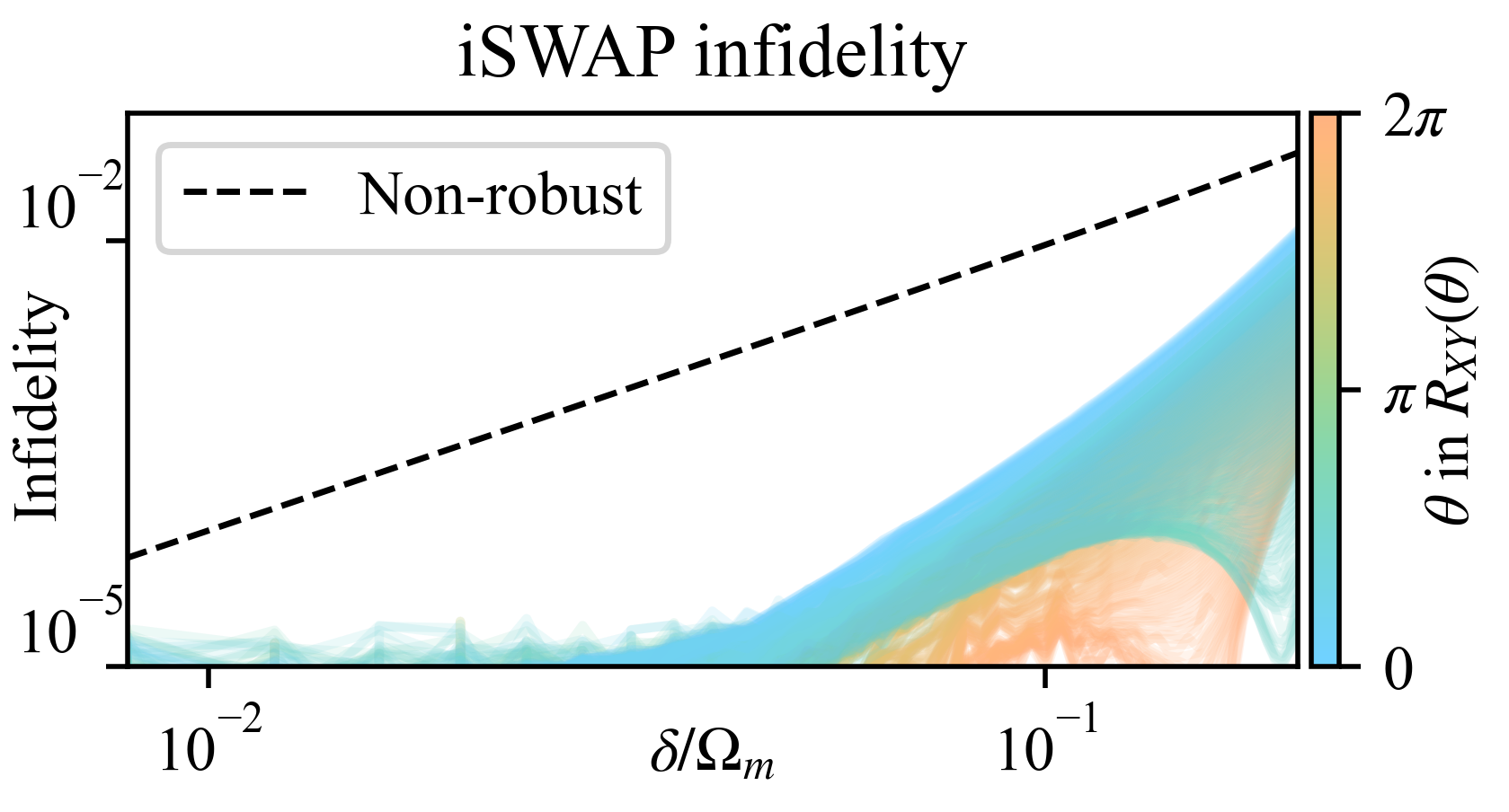}
  \caption{Infidelity-noise graph when using single control pulse sequence for parametric \iswap gate, or the \(R_{XY}(\theta)\) gate.}
  \label{fig:iswap-fidelity}
\end{figure}

Since both \(H_\text{c}\) and \(H_\text{n}\) interact with the subspace spanned by \(\{\ket{01}, \ket{10}\}\) in the same way as \(\sigma_x\) and \(\sigma_z\) interact with \(\{\ket{0}, \ket{1}\}\), we can use single-qubit \(R_x(\theta)\) pulses for  \(\Omega(t)\) in \autoref{eq:Hc-for-RXY} to implement \(R_{XY}(\theta)\).
The infidelity-noise graph of \(R_{XY}(\theta)\) using 2nd order robust pulses obtained in \autoref{sec:results-multiple-noise-sources}, is shown in \autoref{fig:iswap-fidelity}.
We can see that they are fundamentally the same with the single-qubit infidelity curves, except that the calculated infidelity is a little smaller now. This is because we are considering four levels, namely \(\ket{00}, \ket{01}, \ket{10}, \ket{11}\), only half of which are affected by the noise.

We emphasize that the RIPV algorithm is not confined to two-dimensional Hilbert space dynamics. By adjusting the calculation of rotation angles and incorporating undesired rotations as constraints, the RIPV algorithm can be adapted for general multi-qubit gates. This allows us to start from a robust pulse sequence for a general multi-qubit gate, which can then be used to generate a continuous series of pulses for parametric multi-qubit gates.

\section{Discussions}

This study inaugurates a novel paradigm in the engineering of quantum gates. As a result, it elicits numerous intriguing inquiries and presents substantial opportunities for future research. Some of these issues are subsequently addressed as follows:

\paragraph*{Further examination of QCRL properties.}

Since this is the first definition and discussion of the QCRL, many questions remain unanswered. The initial step before traversing level sets is optimizing the control’s robustness. This naturally raises questions similar to the optimization in QCL, such as the existence of local traps, the reachable set from a given control, the critical topology, and the computational complexity of optimization algorithms.
Additionally, there are questions specific to the QCRL. For instance, while a system may be controllable, it’s another matter to determine whether we can find the corresponding control configuration within the level set of the QCRL. We have observed that by traversing the level sets, we can identify arbitrary rotational angles \(R_x(\theta)\), but there is no guarantee that this applies to every quantum gate. For the QCRL, we still do not know: (1) whether a control lies within the level set (referred to as accessibility), and (2) the condition under which the level set is connected (referred to as connectivity). We probably need to look into the details of how the variation process happens in the QCRL and also examine the properties of the QCRL itself.

\begin{figure}[ht]
  \centering
  \includegraphics[width=.3\textwidth]{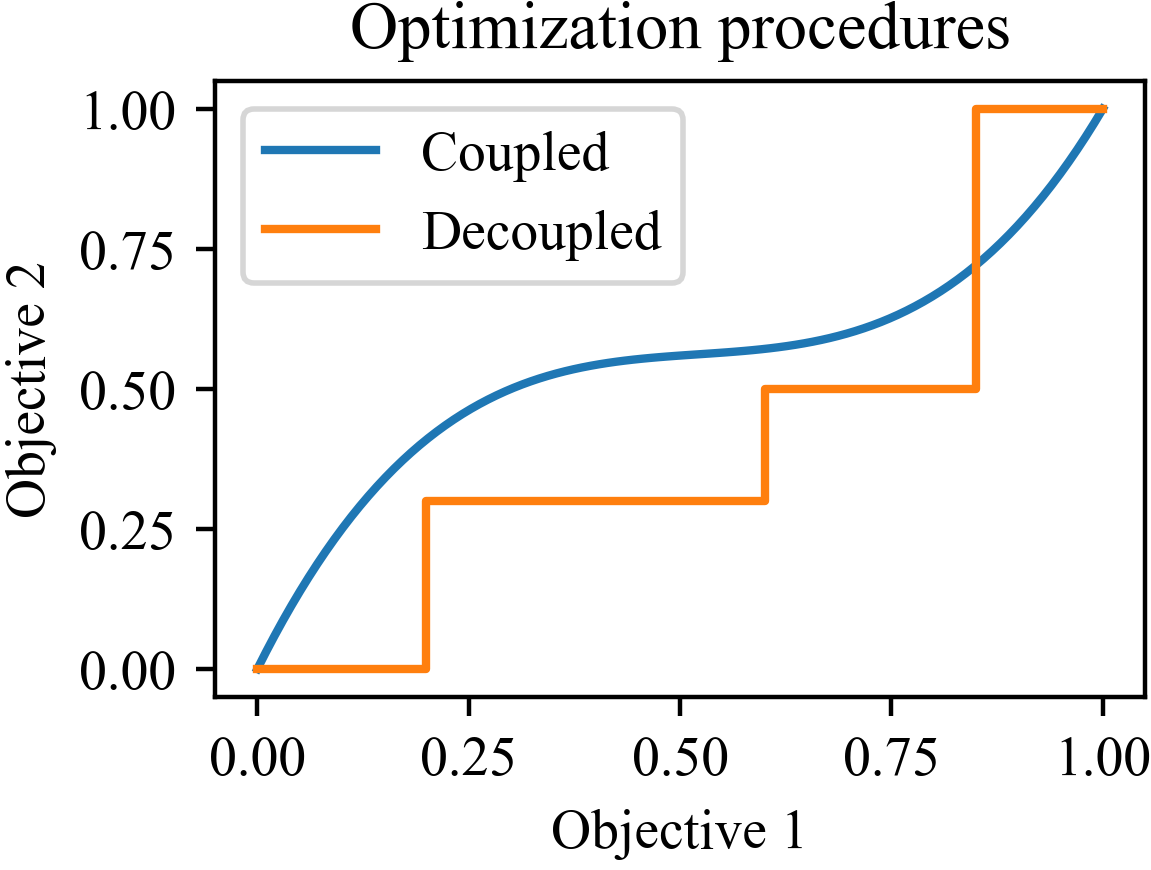}
  \caption{Decoupling two objectives.}
  \label{fig:decouple-obj}
\end{figure}

\paragraph*{Multi-objective optimization.}
Optimization of multiple objectives is challenging with traditional methods, especially when an aggregated objective function is used. This approach often leads to both objectives changing simultaneously, without control over which one changes faster or slower. This issue is illustrated in \autoref{fig:decouple-obj} by the blue curve. However, by traversing a level set of one landscape using the GOV algorithm, it is possible to optimize one objective while holding the other constant (the orange zigzag line). This decouples the two objectives, allowing for independent optimization or adjustment. By substituting robustness or gate parameters with desired properties, this approach can further help optimize factors like gate time, leakage, and energy consumption, etc.

\paragraph*{Incoherent noise models.}
In this work, we assumed the noise can be expressed as a stochastic Hamiltonian \(H_\text{n}\). Although when we treat the noise strength as a stochastic parameter, the resulting noise channel becomes incoherent, it is still valuable and more straightforward to use some open-system descriptions to describe incoherent noise channels. Then we have to change the noisy propagator \(U_\text{scn}\) in \autoref{def:full-robustness} to a non-unitary quantum channel, in general, a completely positive trace-preserving operator.

\paragraph*{Better robustness metric?} The selection of the robustness function remains specific, and there has been no systematic methodology presented for this selection. It is anticipated that further investigation into the QCRL could substantiate that its level set is sufficiently comprehensive to facilitate the implementation of any controllable parametric gate, contingent upon the satisfaction of requisite conditions.

\section{Conclusion}

In this paper, we have presented a novel framework on quantum control tasks, namely the QCRL. Unlike the traditional concept of QCL~\cite{chakrabarti2007quantum}, which is by definition dependent on the specific gate to implement, QCRL emphasizes the effects of noise. The structure of QCRL is largely determined by the robustness functions, either integral or asymptotic, and by the parametrization of control pulses. This framework enables us to find robust control for all tasks within a unified landscape, allowing us to exploit their correlations. Moreover, the level sets of the QCRL offer a rich family of controls, each equally robust against the relevant noise. Although the QCRL demonstrates utility, it presently lacks rigorous mathematical proof concerning the connectivity of level sets, critical points, and analogous topics as explored in the QCL. 

Building on top of QCRL, we have developed a paradigm-shifting algorithm called the RIPV algorithm. As the name suggests, it changes control pulses without undermining robustness. This algorithm is useful to propagate robustness from the control of one gate to a series of others, for example, when implementing a family of parametric gates. We have numerically demonstrated on single-qubit and two-qubit parametric gates, with up to two independent control terms. From the QEEDs and infidelity-versus-noise graphs, we can verify that the \(R_x(\theta)\) fidelity remains at least \(0.999\) with \(10\%\) noise on \(\sigma_z\), or at least \(0.999\) with \(2\%\) noise on three Pauli operators (percentage of pulse maximum amplitude). If we replace robustness or gate parameter with other properties we care about, we can, for example, further optimize gate time, leakage, energy consumption, etc.

At the core of RIPV is the GOV algorithm, a more versatile multi-objective algorithm. Basically, it ensures that, when we vary a point in some landscape, the data point stays in the level set with the freedom of choosing a direction called pre-variation. Choosing the direction that increases (or decreases) the gate parameter gives rise to RIPV. The GOV algorithm brings more than robust quantum control. It allows one to independently change one of many objectives in a multi-objective problem. Through GOV, we can leverage the properties of different landscapes, optimize any one of them without undermining another, and explore their level sets either purposefully or randomly.

\acknowledgments
We thank the inspiring suggestions from H. Rabitz and the fruitful discussion with R. Wu, D. Dong, and Z. Zhang. 

This work was supported by the Key-Area Research and Development Program of Guang-Dong Province (Grant No. 2018B030326001), and the Science, Technology and Innovation Commission of Shenzhen Municipality (JCYJ20170412152620376, KYTDPT20181011104202253), and the Shenzhen Science and Technology Program (KQTD20200820113010023).

\bibliography{ref}

\begin{thebibliography}{48}%
\makeatletter
\providecommand \@ifxundefined [1]{%
 \@ifx{#1\undefined}
}%
\providecommand \@ifnum [1]{%
 \ifnum #1\expandafter \@firstoftwo
 \else \expandafter \@secondoftwo
 \fi
}%
\providecommand \@ifx [1]{%
 \ifx #1\expandafter \@firstoftwo
 \else \expandafter \@secondoftwo
 \fi
}%
\providecommand \natexlab [1]{#1}%
\providecommand \enquote  [1]{``#1''}%
\providecommand \bibnamefont  [1]{#1}%
\providecommand \bibfnamefont [1]{#1}%
\providecommand \citenamefont [1]{#1}%
\providecommand \href@noop [0]{\@secondoftwo}%
\providecommand \href [0]{\begingroup \@sanitize@url \@href}%
\providecommand \@href[1]{\@@startlink{#1}\@@href}%
\providecommand \@@href[1]{\endgroup#1\@@endlink}%
\providecommand \@sanitize@url [0]{\catcode `\\12\catcode `\$12\catcode `\&12\catcode `\#12\catcode `\^12\catcode `\_12\catcode `\%12\relax}%
\providecommand \@@startlink[1]{}%
\providecommand \@@endlink[0]{}%
\providecommand \url  [0]{\begingroup\@sanitize@url \@url }%
\providecommand \@url [1]{\endgroup\@href {#1}{\urlprefix }}%
\providecommand \urlprefix  [0]{URL }%
\providecommand \Eprint [0]{\href }%
\providecommand \doibase [0]{https://doi.org/}%
\providecommand \selectlanguage [0]{\@gobble}%
\providecommand \bibinfo  [0]{\@secondoftwo}%
\providecommand \bibfield  [0]{\@secondoftwo}%
\providecommand \translation [1]{[#1]}%
\providecommand \BibitemOpen [0]{}%
\providecommand \bibitemStop [0]{}%
\providecommand \bibitemNoStop [0]{.\EOS\space}%
\providecommand \EOS [0]{\spacefactor3000\relax}%
\providecommand \BibitemShut  [1]{\csname bibitem#1\endcsname}%
\let\auto@bib@innerbib\@empty
\bibitem [{\citenamefont {Cheng}\ \emph {et~al.}(2023)\citenamefont {Cheng}, \citenamefont {Deng}, \citenamefont {Gu}, \citenamefont {He}, \citenamefont {Hu}, \citenamefont {Huang}, \citenamefont {Li}, \citenamefont {Lin}, \citenamefont {Lu}, \citenamefont {Lu} \emph {et~al.}}]{cheng2023noisy}%
  \BibitemOpen
  \bibfield  {author} {\bibinfo {author} {\bibfnamefont {B.}~\bibnamefont {Cheng}}, \bibinfo {author} {\bibfnamefont {X.-H.}\ \bibnamefont {Deng}}, \bibinfo {author} {\bibfnamefont {X.}~\bibnamefont {Gu}}, \bibinfo {author} {\bibfnamefont {Y.}~\bibnamefont {He}}, \bibinfo {author} {\bibfnamefont {G.}~\bibnamefont {Hu}}, \bibinfo {author} {\bibfnamefont {P.}~\bibnamefont {Huang}}, \bibinfo {author} {\bibfnamefont {J.}~\bibnamefont {Li}}, \bibinfo {author} {\bibfnamefont {B.-C.}\ \bibnamefont {Lin}}, \bibinfo {author} {\bibfnamefont {D.}~\bibnamefont {Lu}}, \bibinfo {author} {\bibfnamefont {Y.}~\bibnamefont {Lu}}, \emph {et~al.},\ }\bibfield  {title} {\bibinfo {title} {Noisy intermediate-scale quantum computers},\ }\href@noop {} {\bibfield  {journal} {\bibinfo  {journal} {Frontiers of Physics}\ }\textbf {\bibinfo {volume} {18}},\ \bibinfo {pages} {21308} (\bibinfo {year} {2023})}\BibitemShut {NoStop}%
\bibitem [{\citenamefont {Campbell}\ \emph {et~al.}(2017)\citenamefont {Campbell}, \citenamefont {Terhal},\ and\ \citenamefont {Vuillot}}]{campbellRoadsFaulttolerantUniversal2017}%
  \BibitemOpen
  \bibfield  {author} {\bibinfo {author} {\bibfnamefont {E.~T.}\ \bibnamefont {Campbell}}, \bibinfo {author} {\bibfnamefont {B.~M.}\ \bibnamefont {Terhal}},\ and\ \bibinfo {author} {\bibfnamefont {C.}~\bibnamefont {Vuillot}},\ }\bibfield  {title} {\bibinfo {title} {Roads towards fault-tolerant universal quantum computation},\ }\href {https://doi.org/10.1038/nature23460} {\bibfield  {journal} {\bibinfo  {journal} {Nature}\ }\textbf {\bibinfo {volume} {549}},\ \bibinfo {pages} {172} (\bibinfo {year} {2017})}\BibitemShut {NoStop}%
\bibitem [{\citenamefont {Koch}\ \emph {et~al.}(2022)\citenamefont {Koch}, \citenamefont {Boscain}, \citenamefont {Calarco}, \citenamefont {Dirr}, \citenamefont {Filipp}, \citenamefont {Glaser}, \citenamefont {Kosloff}, \citenamefont {Montangero}, \citenamefont {Schulte-Herbr{\"u}ggen}, \citenamefont {Sugny} \emph {et~al.}}]{koch2022quantum}%
  \BibitemOpen
  \bibfield  {author} {\bibinfo {author} {\bibfnamefont {C.~P.}\ \bibnamefont {Koch}}, \bibinfo {author} {\bibfnamefont {U.}~\bibnamefont {Boscain}}, \bibinfo {author} {\bibfnamefont {T.}~\bibnamefont {Calarco}}, \bibinfo {author} {\bibfnamefont {G.}~\bibnamefont {Dirr}}, \bibinfo {author} {\bibfnamefont {S.}~\bibnamefont {Filipp}}, \bibinfo {author} {\bibfnamefont {S.~J.}\ \bibnamefont {Glaser}}, \bibinfo {author} {\bibfnamefont {R.}~\bibnamefont {Kosloff}}, \bibinfo {author} {\bibfnamefont {S.}~\bibnamefont {Montangero}}, \bibinfo {author} {\bibfnamefont {T.}~\bibnamefont {Schulte-Herbr{\"u}ggen}}, \bibinfo {author} {\bibfnamefont {D.}~\bibnamefont {Sugny}}, \emph {et~al.},\ }\bibfield  {title} {\bibinfo {title} {Quantum optimal control in quantum technologies. strategic report on current status, visions and goals for research in europe},\ }\href@noop {} {\bibfield  {journal} {\bibinfo  {journal} {EPJ Quantum Technology}\ }\textbf {\bibinfo {volume} {9}},\ \bibinfo {pages} {19} (\bibinfo {year}
  {2022})}\BibitemShut {NoStop}%
\bibitem [{\citenamefont {Chakrabarti}\ and\ \citenamefont {Rabitz}(2007)}]{chakrabarti2007quantum}%
  \BibitemOpen
  \bibfield  {author} {\bibinfo {author} {\bibfnamefont {R.}~\bibnamefont {Chakrabarti}}\ and\ \bibinfo {author} {\bibfnamefont {H.}~\bibnamefont {Rabitz}},\ }\bibfield  {title} {\bibinfo {title} {Quantum control landscapes},\ }\href@noop {} {\bibfield  {journal} {\bibinfo  {journal} {International Reviews in Physical Chemistry}\ }\textbf {\bibinfo {volume} {26}},\ \bibinfo {pages} {671} (\bibinfo {year} {2007})}\BibitemShut {NoStop}%
\bibitem [{\citenamefont {Ge}\ \emph {et~al.}(2022)\citenamefont {Ge}, \citenamefont {Wu},\ and\ \citenamefont {Rabitz}}]{ge2022optimization}%
  \BibitemOpen
  \bibfield  {author} {\bibinfo {author} {\bibfnamefont {X.}~\bibnamefont {Ge}}, \bibinfo {author} {\bibfnamefont {R.-B.}\ \bibnamefont {Wu}},\ and\ \bibinfo {author} {\bibfnamefont {H.}~\bibnamefont {Rabitz}},\ }\bibfield  {title} {\bibinfo {title} {The optimization landscape of hybrid quantum--classical algorithms: From quantum control to nisq applications},\ }\href@noop {} {\bibfield  {journal} {\bibinfo  {journal} {Annual Reviews in Control}\ }\textbf {\bibinfo {volume} {54}},\ \bibinfo {pages} {314} (\bibinfo {year} {2022})}\BibitemShut {NoStop}%
\bibitem [{\citenamefont {Rothman}\ \emph {et~al.}(2006)\citenamefont {Rothman}, \citenamefont {Ho},\ and\ \citenamefont {Rabitz}}]{rothmanExploringLevelSets2006}%
  \BibitemOpen
  \bibfield  {author} {\bibinfo {author} {\bibfnamefont {A.}~\bibnamefont {Rothman}}, \bibinfo {author} {\bibfnamefont {T.-S.}\ \bibnamefont {Ho}},\ and\ \bibinfo {author} {\bibfnamefont {H.}~\bibnamefont {Rabitz}},\ }\bibfield  {title} {\bibinfo {title} {Exploring the level sets of quantum control landscapes},\ }\href {https://doi.org/10.1103/PhysRevA.73.053401} {\bibfield  {journal} {\bibinfo  {journal} {Physical Review A}\ }\textbf {\bibinfo {volume} {73}},\ \bibinfo {pages} {053401} (\bibinfo {year} {2006})}\BibitemShut {NoStop}%
\bibitem [{\citenamefont {Dominy}\ and\ \citenamefont {Rabitz}(2008)}]{dominyExploringFamiliesQuantum2008}%
  \BibitemOpen
  \bibfield  {author} {\bibinfo {author} {\bibfnamefont {J.}~\bibnamefont {Dominy}}\ and\ \bibinfo {author} {\bibfnamefont {H.}~\bibnamefont {Rabitz}},\ }\bibfield  {title} {\bibinfo {title} {Exploring families of quantum controls for generating unitary transformations},\ }\href {https://doi.org/10.1088/1751-8113/41/20/205305} {\bibfield  {journal} {\bibinfo  {journal} {Journal of Physics A: Mathematical and Theoretical}\ }\textbf {\bibinfo {volume} {41}},\ \bibinfo {pages} {205305} (\bibinfo {year} {2008})}\BibitemShut {NoStop}%
\bibitem [{\citenamefont {Chen}\ \emph {et~al.}(2015)\citenamefont {Chen}, \citenamefont {Wu}, \citenamefont {Zhang},\ and\ \citenamefont {Rabitz}}]{chen2015near}%
  \BibitemOpen
  \bibfield  {author} {\bibinfo {author} {\bibfnamefont {Q.-M.}\ \bibnamefont {Chen}}, \bibinfo {author} {\bibfnamefont {R.-B.}\ \bibnamefont {Wu}}, \bibinfo {author} {\bibfnamefont {T.-M.}\ \bibnamefont {Zhang}},\ and\ \bibinfo {author} {\bibfnamefont {H.}~\bibnamefont {Rabitz}},\ }\bibfield  {title} {\bibinfo {title} {Near-time-optimal control for quantum systems},\ }\href@noop {} {\bibfield  {journal} {\bibinfo  {journal} {Physical Review A}\ }\textbf {\bibinfo {volume} {92}},\ \bibinfo {pages} {063415} (\bibinfo {year} {2015})}\BibitemShut {NoStop}%
\bibitem [{\citenamefont {Souza}\ \emph {et~al.}(2011)\citenamefont {Souza}, \citenamefont {Alvarez},\ and\ \citenamefont {Suter}}]{souza2011robust}%
  \BibitemOpen
  \bibfield  {author} {\bibinfo {author} {\bibfnamefont {A.~M.}\ \bibnamefont {Souza}}, \bibinfo {author} {\bibfnamefont {G.~A.}\ \bibnamefont {Alvarez}},\ and\ \bibinfo {author} {\bibfnamefont {D.}~\bibnamefont {Suter}},\ }\bibfield  {title} {\bibinfo {title} {Robust dynamical decoupling for quantum computing and quantum memory},\ }\href@noop {} {\bibfield  {journal} {\bibinfo  {journal} {Physical review letters}\ }\textbf {\bibinfo {volume} {106}},\ \bibinfo {pages} {240501} (\bibinfo {year} {2011})}\BibitemShut {NoStop}%
\bibitem [{\citenamefont {Genov}\ \emph {et~al.}(2014)\citenamefont {Genov}, \citenamefont {Schraft}, \citenamefont {Halfmann},\ and\ \citenamefont {Vitanov}}]{genov2014correction}%
  \BibitemOpen
  \bibfield  {author} {\bibinfo {author} {\bibfnamefont {G.~T.}\ \bibnamefont {Genov}}, \bibinfo {author} {\bibfnamefont {D.}~\bibnamefont {Schraft}}, \bibinfo {author} {\bibfnamefont {T.}~\bibnamefont {Halfmann}},\ and\ \bibinfo {author} {\bibfnamefont {N.~V.}\ \bibnamefont {Vitanov}},\ }\bibfield  {title} {\bibinfo {title} {Correction of arbitrary field errors in population inversion of quantum systems by universal composite pulses},\ }\href@noop {} {\bibfield  {journal} {\bibinfo  {journal} {Physical review letters}\ }\textbf {\bibinfo {volume} {113}},\ \bibinfo {pages} {043001} (\bibinfo {year} {2014})}\BibitemShut {NoStop}%
\bibitem [{\citenamefont {Zhang}\ \emph {et~al.}(2023)\citenamefont {Zhang}, \citenamefont {Kyaw}, \citenamefont {Filipp}, \citenamefont {Kwek}, \citenamefont {Sj{\"o}qvist},\ and\ \citenamefont {Tong}}]{zhang2023geometric}%
  \BibitemOpen
  \bibfield  {author} {\bibinfo {author} {\bibfnamefont {J.}~\bibnamefont {Zhang}}, \bibinfo {author} {\bibfnamefont {T.~H.}\ \bibnamefont {Kyaw}}, \bibinfo {author} {\bibfnamefont {S.}~\bibnamefont {Filipp}}, \bibinfo {author} {\bibfnamefont {L.-C.}\ \bibnamefont {Kwek}}, \bibinfo {author} {\bibfnamefont {E.}~\bibnamefont {Sj{\"o}qvist}},\ and\ \bibinfo {author} {\bibfnamefont {D.}~\bibnamefont {Tong}},\ }\bibfield  {title} {\bibinfo {title} {Geometric and holonomic quantum computation},\ }\href@noop {} {\bibfield  {journal} {\bibinfo  {journal} {Physics Reports}\ }\textbf {\bibinfo {volume} {1027}},\ \bibinfo {pages} {1} (\bibinfo {year} {2023})}\BibitemShut {NoStop}%
\bibitem [{\citenamefont {Khodjasteh}\ and\ \citenamefont {Viola}(2009)}]{khodjasteh2009dynamically}%
  \BibitemOpen
  \bibfield  {author} {\bibinfo {author} {\bibfnamefont {K.}~\bibnamefont {Khodjasteh}}\ and\ \bibinfo {author} {\bibfnamefont {L.}~\bibnamefont {Viola}},\ }\bibfield  {title} {\bibinfo {title} {Dynamically error-corrected gates for universal quantum computation},\ }\href@noop {} {\bibfield  {journal} {\bibinfo  {journal} {Physical review letters}\ }\textbf {\bibinfo {volume} {102}},\ \bibinfo {pages} {080501} (\bibinfo {year} {2009})}\BibitemShut {NoStop}%
\bibitem [{\citenamefont {Zeng}\ \emph {et~al.}(2018)\citenamefont {Zeng}, \citenamefont {Deng}, \citenamefont {Russo},\ and\ \citenamefont {Barnes}}]{zengGeneralSolutionInhomogeneous2018}%
  \BibitemOpen
  \bibfield  {author} {\bibinfo {author} {\bibfnamefont {J.}~\bibnamefont {Zeng}}, \bibinfo {author} {\bibfnamefont {X.-H.}\ \bibnamefont {Deng}}, \bibinfo {author} {\bibfnamefont {A.}~\bibnamefont {Russo}},\ and\ \bibinfo {author} {\bibfnamefont {E.}~\bibnamefont {Barnes}},\ }\bibfield  {title} {\bibinfo {title} {General solution to inhomogeneous dephasing and smooth pulse dynamical decoupling},\ }\href {https://doi.org/10.1088/1367-2630/aaafe9} {\bibfield  {journal} {\bibinfo  {journal} {New Journal of Physics}\ }\textbf {\bibinfo {volume} {20}},\ \bibinfo {pages} {033011} (\bibinfo {year} {2018})}\BibitemShut {NoStop}%
\bibitem [{\citenamefont {Dong}\ and\ \citenamefont {Petersen}(2023)}]{dong2023learning}%
  \BibitemOpen
  \bibfield  {author} {\bibinfo {author} {\bibfnamefont {D.}~\bibnamefont {Dong}}\ and\ \bibinfo {author} {\bibfnamefont {I.~R.}\ \bibnamefont {Petersen}},\ }\href@noop {} {\emph {\bibinfo {title} {Learning and robust control in quantum technology}}}\ (\bibinfo  {publisher} {Springer},\ \bibinfo {year} {2023})\BibitemShut {NoStop}%
\bibitem [{\citenamefont {Cao}\ \emph {et~al.}(2024)\citenamefont {Cao}, \citenamefont {Cui}, \citenamefont {Yung},\ and\ \citenamefont {Wu}}]{cao2024robust}%
  \BibitemOpen
  \bibfield  {author} {\bibinfo {author} {\bibfnamefont {X.}~\bibnamefont {Cao}}, \bibinfo {author} {\bibfnamefont {J.}~\bibnamefont {Cui}}, \bibinfo {author} {\bibfnamefont {M.~H.}\ \bibnamefont {Yung}},\ and\ \bibinfo {author} {\bibfnamefont {R.-B.}\ \bibnamefont {Wu}},\ }\bibfield  {title} {\bibinfo {title} {Robust control of single-qubit gates at the quantum speed limit},\ }\href@noop {} {\bibfield  {journal} {\bibinfo  {journal} {Physical Review A}\ }\textbf {\bibinfo {volume} {110}},\ \bibinfo {pages} {022603} (\bibinfo {year} {2024})}\BibitemShut {NoStop}%
\bibitem [{\citenamefont {Kosut}\ \emph {et~al.}(2022)\citenamefont {Kosut}, \citenamefont {Bhole},\ and\ \citenamefont {Rabitz}}]{kosut2022robust}%
  \BibitemOpen
  \bibfield  {author} {\bibinfo {author} {\bibfnamefont {R.~L.}\ \bibnamefont {Kosut}}, \bibinfo {author} {\bibfnamefont {G.}~\bibnamefont {Bhole}},\ and\ \bibinfo {author} {\bibfnamefont {H.}~\bibnamefont {Rabitz}},\ }\bibfield  {title} {\bibinfo {title} {Robust quantum control: Analysis \& synthesis via averaging},\ }\href@noop {} {\bibfield  {journal} {\bibinfo  {journal} {arXiv preprint arXiv:2208.14193}\ } (\bibinfo {year} {2022})}\BibitemShut {NoStop}%
\bibitem [{\citenamefont {Ramakrishna}\ \emph {et~al.}(1995)\citenamefont {Ramakrishna}, \citenamefont {Salapaka}, \citenamefont {Dahleh}, \citenamefont {Rabitz},\ and\ \citenamefont {Peirce}}]{ramakrishna1995controllability}%
  \BibitemOpen
  \bibfield  {author} {\bibinfo {author} {\bibfnamefont {V.}~\bibnamefont {Ramakrishna}}, \bibinfo {author} {\bibfnamefont {M.~V.}\ \bibnamefont {Salapaka}}, \bibinfo {author} {\bibfnamefont {M.}~\bibnamefont {Dahleh}}, \bibinfo {author} {\bibfnamefont {H.}~\bibnamefont {Rabitz}},\ and\ \bibinfo {author} {\bibfnamefont {A.}~\bibnamefont {Peirce}},\ }\bibfield  {title} {\bibinfo {title} {Controllability of molecular systems},\ }\href@noop {} {\bibfield  {journal} {\bibinfo  {journal} {Physical Review A}\ }\textbf {\bibinfo {volume} {51}},\ \bibinfo {pages} {960} (\bibinfo {year} {1995})}\BibitemShut {NoStop}%
\bibitem [{\citenamefont {Fu}\ \emph {et~al.}(2001)\citenamefont {Fu}, \citenamefont {Schirmer},\ and\ \citenamefont {Solomon}}]{fu2001complete}%
  \BibitemOpen
  \bibfield  {author} {\bibinfo {author} {\bibfnamefont {H.}~\bibnamefont {Fu}}, \bibinfo {author} {\bibfnamefont {S.~G.}\ \bibnamefont {Schirmer}},\ and\ \bibinfo {author} {\bibfnamefont {A.~I.}\ \bibnamefont {Solomon}},\ }\bibfield  {title} {\bibinfo {title} {Complete controllability of finite-level quantum systems},\ }\href@noop {} {\bibfield  {journal} {\bibinfo  {journal} {Journal of Physics A: Mathematical and General}\ }\textbf {\bibinfo {volume} {34}},\ \bibinfo {pages} {1679} (\bibinfo {year} {2001})}\BibitemShut {NoStop}%
\bibitem [{\citenamefont {Schirmer}\ \emph {et~al.}(2001)\citenamefont {Schirmer}, \citenamefont {Fu},\ and\ \citenamefont {Solomon}}]{schirmer2001complete}%
  \BibitemOpen
  \bibfield  {author} {\bibinfo {author} {\bibfnamefont {S.~G.}\ \bibnamefont {Schirmer}}, \bibinfo {author} {\bibfnamefont {H.}~\bibnamefont {Fu}},\ and\ \bibinfo {author} {\bibfnamefont {A.~I.}\ \bibnamefont {Solomon}},\ }\bibfield  {title} {\bibinfo {title} {Complete controllability of quantum systems},\ }\href@noop {} {\bibfield  {journal} {\bibinfo  {journal} {Physical Review A}\ }\textbf {\bibinfo {volume} {63}},\ \bibinfo {pages} {063410} (\bibinfo {year} {2001})}\BibitemShut {NoStop}%
\bibitem [{\citenamefont {Polack}\ \emph {et~al.}(2009)\citenamefont {Polack}, \citenamefont {Suchowski},\ and\ \citenamefont {Tannor}}]{polack2009uncontrollable}%
  \BibitemOpen
  \bibfield  {author} {\bibinfo {author} {\bibfnamefont {T.}~\bibnamefont {Polack}}, \bibinfo {author} {\bibfnamefont {H.}~\bibnamefont {Suchowski}},\ and\ \bibinfo {author} {\bibfnamefont {D.~J.}\ \bibnamefont {Tannor}},\ }\bibfield  {title} {\bibinfo {title} {Uncontrollable quantum systems: A classification scheme based on lie subalgebras},\ }\href@noop {} {\bibfield  {journal} {\bibinfo  {journal} {Physical Review A}\ }\textbf {\bibinfo {volume} {79}},\ \bibinfo {pages} {053403} (\bibinfo {year} {2009})}\BibitemShut {NoStop}%
\bibitem [{\citenamefont {Rabitz}\ \emph {et~al.}(2004)\citenamefont {Rabitz}, \citenamefont {Hsieh},\ and\ \citenamefont {Rosenthal}}]{rabitzQuantumOptimallyControlled2004}%
  \BibitemOpen
  \bibfield  {author} {\bibinfo {author} {\bibfnamefont {H.~A.}\ \bibnamefont {Rabitz}}, \bibinfo {author} {\bibfnamefont {M.~M.}\ \bibnamefont {Hsieh}},\ and\ \bibinfo {author} {\bibfnamefont {C.~M.}\ \bibnamefont {Rosenthal}},\ }\bibfield  {title} {\bibinfo {title} {Quantum {{Optimally Controlled Transition Landscapes}}},\ }\href {https://doi.org/10.1126/science.1093649} {\bibfield  {journal} {\bibinfo  {journal} {Science}\ }\textbf {\bibinfo {volume} {303}},\ \bibinfo {pages} {1998} (\bibinfo {year} {2004})}\BibitemShut {NoStop}%
\bibitem [{\citenamefont {Ho}\ \emph {et~al.}(2009)\citenamefont {Ho}, \citenamefont {Dominy},\ and\ \citenamefont {Rabitz}}]{hoLandscapeUnitaryTransformations2009}%
  \BibitemOpen
  \bibfield  {author} {\bibinfo {author} {\bibfnamefont {T.-S.}\ \bibnamefont {Ho}}, \bibinfo {author} {\bibfnamefont {J.}~\bibnamefont {Dominy}},\ and\ \bibinfo {author} {\bibfnamefont {H.}~\bibnamefont {Rabitz}},\ }\bibfield  {title} {\bibinfo {title} {Landscape of unitary transformations in controlled quantum dynamics},\ }\href {https://doi.org/10.1103/PhysRevA.79.013422} {\bibfield  {journal} {\bibinfo  {journal} {Physical Review A}\ }\textbf {\bibinfo {volume} {79}},\ \bibinfo {pages} {013422} (\bibinfo {year} {2009})}\BibitemShut {NoStop}%
\bibitem [{\citenamefont {Brif}\ \emph {et~al.}(2010)\citenamefont {Brif}, \citenamefont {Chakrabarti},\ and\ \citenamefont {Rabitz}}]{brifControlQuantumPhenomena2010}%
  \BibitemOpen
  \bibfield  {author} {\bibinfo {author} {\bibfnamefont {C.}~\bibnamefont {Brif}}, \bibinfo {author} {\bibfnamefont {R.}~\bibnamefont {Chakrabarti}},\ and\ \bibinfo {author} {\bibfnamefont {H.}~\bibnamefont {Rabitz}},\ }\bibfield  {title} {\bibinfo {title} {Control of quantum phenomena: Past, present and future},\ }\href {https://doi.org/10.1088/1367-2630/12/7/075008} {\bibfield  {journal} {\bibinfo  {journal} {New Journal of Physics}\ }\textbf {\bibinfo {volume} {12}},\ \bibinfo {pages} {075008} (\bibinfo {year} {2010})}\BibitemShut {NoStop}%
\bibitem [{\citenamefont {Pechen}\ and\ \citenamefont {Tannor}(2011)}]{pechenAreThereTraps2011}%
  \BibitemOpen
  \bibfield  {author} {\bibinfo {author} {\bibfnamefont {A.~N.}\ \bibnamefont {Pechen}}\ and\ \bibinfo {author} {\bibfnamefont {D.~J.}\ \bibnamefont {Tannor}},\ }\bibfield  {title} {\bibinfo {title} {Are there {{Traps}} in {{Quantum Control Landscapes}}?},\ }\href {https://doi.org/10.1103/PhysRevLett.106.120402} {\bibfield  {journal} {\bibinfo  {journal} {Physical Review Letters}\ }\textbf {\bibinfo {volume} {106}},\ \bibinfo {pages} {120402} (\bibinfo {year} {2011})}\BibitemShut {NoStop}%
\bibitem [{\citenamefont {De~Fouquieres}\ and\ \citenamefont {Schirmer}(2013)}]{defouquieresCloserLookQuantum2013}%
  \BibitemOpen
  \bibfield  {author} {\bibinfo {author} {\bibfnamefont {P.}~\bibnamefont {De~Fouquieres}}\ and\ \bibinfo {author} {\bibfnamefont {S.~G.}\ \bibnamefont {Schirmer}},\ }\bibfield  {title} {\bibinfo {title} {A {{Closer Look}} at {{Quantum Control Landscapes}} \& their {{Implication}} for {{Control Optimization}}},\ }\href {https://doi.org/10.1142/S0219025713500215} {\bibfield  {journal} {\bibinfo  {journal} {Infinite Dimensional Analysis, Quantum Probability and Related Topics}\ }\textbf {\bibinfo {volume} {16}},\ \bibinfo {pages} {1350021} (\bibinfo {year} {2013})}\BibitemShut {NoStop}%
\bibitem [{\citenamefont {Birtea}\ \emph {et~al.}(2022)\citenamefont {Birtea}, \citenamefont {Ca{\c s}u},\ and\ \citenamefont {Com{\u a}nescu}}]{birteaConstraintOptimizationSU2022}%
  \BibitemOpen
  \bibfield  {author} {\bibinfo {author} {\bibfnamefont {P.}~\bibnamefont {Birtea}}, \bibinfo {author} {\bibfnamefont {I.}~\bibnamefont {Ca{\c s}u}},\ and\ \bibinfo {author} {\bibfnamefont {D.}~\bibnamefont {Com{\u a}nescu}},\ }\bibfield  {title} {\bibinfo {title} {Constraint optimization and {{SU}}({{N}}) quantum control landscapes},\ }\href {https://doi.org/10.1088/1751-8121/ac5189} {\bibfield  {journal} {\bibinfo  {journal} {Journal of Physics A: Mathematical and Theoretical}\ }\textbf {\bibinfo {volume} {55}},\ \bibinfo {pages} {115301} (\bibinfo {year} {2022})}\BibitemShut {NoStop}%
\bibitem [{\citenamefont {Rabitz}\ \emph {et~al.}(2012)\citenamefont {Rabitz}, \citenamefont {Ho}, \citenamefont {Long}, \citenamefont {Wu},\ and\ \citenamefont {Brif}}]{rabitzCommentAreThere2012}%
  \BibitemOpen
  \bibfield  {author} {\bibinfo {author} {\bibfnamefont {H.}~\bibnamefont {Rabitz}}, \bibinfo {author} {\bibfnamefont {T.-S.}\ \bibnamefont {Ho}}, \bibinfo {author} {\bibfnamefont {R.}~\bibnamefont {Long}}, \bibinfo {author} {\bibfnamefont {R.}~\bibnamefont {Wu}},\ and\ \bibinfo {author} {\bibfnamefont {C.}~\bibnamefont {Brif}},\ }\bibfield  {title} {\bibinfo {title} {Comment on ``{{Are There Traps}} in {{Quantum Control Landscapes}}?''},\ }\href {https://doi.org/10.1103/PhysRevLett.108.198901} {\bibfield  {journal} {\bibinfo  {journal} {Physical Review Letters}\ }\textbf {\bibinfo {volume} {108}},\ \bibinfo {pages} {198901} (\bibinfo {year} {2012})}\BibitemShut {NoStop}%
\bibitem [{\citenamefont {Wu}\ \emph {et~al.}(2012)\citenamefont {Wu}, \citenamefont {Dominy}, \citenamefont {Ho},\ and\ \citenamefont {Rabitz}}]{wuSingularitiesQuantumControl2012}%
  \BibitemOpen
  \bibfield  {author} {\bibinfo {author} {\bibfnamefont {R.}~\bibnamefont {Wu}}, \bibinfo {author} {\bibfnamefont {J.}~\bibnamefont {Dominy}}, \bibinfo {author} {\bibfnamefont {T.-S.}\ \bibnamefont {Ho}},\ and\ \bibinfo {author} {\bibfnamefont {H.}~\bibnamefont {Rabitz}},\ }\bibfield  {title} {\bibinfo {title} {Singularities of {{Quantum Control Landscapes}}},\ }\href {https://doi.org/10.1103/PhysRevA.86.013405} {\bibfield  {journal} {\bibinfo  {journal} {Physical Review A}\ }\textbf {\bibinfo {volume} {86}},\ \bibinfo {pages} {013405} (\bibinfo {year} {2012})},\ \Eprint {https://arxiv.org/abs/0907.2354} {arxiv:0907.2354 [quant-ph]} \BibitemShut {NoStop}%
\bibitem [{\citenamefont {Russell}\ \emph {et~al.}(2017)\citenamefont {Russell}, \citenamefont {Rabitz},\ and\ \citenamefont {Wu}}]{russellControlLandscapesAre2017}%
  \BibitemOpen
  \bibfield  {author} {\bibinfo {author} {\bibfnamefont {B.}~\bibnamefont {Russell}}, \bibinfo {author} {\bibfnamefont {H.}~\bibnamefont {Rabitz}},\ and\ \bibinfo {author} {\bibfnamefont {R.-B.}\ \bibnamefont {Wu}},\ }\bibfield  {title} {\bibinfo {title} {Control landscapes are almost always trap free: A geometric assessment},\ }\href {https://doi.org/10.1088/1751-8121/aa6b77} {\bibfield  {journal} {\bibinfo  {journal} {Journal of Physics A: Mathematical and Theoretical}\ }\textbf {\bibinfo {volume} {50}},\ \bibinfo {pages} {205302} (\bibinfo {year} {2017})}\BibitemShut {NoStop}%
\bibitem [{\citenamefont {Pechen}\ \emph {et~al.}(2008)\citenamefont {Pechen}, \citenamefont {Prokhorenko}, \citenamefont {Wu},\ and\ \citenamefont {Rabitz}}]{pechen2008control}%
  \BibitemOpen
  \bibfield  {author} {\bibinfo {author} {\bibfnamefont {A.}~\bibnamefont {Pechen}}, \bibinfo {author} {\bibfnamefont {D.}~\bibnamefont {Prokhorenko}}, \bibinfo {author} {\bibfnamefont {R.}~\bibnamefont {Wu}},\ and\ \bibinfo {author} {\bibfnamefont {H.}~\bibnamefont {Rabitz}},\ }\bibfield  {title} {\bibinfo {title} {Control landscapes for two-level open quantum systems},\ }\href@noop {} {\bibfield  {journal} {\bibinfo  {journal} {Journal of Physics A: Mathematical and Theoretical}\ }\textbf {\bibinfo {volume} {41}},\ \bibinfo {pages} {045205} (\bibinfo {year} {2008})}\BibitemShut {NoStop}%
\bibitem [{\citenamefont {Wu}\ \emph {et~al.}(2019)\citenamefont {Wu}, \citenamefont {Ding}, \citenamefont {Dong},\ and\ \citenamefont {Wang}}]{wu2019learning}%
  \BibitemOpen
  \bibfield  {author} {\bibinfo {author} {\bibfnamefont {R.-B.}\ \bibnamefont {Wu}}, \bibinfo {author} {\bibfnamefont {H.}~\bibnamefont {Ding}}, \bibinfo {author} {\bibfnamefont {D.}~\bibnamefont {Dong}},\ and\ \bibinfo {author} {\bibfnamefont {X.}~\bibnamefont {Wang}},\ }\bibfield  {title} {\bibinfo {title} {Learning robust and high-precision quantum controls},\ }\href@noop {} {\bibfield  {journal} {\bibinfo  {journal} {Physical Review A}\ }\textbf {\bibinfo {volume} {99}},\ \bibinfo {pages} {042327} (\bibinfo {year} {2019})}\BibitemShut {NoStop}%
\bibitem [{\citenamefont {Clarke}\ and\ \citenamefont {Wilhelm}(2008)}]{clarkeSuperconductingQuantumBits2008}%
  \BibitemOpen
  \bibfield  {author} {\bibinfo {author} {\bibfnamefont {J.}~\bibnamefont {Clarke}}\ and\ \bibinfo {author} {\bibfnamefont {F.~K.}\ \bibnamefont {Wilhelm}},\ }\bibfield  {title} {\bibinfo {title} {Superconducting quantum bits},\ }\href {https://doi.org/10.1038/nature07128} {\bibfield  {journal} {\bibinfo  {journal} {Nature}\ }\textbf {\bibinfo {volume} {453}},\ \bibinfo {pages} {1031} (\bibinfo {year} {2008})}\BibitemShut {NoStop}%
\bibitem [{\citenamefont {Krantz}\ \emph {et~al.}(2019)\citenamefont {Krantz}, \citenamefont {Kjaergaard}, \citenamefont {Yan}, \citenamefont {Orlando}, \citenamefont {Gustavsson},\ and\ \citenamefont {Oliver}}]{krantz2019quantum}%
  \BibitemOpen
  \bibfield  {author} {\bibinfo {author} {\bibfnamefont {P.}~\bibnamefont {Krantz}}, \bibinfo {author} {\bibfnamefont {M.}~\bibnamefont {Kjaergaard}}, \bibinfo {author} {\bibfnamefont {F.}~\bibnamefont {Yan}}, \bibinfo {author} {\bibfnamefont {T.~P.}\ \bibnamefont {Orlando}}, \bibinfo {author} {\bibfnamefont {S.}~\bibnamefont {Gustavsson}},\ and\ \bibinfo {author} {\bibfnamefont {W.~D.}\ \bibnamefont {Oliver}},\ }\bibfield  {title} {\bibinfo {title} {A quantum engineer's guide to superconducting qubits},\ }\href@noop {} {\bibfield  {journal} {\bibinfo  {journal} {Applied physics reviews}\ }\textbf {\bibinfo {volume} {6}} (\bibinfo {year} {2019})}\BibitemShut {NoStop}%
\bibitem [{\citenamefont {Koch}\ \emph {et~al.}(2007)\citenamefont {Koch}, \citenamefont {Yu}, \citenamefont {Gambetta}, \citenamefont {Houck}, \citenamefont {Schuster}, \citenamefont {Majer}, \citenamefont {Blais}, \citenamefont {Devoret}, \citenamefont {Girvin},\ and\ \citenamefont {Schoelkopf}}]{kochChargeinsensitiveQubitDesign2007}%
  \BibitemOpen
  \bibfield  {author} {\bibinfo {author} {\bibfnamefont {J.}~\bibnamefont {Koch}}, \bibinfo {author} {\bibfnamefont {T.~M.}\ \bibnamefont {Yu}}, \bibinfo {author} {\bibfnamefont {J.}~\bibnamefont {Gambetta}}, \bibinfo {author} {\bibfnamefont {A.~A.}\ \bibnamefont {Houck}}, \bibinfo {author} {\bibfnamefont {D.~I.}\ \bibnamefont {Schuster}}, \bibinfo {author} {\bibfnamefont {J.}~\bibnamefont {Majer}}, \bibinfo {author} {\bibfnamefont {A.}~\bibnamefont {Blais}}, \bibinfo {author} {\bibfnamefont {M.~H.}\ \bibnamefont {Devoret}}, \bibinfo {author} {\bibfnamefont {S.~M.}\ \bibnamefont {Girvin}},\ and\ \bibinfo {author} {\bibfnamefont {R.~J.}\ \bibnamefont {Schoelkopf}},\ }\bibfield  {title} {\bibinfo {title} {Charge-insensitive qubit design derived from the {{Cooper}} pair box},\ }\href {https://doi.org/10.1103/PhysRevA.76.042319} {\bibfield  {journal} {\bibinfo  {journal} {Physical Review A}\ }\textbf {\bibinfo {volume} {76}},\ \bibinfo {pages} {042319} (\bibinfo {year} {2007})}\BibitemShut {NoStop}%
\bibitem [{\citenamefont {Sarovar}\ \emph {et~al.}(2020)\citenamefont {Sarovar}, \citenamefont {Proctor}, \citenamefont {Rudinger}, \citenamefont {Young}, \citenamefont {Nielsen},\ and\ \citenamefont {Blume-Kohout}}]{sarovar2020detecting}%
  \BibitemOpen
  \bibfield  {author} {\bibinfo {author} {\bibfnamefont {M.}~\bibnamefont {Sarovar}}, \bibinfo {author} {\bibfnamefont {T.}~\bibnamefont {Proctor}}, \bibinfo {author} {\bibfnamefont {K.}~\bibnamefont {Rudinger}}, \bibinfo {author} {\bibfnamefont {K.}~\bibnamefont {Young}}, \bibinfo {author} {\bibfnamefont {E.}~\bibnamefont {Nielsen}},\ and\ \bibinfo {author} {\bibfnamefont {R.}~\bibnamefont {Blume-Kohout}},\ }\bibfield  {title} {\bibinfo {title} {Detecting crosstalk errors in quantum information processors},\ }\href@noop {} {\bibfield  {journal} {\bibinfo  {journal} {Quantum}\ }\textbf {\bibinfo {volume} {4}},\ \bibinfo {pages} {321} (\bibinfo {year} {2020})}\BibitemShut {NoStop}%
\bibitem [{\citenamefont {Rudinger}\ \emph {et~al.}(2021)\citenamefont {Rudinger}, \citenamefont {Hogle}, \citenamefont {Naik}, \citenamefont {Hashim}, \citenamefont {Lobser}, \citenamefont {Santiago}, \citenamefont {Grace}, \citenamefont {Nielsen}, \citenamefont {Proctor}, \citenamefont {Seritan} \emph {et~al.}}]{rudinger2021experimental}%
  \BibitemOpen
  \bibfield  {author} {\bibinfo {author} {\bibfnamefont {K.}~\bibnamefont {Rudinger}}, \bibinfo {author} {\bibfnamefont {C.~W.}\ \bibnamefont {Hogle}}, \bibinfo {author} {\bibfnamefont {R.~K.}\ \bibnamefont {Naik}}, \bibinfo {author} {\bibfnamefont {A.}~\bibnamefont {Hashim}}, \bibinfo {author} {\bibfnamefont {D.}~\bibnamefont {Lobser}}, \bibinfo {author} {\bibfnamefont {D.~I.}\ \bibnamefont {Santiago}}, \bibinfo {author} {\bibfnamefont {M.~D.}\ \bibnamefont {Grace}}, \bibinfo {author} {\bibfnamefont {E.}~\bibnamefont {Nielsen}}, \bibinfo {author} {\bibfnamefont {T.}~\bibnamefont {Proctor}}, \bibinfo {author} {\bibfnamefont {S.}~\bibnamefont {Seritan}}, \emph {et~al.},\ }\bibfield  {title} {\bibinfo {title} {Experimental characterization of crosstalk errors with simultaneous gate set tomography},\ }\href@noop {} {\bibfield  {journal} {\bibinfo  {journal} {PRX Quantum}\ }\textbf {\bibinfo {volume} {2}},\ \bibinfo {pages} {040338} (\bibinfo {year} {2021})}\BibitemShut {NoStop}%
\bibitem [{\citenamefont {Yi}\ \emph {et~al.}(2024)\citenamefont {Yi}, \citenamefont {Hai}, \citenamefont {Luo}, \citenamefont {Chu}, \citenamefont {Zhang}, \citenamefont {Zhou}, \citenamefont {Song}, \citenamefont {Liu}, \citenamefont {Yan}, \citenamefont {Deng} \emph {et~al.}}]{yi2024robust}%
  \BibitemOpen
  \bibfield  {author} {\bibinfo {author} {\bibfnamefont {K.}~\bibnamefont {Yi}}, \bibinfo {author} {\bibfnamefont {Y.-J.}\ \bibnamefont {Hai}}, \bibinfo {author} {\bibfnamefont {K.}~\bibnamefont {Luo}}, \bibinfo {author} {\bibfnamefont {J.}~\bibnamefont {Chu}}, \bibinfo {author} {\bibfnamefont {L.}~\bibnamefont {Zhang}}, \bibinfo {author} {\bibfnamefont {Y.}~\bibnamefont {Zhou}}, \bibinfo {author} {\bibfnamefont {Y.}~\bibnamefont {Song}}, \bibinfo {author} {\bibfnamefont {S.}~\bibnamefont {Liu}}, \bibinfo {author} {\bibfnamefont {T.}~\bibnamefont {Yan}}, \bibinfo {author} {\bibfnamefont {X.-H.}\ \bibnamefont {Deng}}, \emph {et~al.},\ }\bibfield  {title} {\bibinfo {title} {Robust quantum gates against correlated noise in integrated quantum chips},\ }\href@noop {} {\bibfield  {journal} {\bibinfo  {journal} {Physical Review Letters}\ }\textbf {\bibinfo {volume} {132}},\ \bibinfo {pages} {250604} (\bibinfo {year} {2024})}\BibitemShut {NoStop}%
\bibitem [{\citenamefont {Bylander}\ \emph {et~al.}(2011)\citenamefont {Bylander}, \citenamefont {Gustavsson}, \citenamefont {Yan}, \citenamefont {Yoshihara}, \citenamefont {Harrabi}, \citenamefont {Fitch}, \citenamefont {Cory}, \citenamefont {Nakamura}, \citenamefont {Tsai},\ and\ \citenamefont {Oliver}}]{Bylander2011}%
  \BibitemOpen
  \bibfield  {author} {\bibinfo {author} {\bibfnamefont {J.}~\bibnamefont {Bylander}}, \bibinfo {author} {\bibfnamefont {S.}~\bibnamefont {Gustavsson}}, \bibinfo {author} {\bibfnamefont {F.}~\bibnamefont {Yan}}, \bibinfo {author} {\bibfnamefont {F.}~\bibnamefont {Yoshihara}}, \bibinfo {author} {\bibfnamefont {K.}~\bibnamefont {Harrabi}}, \bibinfo {author} {\bibfnamefont {G.}~\bibnamefont {Fitch}}, \bibinfo {author} {\bibfnamefont {D.~G.}\ \bibnamefont {Cory}}, \bibinfo {author} {\bibfnamefont {Y.}~\bibnamefont {Nakamura}}, \bibinfo {author} {\bibfnamefont {J.-S.}\ \bibnamefont {Tsai}},\ and\ \bibinfo {author} {\bibfnamefont {W.~D.}\ \bibnamefont {Oliver}},\ }\bibfield  {title} {\bibinfo {title} {Noise spectroscopy through dynamical decoupling with a superconducting flux qubit},\ }\href {https://doi.org/10.1038/nphys1994} {\bibfield  {journal} {\bibinfo  {journal} {Nature Physics}\ }\textbf {\bibinfo {volume} {7}},\ \bibinfo {pages} {565} (\bibinfo {year} {2011})}\BibitemShut {NoStop}%
\bibitem [{\citenamefont {Brown}\ \emph {et~al.}(2004)\citenamefont {Brown}, \citenamefont {Harrow},\ and\ \citenamefont {Chuang}}]{brownArbitrarilyAccurateComposite2004}%
  \BibitemOpen
  \bibfield  {author} {\bibinfo {author} {\bibfnamefont {K.~R.}\ \bibnamefont {Brown}}, \bibinfo {author} {\bibfnamefont {A.~W.}\ \bibnamefont {Harrow}},\ and\ \bibinfo {author} {\bibfnamefont {I.~L.}\ \bibnamefont {Chuang}},\ }\bibfield  {title} {\bibinfo {title} {Arbitrarily accurate composite pulse sequences},\ }\href {https://doi.org/10.1103/PhysRevA.70.052318} {\bibfield  {journal} {\bibinfo  {journal} {Physical Review A}\ }\textbf {\bibinfo {volume} {70}},\ \bibinfo {pages} {052318} (\bibinfo {year} {2004})}\BibitemShut {NoStop}%
\bibitem [{\citenamefont {Shi}\ \emph {et~al.}(2024)\citenamefont {Shi}, \citenamefont {Ding}, \citenamefont {Chen}, \citenamefont {Song}, \citenamefont {Xia}, \citenamefont {Yi},\ and\ \citenamefont {Nori}}]{shi2024supervised}%
  \BibitemOpen
  \bibfield  {author} {\bibinfo {author} {\bibfnamefont {Z.-C.}\ \bibnamefont {Shi}}, \bibinfo {author} {\bibfnamefont {J.-T.}\ \bibnamefont {Ding}}, \bibinfo {author} {\bibfnamefont {Y.-H.}\ \bibnamefont {Chen}}, \bibinfo {author} {\bibfnamefont {J.}~\bibnamefont {Song}}, \bibinfo {author} {\bibfnamefont {Y.}~\bibnamefont {Xia}}, \bibinfo {author} {\bibfnamefont {X.}~\bibnamefont {Yi}},\ and\ \bibinfo {author} {\bibfnamefont {F.}~\bibnamefont {Nori}},\ }\bibfield  {title} {\bibinfo {title} {Supervised learning for robust quantum control in composite-pulse systems},\ }\href@noop {} {\bibfield  {journal} {\bibinfo  {journal} {Physical Review Applied}\ }\textbf {\bibinfo {volume} {21}},\ \bibinfo {pages} {044012} (\bibinfo {year} {2024})}\BibitemShut {NoStop}%
\bibitem [{\citenamefont {Zhang}\ \emph {et~al.}(2024)\citenamefont {Zhang}, \citenamefont {Miao}, \citenamefont {Chen},\ and\ \citenamefont {Deng}}]{zhang2024smolyak}%
  \BibitemOpen
  \bibfield  {author} {\bibinfo {author} {\bibfnamefont {Z.}~\bibnamefont {Zhang}}, \bibinfo {author} {\bibfnamefont {Z.}~\bibnamefont {Miao}}, \bibinfo {author} {\bibfnamefont {Y.}~\bibnamefont {Chen}},\ and\ \bibinfo {author} {\bibfnamefont {X.-H.}\ \bibnamefont {Deng}},\ }\bibfield  {title} {\bibinfo {title} {Smolyak algorithm assisted robust control for quantum systems with uncertainties},\ }\href@noop {} {\bibfield  {journal} {\bibinfo  {journal} {arXiv preprint arXiv:2410.14286}\ } (\bibinfo {year} {2024})}\BibitemShut {NoStop}%
\bibitem [{\citenamefont {Li}\ \emph {et~al.}(2024)\citenamefont {Li}, \citenamefont {Calarco},\ and\ \citenamefont {Motzoi}}]{li2024experimental}%
  \BibitemOpen
  \bibfield  {author} {\bibinfo {author} {\bibfnamefont {B.}~\bibnamefont {Li}}, \bibinfo {author} {\bibfnamefont {T.}~\bibnamefont {Calarco}},\ and\ \bibinfo {author} {\bibfnamefont {F.}~\bibnamefont {Motzoi}},\ }\bibfield  {title} {\bibinfo {title} {Experimental error suppression in cross-resonance gates via multi-derivative pulse shaping},\ }\href@noop {} {\bibfield  {journal} {\bibinfo  {journal} {npj Quantum Information}\ }\textbf {\bibinfo {volume} {10}},\ \bibinfo {pages} {66} (\bibinfo {year} {2024})}\BibitemShut {NoStop}%
\bibitem [{\citenamefont {Barnes}\ \emph {et~al.}(2022)\citenamefont {Barnes}, \citenamefont {{Calderon-Vargas}}, \citenamefont {Dong}, \citenamefont {Li}, \citenamefont {Zeng},\ and\ \citenamefont {Zhuang}}]{barnesDynamicallyCorrectedGates2022}%
  \BibitemOpen
  \bibfield  {author} {\bibinfo {author} {\bibfnamefont {E.}~\bibnamefont {Barnes}}, \bibinfo {author} {\bibfnamefont {F.~A.}\ \bibnamefont {{Calderon-Vargas}}}, \bibinfo {author} {\bibfnamefont {W.}~\bibnamefont {Dong}}, \bibinfo {author} {\bibfnamefont {B.}~\bibnamefont {Li}}, \bibinfo {author} {\bibfnamefont {J.}~\bibnamefont {Zeng}},\ and\ \bibinfo {author} {\bibfnamefont {F.}~\bibnamefont {Zhuang}},\ }\bibfield  {title} {\bibinfo {title} {Dynamically corrected gates from geometric space curves},\ }\href {https://doi.org/10.1088/2058-9565/ac4421} {\bibfield  {journal} {\bibinfo  {journal} {Quantum Science and Technology}\ }\textbf {\bibinfo {volume} {7}},\ \bibinfo {pages} {023001} (\bibinfo {year} {2022})}\BibitemShut {NoStop}%
\bibitem [{\citenamefont {Dong}\ \emph {et~al.}(2021)\citenamefont {Dong}, \citenamefont {Zhuang}, \citenamefont {Economou},\ and\ \citenamefont {Barnes}}]{dong2021doubly}%
  \BibitemOpen
  \bibfield  {author} {\bibinfo {author} {\bibfnamefont {W.}~\bibnamefont {Dong}}, \bibinfo {author} {\bibfnamefont {F.}~\bibnamefont {Zhuang}}, \bibinfo {author} {\bibfnamefont {S.~E.}\ \bibnamefont {Economou}},\ and\ \bibinfo {author} {\bibfnamefont {E.}~\bibnamefont {Barnes}},\ }\bibfield  {title} {\bibinfo {title} {Doubly geometric quantum control},\ }\href@noop {} {\bibfield  {journal} {\bibinfo  {journal} {PRX Quantum}\ }\textbf {\bibinfo {volume} {2}},\ \bibinfo {pages} {030333} (\bibinfo {year} {2021})}\BibitemShut {NoStop}%
\bibitem [{\citenamefont {Hai}\ \emph {et~al.}(2023)\citenamefont {Hai}, \citenamefont {Li}, \citenamefont {Zeng}, \citenamefont {Yu},\ and\ \citenamefont {Deng}}]{haiUniversalRobustQuantum2023}%
  \BibitemOpen
  \bibfield  {author} {\bibinfo {author} {\bibfnamefont {Y.-J.}\ \bibnamefont {Hai}}, \bibinfo {author} {\bibfnamefont {J.}~\bibnamefont {Li}}, \bibinfo {author} {\bibfnamefont {J.}~\bibnamefont {Zeng}}, \bibinfo {author} {\bibfnamefont {D.}~\bibnamefont {Yu}},\ and\ \bibinfo {author} {\bibfnamefont {X.-H.}\ \bibnamefont {Deng}},\ }\href@noop {} {\bibinfo {title} {Universal robust quantum gates by geometric correspondence of noisy quantum evolution}} (\bibinfo {year} {2023}),\ \Eprint {https://arxiv.org/abs/2210.14521} {arxiv:2210.14521 [quant-ph]} \BibitemShut {NoStop}%
\bibitem [{\citenamefont {Sauvage}\ and\ \citenamefont {Mintert}(2022)}]{sauvage2022optimal}%
  \BibitemOpen
  \bibfield  {author} {\bibinfo {author} {\bibfnamefont {F.}~\bibnamefont {Sauvage}}\ and\ \bibinfo {author} {\bibfnamefont {F.}~\bibnamefont {Mintert}},\ }\bibfield  {title} {\bibinfo {title} {Optimal control of families of quantum gates},\ }\href@noop {} {\bibfield  {journal} {\bibinfo  {journal} {Physical Review Letters}\ }\textbf {\bibinfo {volume} {129}},\ \bibinfo {pages} {050507} (\bibinfo {year} {2022})}\BibitemShut {NoStop}%
\bibitem [{\citenamefont {Barnes}\ and\ \citenamefont {Zeng}(2022)}]{barnes2022generating}%
  \BibitemOpen
  \bibfield  {author} {\bibinfo {author} {\bibfnamefont {E.}~\bibnamefont {Barnes}}\ and\ \bibinfo {author} {\bibfnamefont {J.}~\bibnamefont {Zeng}},\ }\href@noop {} {\bibinfo {title} {Generating error-resistant quantum control pulses from geometrical curves}} (\bibinfo {year} {2022}),\ \bibinfo {note} {uS Patent App. 17/765,875}\BibitemShut {NoStop}%
\bibitem [{\citenamefont {Foxen}\ \emph {et~al.}(2020)\citenamefont {Foxen}, \citenamefont {Neill}, \citenamefont {Dunsworth}, \citenamefont {Roushan}, \citenamefont {Chiaro}, \citenamefont {Megrant}, \citenamefont {Kelly}, \citenamefont {Chen}, \citenamefont {Satzinger}, \citenamefont {Barends}, \citenamefont {Arute}, \citenamefont {Arya}, \citenamefont {Babbush}, \citenamefont {Bacon}, \citenamefont {Bardin}, \citenamefont {Boixo}, \citenamefont {Buell}, \citenamefont {Burkett}, \citenamefont {Chen}, \citenamefont {Collins}, \citenamefont {Farhi}, \citenamefont {Fowler}, \citenamefont {Gidney}, \citenamefont {Giustina}, \citenamefont {Graff}, \citenamefont {Harrigan}, \citenamefont {Huang}, \citenamefont {Isakov}, \citenamefont {Jeffrey}, \citenamefont {Jiang}, \citenamefont {Kafri}, \citenamefont {Kechedzhi}, \citenamefont {Klimov}, \citenamefont {Korotkov}, \citenamefont {Kostritsa}, \citenamefont {Landhuis}, \citenamefont {Lucero}, \citenamefont {McClean}, \citenamefont {McEwen}, \citenamefont {Mi},
  \citenamefont {Mohseni}, \citenamefont {Mutus}, \citenamefont {Naaman}, \citenamefont {Neeley}, \citenamefont {Niu}, \citenamefont {Petukhov}, \citenamefont {Quintana}, \citenamefont {Rubin}, \citenamefont {Sank}, \citenamefont {Smelyanskiy}, \citenamefont {Vainsencher}, \citenamefont {White}, \citenamefont {Yao}, \citenamefont {Yeh}, \citenamefont {Zalcman}, \citenamefont {Neven},\ and\ \citenamefont {Martinis}}]{demoContinuousTwoQubitGates}%
  \BibitemOpen
  \bibfield  {author} {\bibinfo {author} {\bibfnamefont {B.}~\bibnamefont {Foxen}}, \bibinfo {author} {\bibfnamefont {C.}~\bibnamefont {Neill}}, \bibinfo {author} {\bibfnamefont {A.}~\bibnamefont {Dunsworth}}, \bibinfo {author} {\bibfnamefont {P.}~\bibnamefont {Roushan}}, \bibinfo {author} {\bibfnamefont {B.}~\bibnamefont {Chiaro}}, \bibinfo {author} {\bibfnamefont {A.}~\bibnamefont {Megrant}}, \bibinfo {author} {\bibfnamefont {J.}~\bibnamefont {Kelly}}, \bibinfo {author} {\bibfnamefont {Z.}~\bibnamefont {Chen}}, \bibinfo {author} {\bibfnamefont {K.}~\bibnamefont {Satzinger}}, \bibinfo {author} {\bibfnamefont {R.}~\bibnamefont {Barends}}, \bibinfo {author} {\bibfnamefont {F.}~\bibnamefont {Arute}}, \bibinfo {author} {\bibfnamefont {K.}~\bibnamefont {Arya}}, \bibinfo {author} {\bibfnamefont {R.}~\bibnamefont {Babbush}}, \bibinfo {author} {\bibfnamefont {D.}~\bibnamefont {Bacon}}, \bibinfo {author} {\bibfnamefont {J.~C.}\ \bibnamefont {Bardin}}, \bibinfo {author} {\bibfnamefont {S.}~\bibnamefont {Boixo}},
  \bibinfo {author} {\bibfnamefont {D.}~\bibnamefont {Buell}}, \bibinfo {author} {\bibfnamefont {B.}~\bibnamefont {Burkett}}, \bibinfo {author} {\bibfnamefont {Y.}~\bibnamefont {Chen}}, \bibinfo {author} {\bibfnamefont {R.}~\bibnamefont {Collins}}, \bibinfo {author} {\bibfnamefont {E.}~\bibnamefont {Farhi}}, \bibinfo {author} {\bibfnamefont {A.}~\bibnamefont {Fowler}}, \bibinfo {author} {\bibfnamefont {C.}~\bibnamefont {Gidney}}, \bibinfo {author} {\bibfnamefont {M.}~\bibnamefont {Giustina}}, \bibinfo {author} {\bibfnamefont {R.}~\bibnamefont {Graff}}, \bibinfo {author} {\bibfnamefont {M.}~\bibnamefont {Harrigan}}, \bibinfo {author} {\bibfnamefont {T.}~\bibnamefont {Huang}}, \bibinfo {author} {\bibfnamefont {S.~V.}\ \bibnamefont {Isakov}}, \bibinfo {author} {\bibfnamefont {E.}~\bibnamefont {Jeffrey}}, \bibinfo {author} {\bibfnamefont {Z.}~\bibnamefont {Jiang}}, \bibinfo {author} {\bibfnamefont {D.}~\bibnamefont {Kafri}}, \bibinfo {author} {\bibfnamefont {K.}~\bibnamefont {Kechedzhi}}, \bibinfo {author}
  {\bibfnamefont {P.}~\bibnamefont {Klimov}}, \bibinfo {author} {\bibfnamefont {A.}~\bibnamefont {Korotkov}}, \bibinfo {author} {\bibfnamefont {F.}~\bibnamefont {Kostritsa}}, \bibinfo {author} {\bibfnamefont {D.}~\bibnamefont {Landhuis}}, \bibinfo {author} {\bibfnamefont {E.}~\bibnamefont {Lucero}}, \bibinfo {author} {\bibfnamefont {J.}~\bibnamefont {McClean}}, \bibinfo {author} {\bibfnamefont {M.}~\bibnamefont {McEwen}}, \bibinfo {author} {\bibfnamefont {X.}~\bibnamefont {Mi}}, \bibinfo {author} {\bibfnamefont {M.}~\bibnamefont {Mohseni}}, \bibinfo {author} {\bibfnamefont {J.~Y.}\ \bibnamefont {Mutus}}, \bibinfo {author} {\bibfnamefont {O.}~\bibnamefont {Naaman}}, \bibinfo {author} {\bibfnamefont {M.}~\bibnamefont {Neeley}}, \bibinfo {author} {\bibfnamefont {M.}~\bibnamefont {Niu}}, \bibinfo {author} {\bibfnamefont {A.}~\bibnamefont {Petukhov}}, \bibinfo {author} {\bibfnamefont {C.}~\bibnamefont {Quintana}}, \bibinfo {author} {\bibfnamefont {N.}~\bibnamefont {Rubin}}, \bibinfo {author} {\bibfnamefont
  {D.}~\bibnamefont {Sank}}, \bibinfo {author} {\bibfnamefont {V.}~\bibnamefont {Smelyanskiy}}, \bibinfo {author} {\bibfnamefont {A.}~\bibnamefont {Vainsencher}}, \bibinfo {author} {\bibfnamefont {T.~C.}\ \bibnamefont {White}}, \bibinfo {author} {\bibfnamefont {Z.}~\bibnamefont {Yao}}, \bibinfo {author} {\bibfnamefont {P.}~\bibnamefont {Yeh}}, \bibinfo {author} {\bibfnamefont {A.}~\bibnamefont {Zalcman}}, \bibinfo {author} {\bibfnamefont {H.}~\bibnamefont {Neven}},\ and\ \bibinfo {author} {\bibfnamefont {J.~M.}\ \bibnamefont {Martinis}} (\bibinfo {collaboration} {Google AI Quantum}),\ }\bibfield  {title} {\bibinfo {title} {Demonstrating a continuous set of two-qubit gates for near-term quantum algorithms},\ }\href {https://doi.org/10.1103/PhysRevLett.125.120504} {\bibfield  {journal} {\bibinfo  {journal} {Phys. Rev. Lett.}\ }\textbf {\bibinfo {volume} {125}},\ \bibinfo {pages} {120504} (\bibinfo {year} {2020})}\BibitemShut {NoStop}%
\end{thebibliography}%

\end{document}